\documentclass[longauth]{aa}

\usepackage{graphicx}
\usepackage{natbib}
\usepackage{scalerel}
\usepackage{mathtools}
\usepackage{lineno}

\usepackage[table]{xcolor}

\usepackage{txfonts}
\usepackage[pdfencoding=auto,psdextra]{hyperref}
\hypersetup{
    colorlinks=true,
    linkcolor=blue,
    filecolor=magenta,      
    urlcolor=blue,
    citecolor=blue
}

\makeatletter
\renewcommand*\aa@pageof{, page \thepage{} of \pageref*{LastPage}}
\makeatother

\usepackage[utf8]{inputenc}

\usepackage{euclid}

\usepackage{orcidlink}

\usepackage[normalem]{ulem}

\begin{document}
%
%

\title{\Euclid preparation}
\subtitle{LXXXIX. Accurate and precise data-driven angular power spectrum covariances}



\DeclareRobustCommand{\orcid}[1]{\orcidlink{#1}}   		   
\author{Euclid Collaboration: K.~Naidoo\orcid{0000-0002-9182-1802}\thanks{\email{krishna.naidoo.11@ucl.ac.uk}}\inst{\ref{aff1},\ref{aff2}}
\and J.~Ruiz-Zapatero\orcid{0000-0002-7951-4391}\inst{\ref{aff1}}
\and N.~Tessore\orcid{0000-0002-9696-7931}\inst{\ref{aff1}}
\and B.~Joachimi\orcid{0000-0001-7494-1303}\inst{\ref{aff1}}
\and A.~Loureiro\orcid{0000-0002-4371-0876}\inst{\ref{aff3},\ref{aff4}}
\and N.~Aghanim\orcid{0000-0002-6688-8992}\inst{\ref{aff5}}
\and B.~Altieri\orcid{0000-0003-3936-0284}\inst{\ref{aff6}}
\and A.~Amara\inst{\ref{aff7}}
\and L.~Amendola\orcid{0000-0002-0835-233X}\inst{\ref{aff8}}
\and S.~Andreon\orcid{0000-0002-2041-8784}\inst{\ref{aff9}}
\and N.~Auricchio\orcid{0000-0003-4444-8651}\inst{\ref{aff10}}
\and C.~Baccigalupi\orcid{0000-0002-8211-1630}\inst{\ref{aff11},\ref{aff12},\ref{aff13},\ref{aff14}}
\and D.~Bagot\inst{\ref{aff15}}
\and M.~Baldi\orcid{0000-0003-4145-1943}\inst{\ref{aff16},\ref{aff10},\ref{aff17}}
\and S.~Bardelli\orcid{0000-0002-8900-0298}\inst{\ref{aff10}}
\and P.~Battaglia\orcid{0000-0002-7337-5909}\inst{\ref{aff10}}
\and A.~Biviano\orcid{0000-0002-0857-0732}\inst{\ref{aff12},\ref{aff11}}
\and E.~Branchini\orcid{0000-0002-0808-6908}\inst{\ref{aff18},\ref{aff19},\ref{aff9}}
\and M.~Brescia\orcid{0000-0001-9506-5680}\inst{\ref{aff20},\ref{aff21}}
\and S.~Camera\orcid{0000-0003-3399-3574}\inst{\ref{aff22},\ref{aff23},\ref{aff24}}
\and V.~Capobianco\orcid{0000-0002-3309-7692}\inst{\ref{aff24}}
\and C.~Carbone\orcid{0000-0003-0125-3563}\inst{\ref{aff25}}
\and V.~F.~Cardone\inst{\ref{aff26},\ref{aff27}}
\and J.~Carretero\orcid{0000-0002-3130-0204}\inst{\ref{aff28},\ref{aff29}}
\and M.~Castellano\orcid{0000-0001-9875-8263}\inst{\ref{aff26}}
\and G.~Castignani\orcid{0000-0001-6831-0687}\inst{\ref{aff10}}
\and S.~Cavuoti\orcid{0000-0002-3787-4196}\inst{\ref{aff21},\ref{aff30}}
\and K.~C.~Chambers\orcid{0000-0001-6965-7789}\inst{\ref{aff31}}
\and A.~Cimatti\inst{\ref{aff32}}
\and C.~Colodro-Conde\inst{\ref{aff33}}
\and G.~Congedo\orcid{0000-0003-2508-0046}\inst{\ref{aff34}}
\and L.~Conversi\orcid{0000-0002-6710-8476}\inst{\ref{aff35},\ref{aff6}}
\and Y.~Copin\orcid{0000-0002-5317-7518}\inst{\ref{aff36}}
\and F.~Courbin\orcid{0000-0003-0758-6510}\inst{\ref{aff37},\ref{aff38}}
\and H.~M.~Courtois\orcid{0000-0003-0509-1776}\inst{\ref{aff39}}
\and A.~Da~Silva\orcid{0000-0002-6385-1609}\inst{\ref{aff40},\ref{aff41}}
\and H.~Degaudenzi\orcid{0000-0002-5887-6799}\inst{\ref{aff42}}
\and G.~De~Lucia\orcid{0000-0002-6220-9104}\inst{\ref{aff12}}
\and F.~Dubath\orcid{0000-0002-6533-2810}\inst{\ref{aff42}}
\and X.~Dupac\inst{\ref{aff6}}
\and S.~Dusini\orcid{0000-0002-1128-0664}\inst{\ref{aff43}}
\and S.~Escoffier\orcid{0000-0002-2847-7498}\inst{\ref{aff44}}
\and M.~Farina\orcid{0000-0002-3089-7846}\inst{\ref{aff45}}
\and R.~Farinelli\inst{\ref{aff10}}
\and S.~Farrens\orcid{0000-0002-9594-9387}\inst{\ref{aff46}}
\and F.~Faustini\orcid{0000-0001-6274-5145}\inst{\ref{aff26},\ref{aff47}}
\and S.~Ferriol\inst{\ref{aff36}}
\and F.~Finelli\orcid{0000-0002-6694-3269}\inst{\ref{aff10},\ref{aff48}}
\and P.~Fosalba\orcid{0000-0002-1510-5214}\inst{\ref{aff49},\ref{aff50}}
\and M.~Frailis\orcid{0000-0002-7400-2135}\inst{\ref{aff12}}
\and E.~Franceschi\orcid{0000-0002-0585-6591}\inst{\ref{aff10}}
\and M.~Fumana\orcid{0000-0001-6787-5950}\inst{\ref{aff25}}
\and S.~Galeotta\orcid{0000-0002-3748-5115}\inst{\ref{aff12}}
\and K.~George\orcid{0000-0002-1734-8455}\inst{\ref{aff51}}
\and B.~Gillis\orcid{0000-0002-4478-1270}\inst{\ref{aff34}}
\and C.~Giocoli\orcid{0000-0002-9590-7961}\inst{\ref{aff10},\ref{aff17}}
\and J.~Gracia-Carpio\inst{\ref{aff52}}
\and A.~Grazian\orcid{0000-0002-5688-0663}\inst{\ref{aff53}}
\and F.~Grupp\inst{\ref{aff52},\ref{aff51}}
\and W.~Holmes\inst{\ref{aff54}}
\and F.~Hormuth\inst{\ref{aff55}}
\and A.~Hornstrup\orcid{0000-0002-3363-0936}\inst{\ref{aff56},\ref{aff57}}
\and K.~Jahnke\orcid{0000-0003-3804-2137}\inst{\ref{aff58}}
\and M.~Jhabvala\inst{\ref{aff59}}
\and E.~Keih\"anen\orcid{0000-0003-1804-7715}\inst{\ref{aff60}}
\and S.~Kermiche\orcid{0000-0002-0302-5735}\inst{\ref{aff44}}
\and A.~Kiessling\orcid{0000-0002-2590-1273}\inst{\ref{aff54}}
\and M.~Kilbinger\orcid{0000-0001-9513-7138}\inst{\ref{aff46}}
\and B.~Kubik\orcid{0009-0006-5823-4880}\inst{\ref{aff36}}
\and M.~K\"ummel\orcid{0000-0003-2791-2117}\inst{\ref{aff51}}
\and M.~Kunz\orcid{0000-0002-3052-7394}\inst{\ref{aff61}}
\and H.~Kurki-Suonio\orcid{0000-0002-4618-3063}\inst{\ref{aff62},\ref{aff63}}
\and A.~M.~C.~Le~Brun\orcid{0000-0002-0936-4594}\inst{\ref{aff64}}
\and S.~Ligori\orcid{0000-0003-4172-4606}\inst{\ref{aff24}}
\and P.~B.~Lilje\orcid{0000-0003-4324-7794}\inst{\ref{aff65}}
\and V.~Lindholm\orcid{0000-0003-2317-5471}\inst{\ref{aff62},\ref{aff63}}
\and I.~Lloro\orcid{0000-0001-5966-1434}\inst{\ref{aff66}}
\and G.~Mainetti\orcid{0000-0003-2384-2377}\inst{\ref{aff67}}
\and D.~Maino\inst{\ref{aff68},\ref{aff25},\ref{aff69}}
\and E.~Maiorano\orcid{0000-0003-2593-4355}\inst{\ref{aff10}}
\and O.~Mansutti\orcid{0000-0001-5758-4658}\inst{\ref{aff12}}
\and S.~Marcin\inst{\ref{aff70}}
\and O.~Marggraf\orcid{0000-0001-7242-3852}\inst{\ref{aff71}}
\and M.~Martinelli\orcid{0000-0002-6943-7732}\inst{\ref{aff26},\ref{aff27}}
\and N.~Martinet\orcid{0000-0003-2786-7790}\inst{\ref{aff72}}
\and F.~Marulli\orcid{0000-0002-8850-0303}\inst{\ref{aff73},\ref{aff10},\ref{aff17}}
\and R.~Massey\orcid{0000-0002-6085-3780}\inst{\ref{aff74}}
\and E.~Medinaceli\orcid{0000-0002-4040-7783}\inst{\ref{aff10}}
\and S.~Mei\orcid{0000-0002-2849-559X}\inst{\ref{aff75},\ref{aff76}}
\and Y.~Mellier\inst{\ref{aff77},\ref{aff78}}
\and M.~Meneghetti\orcid{0000-0003-1225-7084}\inst{\ref{aff10},\ref{aff17}}
\and E.~Merlin\orcid{0000-0001-6870-8900}\inst{\ref{aff26}}
\and G.~Meylan\inst{\ref{aff79}}
\and A.~Mora\orcid{0000-0002-1922-8529}\inst{\ref{aff80}}
\and L.~Moscardini\orcid{0000-0002-3473-6716}\inst{\ref{aff73},\ref{aff10},\ref{aff17}}
\and C.~Neissner\orcid{0000-0001-8524-4968}\inst{\ref{aff81},\ref{aff29}}
\and S.-M.~Niemi\orcid{0009-0005-0247-0086}\inst{\ref{aff82}}
\and C.~Padilla\orcid{0000-0001-7951-0166}\inst{\ref{aff81}}
\and S.~Paltani\orcid{0000-0002-8108-9179}\inst{\ref{aff42}}
\and F.~Pasian\orcid{0000-0002-4869-3227}\inst{\ref{aff12}}
\and K.~Pedersen\inst{\ref{aff83}}
\and W.~J.~Percival\orcid{0000-0002-0644-5727}\inst{\ref{aff84},\ref{aff85},\ref{aff86}}
\and V.~Pettorino\inst{\ref{aff82}}
\and S.~Pires\orcid{0000-0002-0249-2104}\inst{\ref{aff46}}
\and G.~Polenta\orcid{0000-0003-4067-9196}\inst{\ref{aff47}}
\and M.~Poncet\inst{\ref{aff15}}
\and L.~A.~Popa\inst{\ref{aff87}}
\and F.~Raison\orcid{0000-0002-7819-6918}\inst{\ref{aff52}}
\and R.~Rebolo\orcid{0000-0003-3767-7085}\inst{\ref{aff33},\ref{aff88},\ref{aff89}}
\and A.~Renzi\orcid{0000-0001-9856-1970}\inst{\ref{aff90},\ref{aff43}}
\and J.~Rhodes\orcid{0000-0002-4485-8549}\inst{\ref{aff54}}
\and G.~Riccio\inst{\ref{aff21}}
\and E.~Romelli\orcid{0000-0003-3069-9222}\inst{\ref{aff12}}
\and M.~Roncarelli\orcid{0000-0001-9587-7822}\inst{\ref{aff10}}
\and C.~Rosset\orcid{0000-0003-0286-2192}\inst{\ref{aff75}}
\and R.~Saglia\orcid{0000-0003-0378-7032}\inst{\ref{aff51},\ref{aff52}}
\and Z.~Sakr\orcid{0000-0002-4823-3757}\inst{\ref{aff8},\ref{aff91},\ref{aff92}}
\and A.~G.~S\'anchez\orcid{0000-0003-1198-831X}\inst{\ref{aff52}}
\and D.~Sapone\orcid{0000-0001-7089-4503}\inst{\ref{aff93}}
\and B.~Sartoris\orcid{0000-0003-1337-5269}\inst{\ref{aff51},\ref{aff12}}
\and P.~Schneider\orcid{0000-0001-8561-2679}\inst{\ref{aff71}}
\and T.~Schrabback\orcid{0000-0002-6987-7834}\inst{\ref{aff94}}
\and A.~Secroun\orcid{0000-0003-0505-3710}\inst{\ref{aff44}}
\and E.~Sefusatti\orcid{0000-0003-0473-1567}\inst{\ref{aff12},\ref{aff11},\ref{aff13}}
\and G.~Seidel\orcid{0000-0003-2907-353X}\inst{\ref{aff58}}
\and M.~Seiffert\orcid{0000-0002-7536-9393}\inst{\ref{aff54}}
\and S.~Serrano\orcid{0000-0002-0211-2861}\inst{\ref{aff49},\ref{aff95},\ref{aff50}}
\and P.~Simon\inst{\ref{aff71}}
\and C.~Sirignano\orcid{0000-0002-0995-7146}\inst{\ref{aff90},\ref{aff43}}
\and G.~Sirri\orcid{0000-0003-2626-2853}\inst{\ref{aff17}}
\and A.~Spurio~Mancini\orcid{0000-0001-5698-0990}\inst{\ref{aff96}}
\and L.~Stanco\orcid{0000-0002-9706-5104}\inst{\ref{aff43}}
\and J.~Steinwagner\orcid{0000-0001-7443-1047}\inst{\ref{aff52}}
\and P.~Tallada-Cresp\'{i}\orcid{0000-0002-1336-8328}\inst{\ref{aff28},\ref{aff29}}
\and D.~Tavagnacco\orcid{0000-0001-7475-9894}\inst{\ref{aff12}}
\and A.~N.~Taylor\inst{\ref{aff34}}
\and I.~Tereno\orcid{0000-0002-4537-6218}\inst{\ref{aff40},\ref{aff97}}
\and S.~Toft\orcid{0000-0003-3631-7176}\inst{\ref{aff98},\ref{aff99}}
\and R.~Toledo-Moreo\orcid{0000-0002-2997-4859}\inst{\ref{aff100}}
\and F.~Torradeflot\orcid{0000-0003-1160-1517}\inst{\ref{aff29},\ref{aff28}}
\and I.~Tutusaus\orcid{0000-0002-3199-0399}\inst{\ref{aff91}}
\and L.~Valenziano\orcid{0000-0002-1170-0104}\inst{\ref{aff10},\ref{aff48}}
\and J.~Valiviita\orcid{0000-0001-6225-3693}\inst{\ref{aff62},\ref{aff63}}
\and T.~Vassallo\orcid{0000-0001-6512-6358}\inst{\ref{aff51},\ref{aff12}}
\and G.~Verdoes~Kleijn\orcid{0000-0001-5803-2580}\inst{\ref{aff101}}
\and A.~Veropalumbo\orcid{0000-0003-2387-1194}\inst{\ref{aff9},\ref{aff19},\ref{aff18}}
\and Y.~Wang\orcid{0000-0002-4749-2984}\inst{\ref{aff102}}
\and J.~Weller\orcid{0000-0002-8282-2010}\inst{\ref{aff51},\ref{aff52}}
\and G.~Zamorani\orcid{0000-0002-2318-301X}\inst{\ref{aff10}}
\and F.~M.~Zerbi\inst{\ref{aff9}}
\and E.~Zucca\orcid{0000-0002-5845-8132}\inst{\ref{aff10}}
\and V.~Allevato\orcid{0000-0001-7232-5152}\inst{\ref{aff21}}
\and M.~Ballardini\orcid{0000-0003-4481-3559}\inst{\ref{aff103},\ref{aff104},\ref{aff10}}
\and M.~Bolzonella\orcid{0000-0003-3278-4607}\inst{\ref{aff10}}
\and E.~Bozzo\orcid{0000-0002-8201-1525}\inst{\ref{aff42}}
\and C.~Burigana\orcid{0000-0002-3005-5796}\inst{\ref{aff105},\ref{aff48}}
\and R.~Cabanac\orcid{0000-0001-6679-2600}\inst{\ref{aff91}}
\and M.~Calabrese\orcid{0000-0002-2637-2422}\inst{\ref{aff106},\ref{aff25}}
\and A.~Cappi\inst{\ref{aff10},\ref{aff107}}
\and D.~Di~Ferdinando\inst{\ref{aff17}}
\and J.~A.~Escartin~Vigo\inst{\ref{aff52}}
\and L.~Gabarra\orcid{0000-0002-8486-8856}\inst{\ref{aff108}}
\and J.~Mart\'{i}n-Fleitas\orcid{0000-0002-8594-569X}\inst{\ref{aff109}}
\and S.~Matthew\orcid{0000-0001-8448-1697}\inst{\ref{aff34}}
\and N.~Mauri\orcid{0000-0001-8196-1548}\inst{\ref{aff32},\ref{aff17}}
\and R.~B.~Metcalf\orcid{0000-0003-3167-2574}\inst{\ref{aff73},\ref{aff10}}
\and A.~Pezzotta\orcid{0000-0003-0726-2268}\inst{\ref{aff110},\ref{aff52}}
\and M.~P\"ontinen\orcid{0000-0001-5442-2530}\inst{\ref{aff62}}
\and I.~Risso\orcid{0000-0003-2525-7761}\inst{\ref{aff111}}
\and V.~Scottez\orcid{0009-0008-3864-940X}\inst{\ref{aff77},\ref{aff112}}
\and M.~Sereno\orcid{0000-0003-0302-0325}\inst{\ref{aff10},\ref{aff17}}
\and M.~Tenti\orcid{0000-0002-4254-5901}\inst{\ref{aff17}}
\and M.~Viel\orcid{0000-0002-2642-5707}\inst{\ref{aff11},\ref{aff12},\ref{aff14},\ref{aff13},\ref{aff113}}
\and M.~Wiesmann\orcid{0009-0000-8199-5860}\inst{\ref{aff65}}
\and Y.~Akrami\orcid{0000-0002-2407-7956}\inst{\ref{aff114},\ref{aff115}}
\and I.~T.~Andika\orcid{0000-0001-6102-9526}\inst{\ref{aff116},\ref{aff117}}
\and S.~Anselmi\orcid{0000-0002-3579-9583}\inst{\ref{aff43},\ref{aff90},\ref{aff118}}
\and M.~Archidiacono\orcid{0000-0003-4952-9012}\inst{\ref{aff68},\ref{aff69}}
\and F.~Atrio-Barandela\orcid{0000-0002-2130-2513}\inst{\ref{aff119}}
\and A.~Balaguera-Antolinez\orcid{0000-0001-5028-3035}\inst{\ref{aff33},\ref{aff120}}
\and D.~Bertacca\orcid{0000-0002-2490-7139}\inst{\ref{aff90},\ref{aff53},\ref{aff43}}
\and M.~Bethermin\orcid{0000-0002-3915-2015}\inst{\ref{aff121}}
\and A.~Blanchard\orcid{0000-0001-8555-9003}\inst{\ref{aff91}}
\and L.~Blot\orcid{0000-0002-9622-7167}\inst{\ref{aff122},\ref{aff64}}
\and S.~Borgani\orcid{0000-0001-6151-6439}\inst{\ref{aff123},\ref{aff11},\ref{aff12},\ref{aff13},\ref{aff113}}
\and M.~L.~Brown\orcid{0000-0002-0370-8077}\inst{\ref{aff124}}
\and S.~Bruton\orcid{0000-0002-6503-5218}\inst{\ref{aff125}}
\and A.~Calabro\orcid{0000-0003-2536-1614}\inst{\ref{aff26}}
\and B.~Camacho~Quevedo\orcid{0000-0002-8789-4232}\inst{\ref{aff11},\ref{aff14},\ref{aff12},\ref{aff49},\ref{aff50}}
\and F.~Caro\inst{\ref{aff26}}
\and C.~S.~Carvalho\inst{\ref{aff97}}
\and T.~Castro\orcid{0000-0002-6292-3228}\inst{\ref{aff12},\ref{aff13},\ref{aff11},\ref{aff113}}
\and F.~Cogato\orcid{0000-0003-4632-6113}\inst{\ref{aff73},\ref{aff10}}
\and S.~Conseil\orcid{0000-0002-3657-4191}\inst{\ref{aff36}}
\and A.~R.~Cooray\orcid{0000-0002-3892-0190}\inst{\ref{aff126}}
\and S.~Davini\orcid{0000-0003-3269-1718}\inst{\ref{aff19}}
\and G.~Desprez\orcid{0000-0001-8325-1742}\inst{\ref{aff101}}
\and A.~D\'iaz-S\'anchez\orcid{0000-0003-0748-4768}\inst{\ref{aff127}}
\and J.~J.~Diaz\orcid{0000-0003-2101-1078}\inst{\ref{aff33}}
\and S.~Di~Domizio\orcid{0000-0003-2863-5895}\inst{\ref{aff18},\ref{aff19}}
\and J.~M.~Diego\orcid{0000-0001-9065-3926}\inst{\ref{aff128}}
\and P.~Dimauro\orcid{0000-0001-7399-2854}\inst{\ref{aff129},\ref{aff26}}
\and A.~Enia\orcid{0000-0002-0200-2857}\inst{\ref{aff16},\ref{aff10}}
\and Y.~Fang\inst{\ref{aff51}}
\and A.~G.~Ferrari\orcid{0009-0005-5266-4110}\inst{\ref{aff17}}
\and P.~G.~Ferreira\orcid{0000-0002-3021-2851}\inst{\ref{aff108}}
\and A.~Finoguenov\orcid{0000-0002-4606-5403}\inst{\ref{aff62}}
\and A.~Fontana\orcid{0000-0003-3820-2823}\inst{\ref{aff26}}
\and A.~Franco\orcid{0000-0002-4761-366X}\inst{\ref{aff130},\ref{aff131},\ref{aff132}}
\and K.~Ganga\orcid{0000-0001-8159-8208}\inst{\ref{aff75}}
\and J.~Garc\'ia-Bellido\orcid{0000-0002-9370-8360}\inst{\ref{aff114}}
\and T.~Gasparetto\orcid{0000-0002-7913-4866}\inst{\ref{aff12}}
\and V.~Gautard\inst{\ref{aff133}}
\and E.~Gaztanaga\orcid{0000-0001-9632-0815}\inst{\ref{aff50},\ref{aff49},\ref{aff2}}
\and F.~Giacomini\orcid{0000-0002-3129-2814}\inst{\ref{aff17}}
\and F.~Gianotti\orcid{0000-0003-4666-119X}\inst{\ref{aff10}}
\and G.~Gozaliasl\orcid{0000-0002-0236-919X}\inst{\ref{aff134},\ref{aff62}}
\and M.~Guidi\orcid{0000-0001-9408-1101}\inst{\ref{aff16},\ref{aff10}}
\and C.~M.~Gutierrez\orcid{0000-0001-7854-783X}\inst{\ref{aff135}}
\and A.~Hall\orcid{0000-0002-3139-8651}\inst{\ref{aff34}}
\and C.~Hern\'andez-Monteagudo\orcid{0000-0001-5471-9166}\inst{\ref{aff89},\ref{aff33}}
\and H.~Hildebrandt\orcid{0000-0002-9814-3338}\inst{\ref{aff136}}
\and J.~Hjorth\orcid{0000-0002-4571-2306}\inst{\ref{aff83}}
\and S.~Joudaki\orcid{0000-0001-8820-673X}\inst{\ref{aff28}}
\and J.~J.~E.~Kajava\orcid{0000-0002-3010-8333}\inst{\ref{aff137},\ref{aff138}}
\and Y.~Kang\orcid{0009-0000-8588-7250}\inst{\ref{aff42}}
\and V.~Kansal\orcid{0000-0002-4008-6078}\inst{\ref{aff139},\ref{aff140}}
\and D.~Karagiannis\orcid{0000-0002-4927-0816}\inst{\ref{aff103},\ref{aff141}}
\and K.~Kiiveri\inst{\ref{aff60}}
\and C.~C.~Kirkpatrick\inst{\ref{aff60}}
\and S.~Kruk\orcid{0000-0001-8010-8879}\inst{\ref{aff6}}
\and M.~Lattanzi\orcid{0000-0003-1059-2532}\inst{\ref{aff104}}
\and L.~Legrand\orcid{0000-0003-0610-5252}\inst{\ref{aff142},\ref{aff143}}
\and M.~Lembo\orcid{0000-0002-5271-5070}\inst{\ref{aff78}}
\and F.~Lepori\orcid{0009-0000-5061-7138}\inst{\ref{aff144}}
\and G.~Leroy\orcid{0009-0004-2523-4425}\inst{\ref{aff145},\ref{aff74}}
\and G.~F.~Lesci\orcid{0000-0002-4607-2830}\inst{\ref{aff73},\ref{aff10}}
\and J.~Lesgourgues\orcid{0000-0001-7627-353X}\inst{\ref{aff146}}
\and L.~Leuzzi\orcid{0009-0006-4479-7017}\inst{\ref{aff10}}
\and T.~I.~Liaudat\orcid{0000-0002-9104-314X}\inst{\ref{aff147}}
\and J.~Macias-Perez\orcid{0000-0002-5385-2763}\inst{\ref{aff148}}
\and G.~Maggio\orcid{0000-0003-4020-4836}\inst{\ref{aff12}}
\and M.~Magliocchetti\orcid{0000-0001-9158-4838}\inst{\ref{aff45}}
\and F.~Mannucci\orcid{0000-0002-4803-2381}\inst{\ref{aff149}}
\and R.~Maoli\orcid{0000-0002-6065-3025}\inst{\ref{aff150},\ref{aff26}}
\and C.~J.~A.~P.~Martins\orcid{0000-0002-4886-9261}\inst{\ref{aff151},\ref{aff152}}
\and L.~Maurin\orcid{0000-0002-8406-0857}\inst{\ref{aff5}}
\and M.~Miluzio\inst{\ref{aff6},\ref{aff153}}
\and P.~Monaco\orcid{0000-0003-2083-7564}\inst{\ref{aff123},\ref{aff12},\ref{aff13},\ref{aff11}}
\and C.~Moretti\orcid{0000-0003-3314-8936}\inst{\ref{aff14},\ref{aff113},\ref{aff12},\ref{aff11},\ref{aff13}}
\and G.~Morgante\inst{\ref{aff10}}
\and S.~Nadathur\orcid{0000-0001-9070-3102}\inst{\ref{aff2}}
\and A.~Navarro-Alsina\orcid{0000-0002-3173-2592}\inst{\ref{aff71}}
\and L.~Pagano\orcid{0000-0003-1820-5998}\inst{\ref{aff103},\ref{aff104}}
\and F.~Passalacqua\orcid{0000-0002-8606-4093}\inst{\ref{aff90},\ref{aff43}}
\and K.~Paterson\orcid{0000-0001-8340-3486}\inst{\ref{aff58}}
\and L.~Patrizii\inst{\ref{aff17}}
\and A.~Pisani\orcid{0000-0002-6146-4437}\inst{\ref{aff44}}
\and D.~Potter\orcid{0000-0002-0757-5195}\inst{\ref{aff144}}
\and S.~Quai\orcid{0000-0002-0449-8163}\inst{\ref{aff73},\ref{aff10}}
\and M.~Radovich\orcid{0000-0002-3585-866X}\inst{\ref{aff53}}
\and P.-F.~Rocci\inst{\ref{aff5}}
\and S.~Sacquegna\orcid{0000-0002-8433-6630}\inst{\ref{aff131},\ref{aff130},\ref{aff132}}
\and M.~Sahl\'en\orcid{0000-0003-0973-4804}\inst{\ref{aff154}}
\and D.~B.~Sanders\orcid{0000-0002-1233-9998}\inst{\ref{aff31}}
\and E.~Sarpa\orcid{0000-0002-1256-655X}\inst{\ref{aff14},\ref{aff113},\ref{aff13}}
\and A.~Schneider\orcid{0000-0001-7055-8104}\inst{\ref{aff144}}
\and D.~Sciotti\orcid{0009-0008-4519-2620}\inst{\ref{aff26},\ref{aff27}}
\and E.~Sellentin\inst{\ref{aff155},\ref{aff156}}
\and L.~C.~Smith\orcid{0000-0002-3259-2771}\inst{\ref{aff157}}
\and K.~Tanidis\orcid{0000-0001-9843-5130}\inst{\ref{aff108}}
\and G.~Testera\inst{\ref{aff19}}
\and R.~Teyssier\orcid{0000-0001-7689-0933}\inst{\ref{aff158}}
\and S.~Tosi\orcid{0000-0002-7275-9193}\inst{\ref{aff18},\ref{aff19},\ref{aff9}}
\and A.~Troja\orcid{0000-0003-0239-4595}\inst{\ref{aff90},\ref{aff43}}
\and M.~Tucci\inst{\ref{aff42}}
\and C.~Valieri\inst{\ref{aff17}}
\and A.~Venhola\orcid{0000-0001-6071-4564}\inst{\ref{aff159}}
\and D.~Vergani\orcid{0000-0003-0898-2216}\inst{\ref{aff10}}
\and G.~Verza\orcid{0000-0002-1886-8348}\inst{\ref{aff160}}
\and P.~Vielzeuf\orcid{0000-0003-2035-9339}\inst{\ref{aff44}}
\and N.~A.~Walton\orcid{0000-0003-3983-8778}\inst{\ref{aff157}}}
										   
\institute{Department of Physics and Astronomy, University College London, Gower Street, London WC1E 6BT, UK\label{aff1}
\and
Institute of Cosmology and Gravitation, University of Portsmouth, Portsmouth PO1 3FX, UK\label{aff2}
\and
Oskar Klein Centre for Cosmoparticle Physics, Department of Physics, Stockholm University, Stockholm, SE-106 91, Sweden\label{aff3}
\and
Astrophysics Group, Blackett Laboratory, Imperial College London, London SW7 2AZ, UK\label{aff4}
\and
Universit\'e Paris-Saclay, CNRS, Institut d'astrophysique spatiale, 91405, Orsay, France\label{aff5}
\and
ESAC/ESA, Camino Bajo del Castillo, s/n., Urb. Villafranca del Castillo, 28692 Villanueva de la Ca\~nada, Madrid, Spain\label{aff6}
\and
School of Mathematics and Physics, University of Surrey, Guildford, Surrey, GU2 7XH, UK\label{aff7}
\and
Institut f\"ur Theoretische Physik, University of Heidelberg, Philosophenweg 16, 69120 Heidelberg, Germany\label{aff8}
\and
INAF-Osservatorio Astronomico di Brera, Via Brera 28, 20122 Milano, Italy\label{aff9}
\and
INAF-Osservatorio di Astrofisica e Scienza dello Spazio di Bologna, Via Piero Gobetti 93/3, 40129 Bologna, Italy\label{aff10}
\and
IFPU, Institute for Fundamental Physics of the Universe, via Beirut 2, 34151 Trieste, Italy\label{aff11}
\and
INAF-Osservatorio Astronomico di Trieste, Via G. B. Tiepolo 11, 34143 Trieste, Italy\label{aff12}
\and
INFN, Sezione di Trieste, Via Valerio 2, 34127 Trieste TS, Italy\label{aff13}
\and
SISSA, International School for Advanced Studies, Via Bonomea 265, 34136 Trieste TS, Italy\label{aff14}
\and
Centre National d'Etudes Spatiales -- Centre spatial de Toulouse, 18 avenue Edouard Belin, 31401 Toulouse Cedex 9, France\label{aff15}
\and
Dipartimento di Fisica e Astronomia, Universit\`a di Bologna, Via Gobetti 93/2, 40129 Bologna, Italy\label{aff16}
\and
INFN-Sezione di Bologna, Viale Berti Pichat 6/2, 40127 Bologna, Italy\label{aff17}
\and
Dipartimento di Fisica, Universit\`a di Genova, Via Dodecaneso 33, 16146, Genova, Italy\label{aff18}
\and
INFN-Sezione di Genova, Via Dodecaneso 33, 16146, Genova, Italy\label{aff19}
\and
Department of Physics "E. Pancini", University Federico II, Via Cinthia 6, 80126, Napoli, Italy\label{aff20}
\and
INAF-Osservatorio Astronomico di Capodimonte, Via Moiariello 16, 80131 Napoli, Italy\label{aff21}
\and
Dipartimento di Fisica, Universit\`a degli Studi di Torino, Via P. Giuria 1, 10125 Torino, Italy\label{aff22}
\and
INFN-Sezione di Torino, Via P. Giuria 1, 10125 Torino, Italy\label{aff23}
\and
INAF-Osservatorio Astrofisico di Torino, Via Osservatorio 20, 10025 Pino Torinese (TO), Italy\label{aff24}
\and
INAF-IASF Milano, Via Alfonso Corti 12, 20133 Milano, Italy\label{aff25}
\and
INAF-Osservatorio Astronomico di Roma, Via Frascati 33, 00078 Monteporzio Catone, Italy\label{aff26}
\and
INFN-Sezione di Roma, Piazzale Aldo Moro, 2 - c/o Dipartimento di Fisica, Edificio G. Marconi, 00185 Roma, Italy\label{aff27}
\and
Centro de Investigaciones Energ\'eticas, Medioambientales y Tecnol\'ogicas (CIEMAT), Avenida Complutense 40, 28040 Madrid, Spain\label{aff28}
\and
Port d'Informaci\'{o} Cient\'{i}fica, Campus UAB, C. Albareda s/n, 08193 Bellaterra (Barcelona), Spain\label{aff29}
\and
INFN section of Naples, Via Cinthia 6, 80126, Napoli, Italy\label{aff30}
\and
Institute for Astronomy, University of Hawaii, 2680 Woodlawn Drive, Honolulu, HI 96822, USA\label{aff31}
\and
Dipartimento di Fisica e Astronomia "Augusto Righi" - Alma Mater Studiorum Universit\`a di Bologna, Viale Berti Pichat 6/2, 40127 Bologna, Italy\label{aff32}
\and
Instituto de Astrof\'{\i}sica de Canarias, V\'{\i}a L\'actea, 38205 La Laguna, Tenerife, Spain\label{aff33}
\and
Institute for Astronomy, University of Edinburgh, Royal Observatory, Blackford Hill, Edinburgh EH9 3HJ, UK\label{aff34}
\and
European Space Agency/ESRIN, Largo Galileo Galilei 1, 00044 Frascati, Roma, Italy\label{aff35}
\and
Universit\'e Claude Bernard Lyon 1, CNRS/IN2P3, IP2I Lyon, UMR 5822, Villeurbanne, F-69100, France\label{aff36}
\and
Institut de Ci\`{e}ncies del Cosmos (ICCUB), Universitat de Barcelona (IEEC-UB), Mart\'{i} i Franqu\`{e}s 1, 08028 Barcelona, Spain\label{aff37}
\and
Instituci\'o Catalana de Recerca i Estudis Avan\c{c}ats (ICREA), Passeig de Llu\'{\i}s Companys 23, 08010 Barcelona, Spain\label{aff38}
\and
UCB Lyon 1, CNRS/IN2P3, IUF, IP2I Lyon, 4 rue Enrico Fermi, 69622 Villeurbanne, France\label{aff39}
\and
Departamento de F\'isica, Faculdade de Ci\^encias, Universidade de Lisboa, Edif\'icio C8, Campo Grande, PT1749-016 Lisboa, Portugal\label{aff40}
\and
Instituto de Astrof\'isica e Ci\^encias do Espa\c{c}o, Faculdade de Ci\^encias, Universidade de Lisboa, Campo Grande, 1749-016 Lisboa, Portugal\label{aff41}
\and
Department of Astronomy, University of Geneva, ch. d'Ecogia 16, 1290 Versoix, Switzerland\label{aff42}
\and
INFN-Padova, Via Marzolo 8, 35131 Padova, Italy\label{aff43}
\and
Aix-Marseille Universit\'e, CNRS/IN2P3, CPPM, Marseille, France\label{aff44}
\and
INAF-Istituto di Astrofisica e Planetologia Spaziali, via del Fosso del Cavaliere, 100, 00100 Roma, Italy\label{aff45}
\and
Universit\'e Paris-Saclay, Universit\'e Paris Cit\'e, CEA, CNRS, AIM, 91191, Gif-sur-Yvette, France\label{aff46}
\and
Space Science Data Center, Italian Space Agency, via del Politecnico snc, 00133 Roma, Italy\label{aff47}
\and
INFN-Bologna, Via Irnerio 46, 40126 Bologna, Italy\label{aff48}
\and
Institut d'Estudis Espacials de Catalunya (IEEC),  Edifici RDIT, Campus UPC, 08860 Castelldefels, Barcelona, Spain\label{aff49}
\and
Institute of Space Sciences (ICE, CSIC), Campus UAB, Carrer de Can Magrans, s/n, 08193 Barcelona, Spain\label{aff50}
\and
Universit\"ats-Sternwarte M\"unchen, Fakult\"at f\"ur Physik, Ludwig-Maximilians-Universit\"at M\"unchen, Scheinerstrasse 1, 81679 M\"unchen, Germany\label{aff51}
\and
Max Planck Institute for Extraterrestrial Physics, Giessenbachstr. 1, 85748 Garching, Germany\label{aff52}
\and
INAF-Osservatorio Astronomico di Padova, Via dell'Osservatorio 5, 35122 Padova, Italy\label{aff53}
\and
Jet Propulsion Laboratory, California Institute of Technology, 4800 Oak Grove Drive, Pasadena, CA, 91109, USA\label{aff54}
\and
Felix Hormuth Engineering, Goethestr. 17, 69181 Leimen, Germany\label{aff55}
\and
Technical University of Denmark, Elektrovej 327, 2800 Kgs. Lyngby, Denmark\label{aff56}
\and
Cosmic Dawn Center (DAWN), Denmark\label{aff57}
\and
Max-Planck-Institut f\"ur Astronomie, K\"onigstuhl 17, 69117 Heidelberg, Germany\label{aff58}
\and
NASA Goddard Space Flight Center, Greenbelt, MD 20771, USA\label{aff59}
\and
Department of Physics and Helsinki Institute of Physics, Gustaf H\"allstr\"omin katu 2, 00014 University of Helsinki, Finland\label{aff60}
\and
Universit\'e de Gen\`eve, D\'epartement de Physique Th\'eorique and Centre for Astroparticle Physics, 24 quai Ernest-Ansermet, CH-1211 Gen\`eve 4, Switzerland\label{aff61}
\and
Department of Physics, P.O. Box 64, 00014 University of Helsinki, Finland\label{aff62}
\and
Helsinki Institute of Physics, Gustaf H{\"a}llstr{\"o}min katu 2, University of Helsinki, Helsinki, Finland\label{aff63}
\and
Laboratoire d'etude de l'Univers et des phenomenes eXtremes, Observatoire de Paris, Universit\'e PSL, Sorbonne Universit\'e, CNRS, 92190 Meudon, France\label{aff64}
\and
Institute of Theoretical Astrophysics, University of Oslo, P.O. Box 1029 Blindern, 0315 Oslo, Norway\label{aff65}
\and
SKA Observatory, Jodrell Bank, Lower Withington, Macclesfield, Cheshire SK11 9FT, UK\label{aff66}
\and
Centre de Calcul de l'IN2P3/CNRS, 21 avenue Pierre de Coubertin 69627 Villeurbanne Cedex, France\label{aff67}
\and
Dipartimento di Fisica "Aldo Pontremoli", Universit\`a degli Studi di Milano, Via Celoria 16, 20133 Milano, Italy\label{aff68}
\and
INFN-Sezione di Milano, Via Celoria 16, 20133 Milano, Italy\label{aff69}
\and
University of Applied Sciences and Arts of Northwestern Switzerland, School of Computer Science, 5210 Windisch, Switzerland\label{aff70}
\and
Universit\"at Bonn, Argelander-Institut f\"ur Astronomie, Auf dem H\"ugel 71, 53121 Bonn, Germany\label{aff71}
\and
Aix-Marseille Universit\'e, CNRS, CNES, LAM, Marseille, France\label{aff72}
\and
Dipartimento di Fisica e Astronomia "Augusto Righi" - Alma Mater Studiorum Universit\`a di Bologna, via Piero Gobetti 93/2, 40129 Bologna, Italy\label{aff73}
\and
Department of Physics, Institute for Computational Cosmology, Durham University, South Road, Durham, DH1 3LE, UK\label{aff74}
\and
Universit\'e Paris Cit\'e, CNRS, Astroparticule et Cosmologie, 75013 Paris, France\label{aff75}
\and
CNRS-UCB International Research Laboratory, Centre Pierre Bin\'etruy, IRL2007, CPB-IN2P3, Berkeley, USA\label{aff76}
\and
Institut d'Astrophysique de Paris, 98bis Boulevard Arago, 75014, Paris, France\label{aff77}
\and
Institut d'Astrophysique de Paris, UMR 7095, CNRS, and Sorbonne Universit\'e, 98 bis boulevard Arago, 75014 Paris, France\label{aff78}
\and
Institute of Physics, Laboratory of Astrophysics, Ecole Polytechnique F\'ed\'erale de Lausanne (EPFL), Observatoire de Sauverny, 1290 Versoix, Switzerland\label{aff79}
\and
Telespazio UK S.L. for European Space Agency (ESA), Camino bajo del Castillo, s/n, Urbanizacion Villafranca del Castillo, Villanueva de la Ca\~nada, 28692 Madrid, Spain\label{aff80}
\and
Institut de F\'{i}sica d'Altes Energies (IFAE), The Barcelona Institute of Science and Technology, Campus UAB, 08193 Bellaterra (Barcelona), Spain\label{aff81}
\and
European Space Agency/ESTEC, Keplerlaan 1, 2201 AZ Noordwijk, The Netherlands\label{aff82}
\and
DARK, Niels Bohr Institute, University of Copenhagen, Jagtvej 155, 2200 Copenhagen, Denmark\label{aff83}
\and
Waterloo Centre for Astrophysics, University of Waterloo, Waterloo, Ontario N2L 3G1, Canada\label{aff84}
\and
Department of Physics and Astronomy, University of Waterloo, Waterloo, Ontario N2L 3G1, Canada\label{aff85}
\and
Perimeter Institute for Theoretical Physics, Waterloo, Ontario N2L 2Y5, Canada\label{aff86}
\and
Institute of Space Science, Str. Atomistilor, nr. 409 M\u{a}gurele, Ilfov, 077125, Romania\label{aff87}
\and
Consejo Superior de Investigaciones Cientificas, Calle Serrano 117, 28006 Madrid, Spain\label{aff88}
\and
Universidad de La Laguna, Departamento de Astrof\'{\i}sica, 38206 La Laguna, Tenerife, Spain\label{aff89}
\and
Dipartimento di Fisica e Astronomia "G. Galilei", Universit\`a di Padova, Via Marzolo 8, 35131 Padova, Italy\label{aff90}
\and
Institut de Recherche en Astrophysique et Plan\'etologie (IRAP), Universit\'e de Toulouse, CNRS, UPS, CNES, 14 Av. Edouard Belin, 31400 Toulouse, France\label{aff91}
\and
Universit\'e St Joseph; Faculty of Sciences, Beirut, Lebanon\label{aff92}
\and
Departamento de F\'isica, FCFM, Universidad de Chile, Blanco Encalada 2008, Santiago, Chile\label{aff93}
\and
Universit\"at Innsbruck, Institut f\"ur Astro- und Teilchenphysik, Technikerstr. 25/8, 6020 Innsbruck, Austria\label{aff94}
\and
Satlantis, University Science Park, Sede Bld 48940, Leioa-Bilbao, Spain\label{aff95}
\and
Department of Physics, Royal Holloway, University of London, TW20 0EX, UK\label{aff96}
\and
Instituto de Astrof\'isica e Ci\^encias do Espa\c{c}o, Faculdade de Ci\^encias, Universidade de Lisboa, Tapada da Ajuda, 1349-018 Lisboa, Portugal\label{aff97}
\and
Cosmic Dawn Center (DAWN)\label{aff98}
\and
Niels Bohr Institute, University of Copenhagen, Jagtvej 128, 2200 Copenhagen, Denmark\label{aff99}
\and
Universidad Polit\'ecnica de Cartagena, Departamento de Electr\'onica y Tecnolog\'ia de Computadoras,  Plaza del Hospital 1, 30202 Cartagena, Spain\label{aff100}
\and
Kapteyn Astronomical Institute, University of Groningen, PO Box 800, 9700 AV Groningen, The Netherlands\label{aff101}
\and
Infrared Processing and Analysis Center, California Institute of Technology, Pasadena, CA 91125, USA\label{aff102}
\and
Dipartimento di Fisica e Scienze della Terra, Universit\`a degli Studi di Ferrara, Via Giuseppe Saragat 1, 44122 Ferrara, Italy\label{aff103}
\and
Istituto Nazionale di Fisica Nucleare, Sezione di Ferrara, Via Giuseppe Saragat 1, 44122 Ferrara, Italy\label{aff104}
\and
INAF, Istituto di Radioastronomia, Via Piero Gobetti 101, 40129 Bologna, Italy\label{aff105}
\and
Astronomical Observatory of the Autonomous Region of the Aosta Valley (OAVdA), Loc. Lignan 39, I-11020, Nus (Aosta Valley), Italy\label{aff106}
\and
Universit\'e C\^{o}te d'Azur, Observatoire de la C\^{o}te d'Azur, CNRS, Laboratoire Lagrange, Bd de l'Observatoire, CS 34229, 06304 Nice cedex 4, France\label{aff107}
\and
Department of Physics, Oxford University, Keble Road, Oxford OX1 3RH, UK\label{aff108}
\and
Aurora Technology for European Space Agency (ESA), Camino bajo del Castillo, s/n, Urbanizacion Villafranca del Castillo, Villanueva de la Ca\~nada, 28692 Madrid, Spain\label{aff109}
\and
INAF - Osservatorio Astronomico di Brera, via Emilio Bianchi 46, 23807 Merate, Italy\label{aff110}
\and
INAF-Osservatorio Astronomico di Brera, Via Brera 28, 20122 Milano, Italy, and INFN-Sezione di Genova, Via Dodecaneso 33, 16146, Genova, Italy\label{aff111}
\and
ICL, Junia, Universit\'e Catholique de Lille, LITL, 59000 Lille, France\label{aff112}
\and
ICSC - Centro Nazionale di Ricerca in High Performance Computing, Big Data e Quantum Computing, Via Magnanelli 2, Bologna, Italy\label{aff113}
\and
Instituto de F\'isica Te\'orica UAM-CSIC, Campus de Cantoblanco, 28049 Madrid, Spain\label{aff114}
\and
CERCA/ISO, Department of Physics, Case Western Reserve University, 10900 Euclid Avenue, Cleveland, OH 44106, USA\label{aff115}
\and
Technical University of Munich, TUM School of Natural Sciences, Physics Department, James-Franck-Str.~1, 85748 Garching, Germany\label{aff116}
\and
Max-Planck-Institut f\"ur Astrophysik, Karl-Schwarzschild-Str.~1, 85748 Garching, Germany\label{aff117}
\and
Laboratoire Univers et Th\'eorie, Observatoire de Paris, Universit\'e PSL, Universit\'e Paris Cit\'e, CNRS, 92190 Meudon, France\label{aff118}
\and
Departamento de F{\'\i}sica Fundamental. Universidad de Salamanca. Plaza de la Merced s/n. 37008 Salamanca, Spain\label{aff119}
\and
Instituto de Astrof\'isica de Canarias (IAC); Departamento de Astrof\'isica, Universidad de La Laguna (ULL), 38200, La Laguna, Tenerife, Spain\label{aff120}
\and
Universit\'e de Strasbourg, CNRS, Observatoire astronomique de Strasbourg, UMR 7550, 67000 Strasbourg, France\label{aff121}
\and
Center for Data-Driven Discovery, Kavli IPMU (WPI), UTIAS, The University of Tokyo, Kashiwa, Chiba 277-8583, Japan\label{aff122}
\and
Dipartimento di Fisica - Sezione di Astronomia, Universit\`a di Trieste, Via Tiepolo 11, 34131 Trieste, Italy\label{aff123}
\and
Jodrell Bank Centre for Astrophysics, Department of Physics and Astronomy, University of Manchester, Oxford Road, Manchester M13 9PL, UK\label{aff124}
\and
California Institute of Technology, 1200 E California Blvd, Pasadena, CA 91125, USA\label{aff125}
\and
Department of Physics \& Astronomy, University of California Irvine, Irvine CA 92697, USA\label{aff126}
\and
Departamento F\'isica Aplicada, Universidad Polit\'ecnica de Cartagena, Campus Muralla del Mar, 30202 Cartagena, Murcia, Spain\label{aff127}
\and
Instituto de F\'isica de Cantabria, Edificio Juan Jord\'a, Avenida de los Castros, 39005 Santander, Spain\label{aff128}
\and
Observatorio Nacional, Rua General Jose Cristino, 77-Bairro Imperial de Sao Cristovao, Rio de Janeiro, 20921-400, Brazil\label{aff129}
\and
INFN, Sezione di Lecce, Via per Arnesano, CP-193, 73100, Lecce, Italy\label{aff130}
\and
Department of Mathematics and Physics E. De Giorgi, University of Salento, Via per Arnesano, CP-I93, 73100, Lecce, Italy\label{aff131}
\and
INAF-Sezione di Lecce, c/o Dipartimento Matematica e Fisica, Via per Arnesano, 73100, Lecce, Italy\label{aff132}
\and
CEA Saclay, DFR/IRFU, Service d'Astrophysique, Bat. 709, 91191 Gif-sur-Yvette, France\label{aff133}
\and
Department of Computer Science, Aalto University, PO Box 15400, Espoo, FI-00 076, Finland\label{aff134}
\and
Instituto de Astrof\'\i sica de Canarias, c/ Via Lactea s/n, La Laguna 38200, Spain. Departamento de Astrof\'\i sica de la Universidad de La Laguna, Avda. Francisco Sanchez, La Laguna, 38200, Spain\label{aff135}
\and
Ruhr University Bochum, Faculty of Physics and Astronomy, Astronomical Institute (AIRUB), German Centre for Cosmological Lensing (GCCL), 44780 Bochum, Germany\label{aff136}
\and
Department of Physics and Astronomy, Vesilinnantie 5, 20014 University of Turku, Finland\label{aff137}
\and
Serco for European Space Agency (ESA), Camino bajo del Castillo, s/n, Urbanizacion Villafranca del Castillo, Villanueva de la Ca\~nada, 28692 Madrid, Spain\label{aff138}
\and
ARC Centre of Excellence for Dark Matter Particle Physics, Melbourne, Australia\label{aff139}
\and
Centre for Astrophysics \& Supercomputing, Swinburne University of Technology,  Hawthorn, Victoria 3122, Australia\label{aff140}
\and
Department of Physics and Astronomy, University of the Western Cape, Bellville, Cape Town, 7535, South Africa\label{aff141}
\and
DAMTP, Centre for Mathematical Sciences, Wilberforce Road, Cambridge CB3 0WA, UK\label{aff142}
\and
Kavli Institute for Cosmology Cambridge, Madingley Road, Cambridge, CB3 0HA, UK\label{aff143}
\and
Department of Astrophysics, University of Zurich, Winterthurerstrasse 190, 8057 Zurich, Switzerland\label{aff144}
\and
Department of Physics, Centre for Extragalactic Astronomy, Durham University, South Road, Durham, DH1 3LE, UK\label{aff145}
\and
Institute for Theoretical Particle Physics and Cosmology (TTK), RWTH Aachen University, 52056 Aachen, Germany\label{aff146}
\and
IRFU, CEA, Universit\'e Paris-Saclay 91191 Gif-sur-Yvette Cedex, France\label{aff147}
\and
Univ. Grenoble Alpes, CNRS, Grenoble INP, LPSC-IN2P3, 53, Avenue des Martyrs, 38000, Grenoble, France\label{aff148}
\and
INAF-Osservatorio Astrofisico di Arcetri, Largo E. Fermi 5, 50125, Firenze, Italy\label{aff149}
\and
Dipartimento di Fisica, Sapienza Universit\`a di Roma, Piazzale Aldo Moro 2, 00185 Roma, Italy\label{aff150}
\and
Centro de Astrof\'{\i}sica da Universidade do Porto, Rua das Estrelas, 4150-762 Porto, Portugal\label{aff151}
\and
Instituto de Astrof\'isica e Ci\^encias do Espa\c{c}o, Universidade do Porto, CAUP, Rua das Estrelas, PT4150-762 Porto, Portugal\label{aff152}
\and
HE Space for European Space Agency (ESA), Camino bajo del Castillo, s/n, Urbanizacion Villafranca del Castillo, Villanueva de la Ca\~nada, 28692 Madrid, Spain\label{aff153}
\and
Theoretical astrophysics, Department of Physics and Astronomy, Uppsala University, Box 516, 751 37 Uppsala, Sweden\label{aff154}
\and
Mathematical Institute, University of Leiden, Einsteinweg 55, 2333 CA Leiden, The Netherlands\label{aff155}
\and
Leiden Observatory, Leiden University, Einsteinweg 55, 2333 CC Leiden, The Netherlands\label{aff156}
\and
Institute of Astronomy, University of Cambridge, Madingley Road, Cambridge CB3 0HA, UK\label{aff157}
\and
Department of Astrophysical Sciences, Peyton Hall, Princeton University, Princeton, NJ 08544, USA\label{aff158}
\and
Space physics and astronomy research unit, University of Oulu, Pentti Kaiteran katu 1, FI-90014 Oulu, Finland\label{aff159}
\and
Center for Computational Astrophysics, Flatiron Institute, 162 5th Avenue, 10010, New York, NY, USA\label{aff160}}    
%
%
%
%

%
%
\abstract{We develop techniques for generating accurate and precise internal covariances for measurements of clustering and weak-lensing angular power spectra. These methods have been designed to produce non-singular and unbiased covariances for \Euclid{}'s large anticipated data vector and will be critical for validation against observational systematic effects. We constructed jackknife segments that are equal in area to a high precision by adapting the binary space partition algorithm to work on arbitrarily shaped regions on the unit sphere. Jackknife estimates of the covariances are internally derived and require no assumptions about cosmology or galaxy population and bias. Our covariance estimation, called DICES (Debiased Internal Covariance Estimation with Shrinkage), first estimated a noisy covariance through conventional delete-1 jackknife resampling. This was followed by linear shrinkage of the empirical correlation matrix towards the Gaussian prediction, rather than linear shrinkage of the covariance matrix. Shrinkage ensures the covariance is non-singular and therefore invertible, which is critical for the estimation of likelihoods and validation. We then applied a delete-2 jackknife bias correction to the diagonal components of the jackknife covariance that removed the general tendency for jackknife error estimates to be biased high. We validated internally derived covariances, which used the jackknife resampling technique, on synthetic \Euclid{}-like lognormal catalogues. We demonstrate that DICES produces accurate, non-singular covariance estimates, with the relative error improving by $33\%$ for the covariance and $48\%$ for the correlation structure in comparison to jackknife estimates. These estimates can be used for highly accurate regression and inference.
}
%
%
%
%
\titlerunning{\Euclid{}: accurate and precise data-driven angular power spectrum covariances}
\authorrunning{Euclid Collaboration: K. Naidoo et al.}

\maketitle

%
%
%
%
   
\section{\label{sc:intro}Introduction}

The Euclid Wide Survey \citep{EuclidSkyOverview, Scaramella-EP1} will map the distribution of billions of galaxies up to redshift $2$ across $14\,000\,\mathrm{deg}^{2}$ of the sky. This unprecedented new view of the Universe promises to further our understanding of the nature of dark energy and to test the assumptions of the standard model of cosmology, a universe dominated by a cosmological constant $\Lambda$ and cold dark matter ($\Lambda$CDM).

For \Euclid{} to deliver on these goals, it will require large-scale structure measurements that are made to a high precision and accuracy and are robust to observational systematic effects. In order to ensure these requirements are met, we will need to validate cosmological measurements against observational systematic effects. Critical for these validation studies are the construction of robust covariances that are independent of cosmological and galaxy bias assumptions. This will enable similar validation studies on weak-lensing measurements from the Kilo Degree Survey \citep[KiDS;][]{Heymans2021}, Dark Energy Survey \citep[DES;][]{Abbott2022} Hyper Suprime-Cam \citep[HSC;][]{Hikage2019}, following the validation studies of \citet{Ross2017} and \citet{Loureiro2022} on the Baryon Oscillation Spectroscopic Survey and KiDS, respectively. This will verify whether measurements are cosmological in origin and therefore ensure fundamental physics interpretations are reliable. 

One of \Euclid{}'s key science goals will be the measurement of clustering and cosmic shear angular power spectra from up to ten tomographic bins \citep{EuclidSkyOverview}, with 30 bins in multipole $\ell$ (between $10\leq\ell\leq3000$) for each auto and cross spectrum for clustering and weak-lensing $E$ and $B$ modes. With this setup, the full data vector is expected to be of the order of $10^{5}$ elements. Typical data-driven techniques for computing internal (i.e.~from the data itself) or external covariances (i.e.~from simulations) will be extremely demanding, both computationally and in terms of computing resources, such as storage. To ensure the covariances are non-singular, external covariances will require more than $10^{5}$ mocks, while internal covariances will require more than $10^{5}$ individual jackknife segments, as the number of mock or jackknife realisations needs to be greater than the length of the data vector. In practice, this requirement will likely be much higher, due to the need to reduce noise in the covariance estimates.

For cosmological measurements, it would seem appropriate to turn to analytic predictions \citep{Alonso2019, garciagarcia2019, Nicola2021}; however, such methods will need to make explicit assumptions about the fiducial cosmology, galaxy population, and the survey (e.g. footprint) -- assumptions that need prior validation and testing. For this reason we have been compelled to use data-driven internal estimates of the covariance matrix, which do no rely on such assumptions and can be relied on in cases where analytic predictions are not possible (such as for survey systematic effects). Furthermore, they allow us to estimate uncertainties critical for validation tests, such as testing for correlations with systematic effects via angular cross spectra where analytic covariance models do not exist.

For this paper, we estimated angular power spectrum covariances using jackknife resampling. Jackknife resampling techniques have been used in cosmology in the past, with a number of studies \citep[such as][]{Escoffier2016,Friedrich2016,Favole2021} finding them to be a reliable technique for covariance estimation. However, \citet{Norberg2009} cautioned the use of internal covariance estimates due to their tendency to be biased high (meaning the diagonals of the covariance are biased to larger values), while \citet{Shirasaki2017} and \citet{Lacasa2017} also raised concerns egarding covariances being underrepresented on large scales, the former due to scales similar to the jackknife regions and the latter as a result of failing to measure supersample covariances. \citet{Efron1981} showed that jackknife covariances, overall, tend to be biased high -- a property that can be removed by estimating the bias or through the computation of correction terms \citep{Mohammad2022}. In this paper, we wish to establish a method for robustly measuring internal covariances using jackknife resampling that is both non-singular and unbiased and is capable of handling \Euclid{}'s large data vector for angular power spectrum measurements of clustering and weak lensing.

The paper is organised as follows: in Sect.~\ref{sc:methods_glass} we describe the construction of our \Euclid{}-like lognormal galaxy catalogues; in Sect.~\ref{sc:methods_mask} we describe a new method for creating jackknife segments on the sky; and in Sect.~\ref{sc:methods_cov} the computation of angular power spectra, jackknife covariances, shrinkage, and bias removal. Furthermore, in Sect.~\ref{sc:results} we explain how we tested the performance of our covariance estimates; in Sect.~\ref{sc:results_cov_accuracy} we explain how we tested the accuracy of our covariance estimates; and, lastly, in Sect.~\ref{sc:conclusion} we summarise our findings and discuss future application to the Euclid Wide Survey.

\begin{figure*}[htbp!]
\centering
\includegraphics[width=0.975\columnwidth]{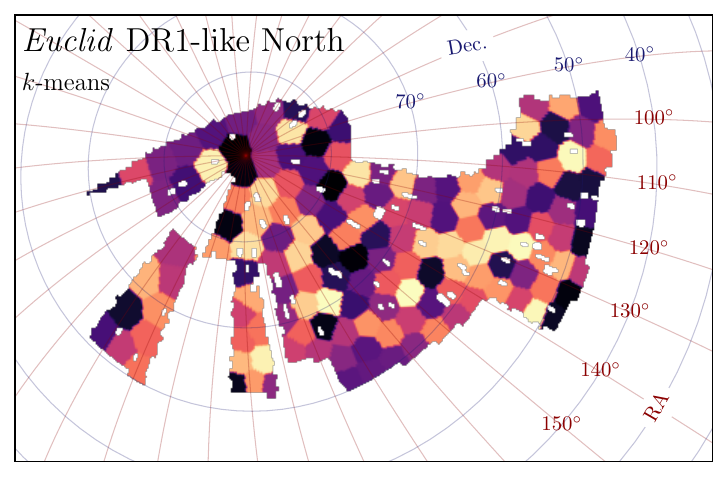}
\includegraphics[width=0.975\columnwidth]{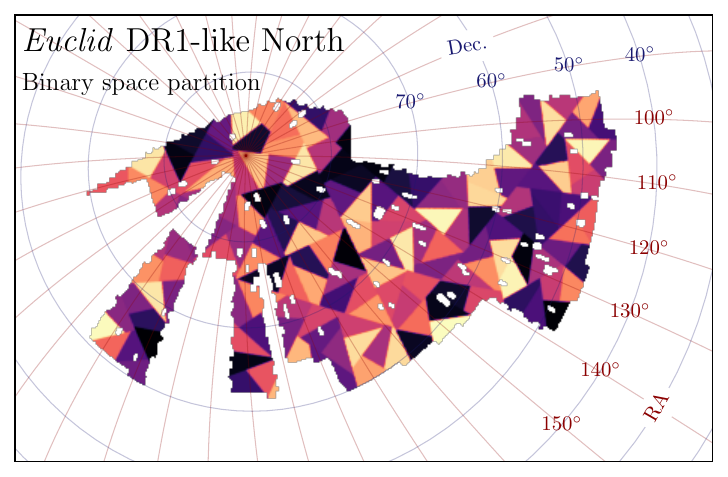}
\caption{The jackknife partition segments shown for the North \Euclid{} DR1-like wide survey footprint using the $k$-means method on the left and the binary space partitioning method on the right. Both maps have been divided into $151$ segments, with each segment assigned a random colour for visibility. The maps are shown in an orthographic projection around the North polar cap.}
\label{fig_kmeans_vs_bpm}
\end{figure*}

\begin{figure*}[htbp!]
\centering
\includegraphics[width=0.975\columnwidth]{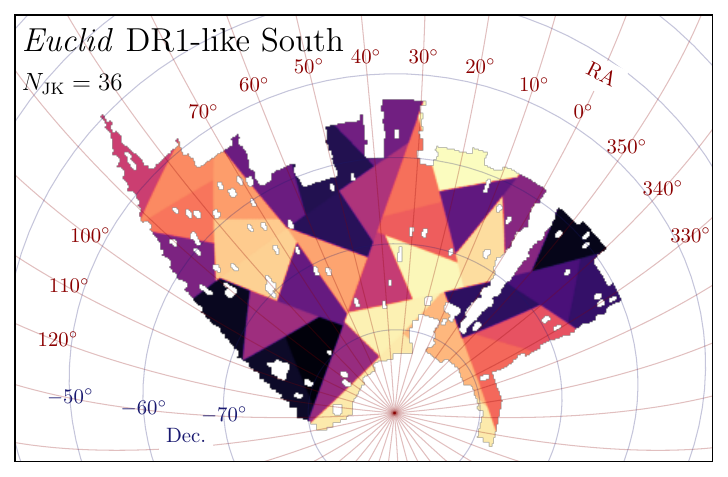}
\includegraphics[width=0.975\columnwidth]{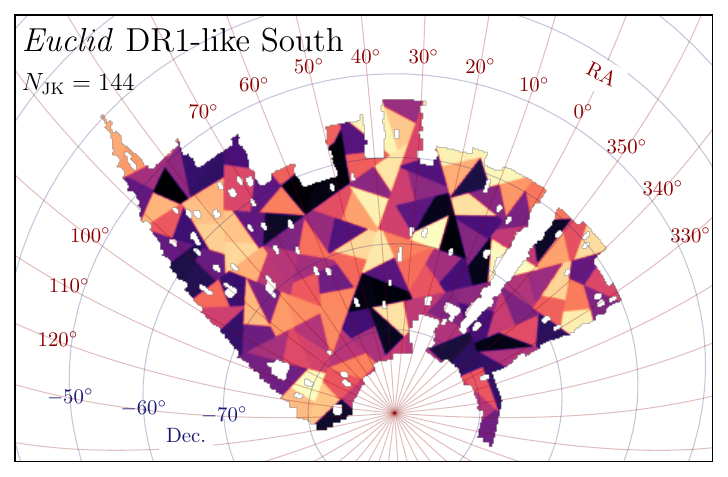}
\caption{The jackknife partition segments shown for the South \Euclid{} DR1-like wide survey footprint using the binary space partition method with $36$ segments on the left and $144$ on the right. Each segment has been assigned a random colour for visibility. The maps are shown in an orthographic projection around the South polar cap.}
\label{fig_bpm_vs_njk}
\end{figure*}

\begin{figure}[htbp!]
\centering
\includegraphics[width=0.95\columnwidth]{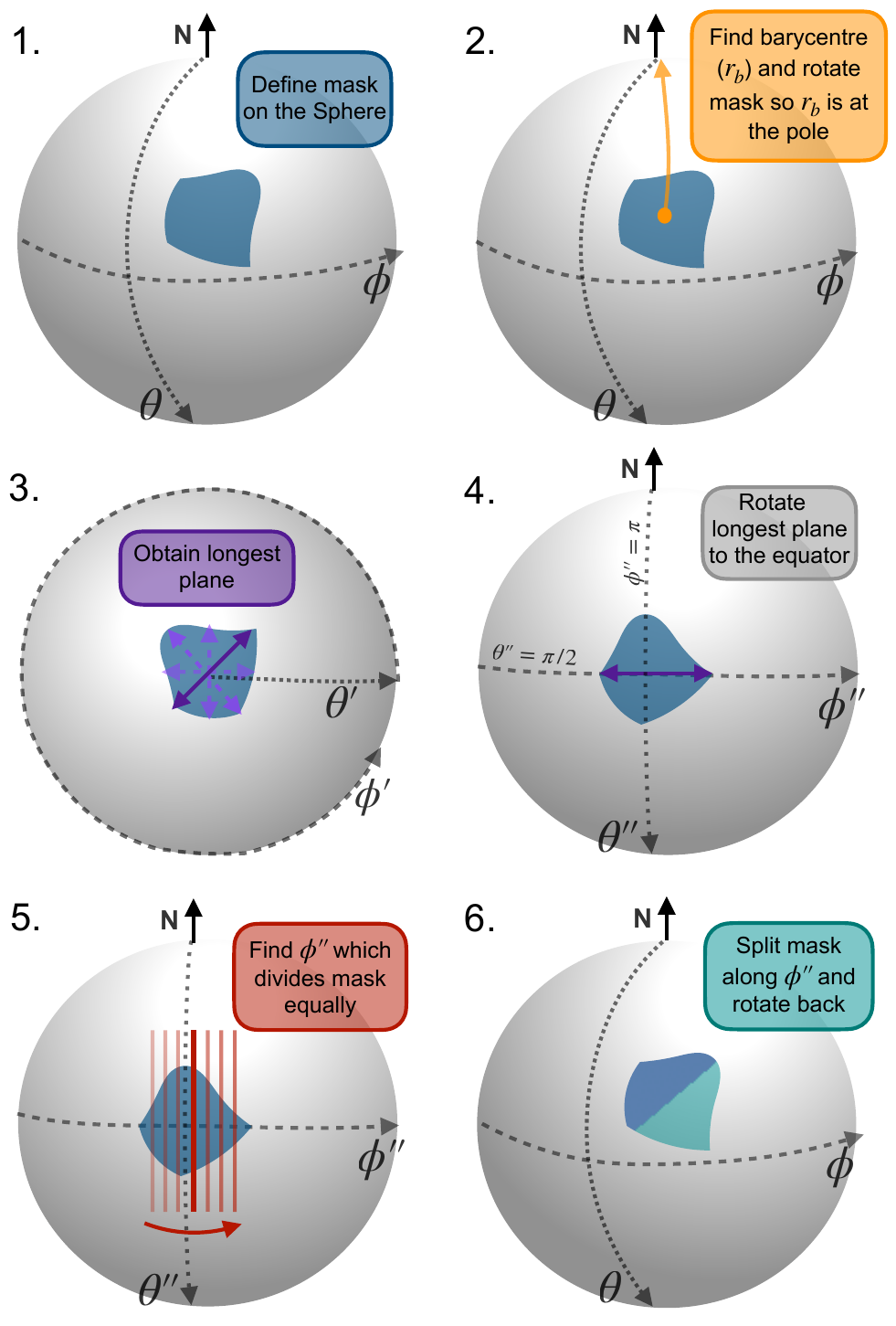}
\caption{A schematic diagram displaying how the BSP algorithm was used to split a single region on the sky into two equal area segments.}
\label{fig_skysegmentor}
\end{figure}

\section{\label{sc:methods_glass}GLASS simulations}

We generate one thousand synthetic \Euclid{}-like galaxy catalogues using GLASS \citep{Tessore2023} -- a package for generating galaxy catalogues from lognormal random fields. Lognormal fields, while not perfect, capture much of the rich covariance structure of real measurements \citep{Hall2022}, and allow us to quickly generate the large number of realisations that is necessary for characterising the population covariance. The simulations are designed to mimic the Euclid Wide Survey following a pre-launch definition of the wide-survey footprint. We assume a fiducial flat $\Lambda$CDM cosmology \citep{Planck2020} with a Hubble constant $H_{0}=67\,\mathrm{km\,s^{-1}\,Mpc^{-1}}$, matter density parameter $\Omega_{\mathrm{m}}=0.319$, baryon density parameter $\Omega_{\mathrm{b}}=0.049$, and a primordial power spectrum with an amplitude $A_{\mathrm{s}}=2\times10^{-9}$ and spectral index $n_{\mathrm{s}}=0.96$.

To ensure that the sample covariance is non-singular and the noise level is low, we limit the size of the data set to only two tomographic bins centred at redshifts $z=0.5$ and $z=1$, with each bin following a Gaussian redshift distribution with standard deviation $0.125$. Furthermore, we adopt a constant galaxy bias of $b_{\mathrm{eff}}=0.8$ with a mean galaxy number density of 1 galaxy per arcmin$^{2}$ for each tomographic bin, approximately matching the number density of tomographic bins from the 13-bin \Euclid{} survey setup \citep{EuclidSkyOverview}. Galaxy intrinsic ellipticities are assigned randomly with a standard deviation of $0.26$ per ellipticity component before being corrected for weak lensing distortions. The catalogues are constructed from lognormal random fields generated in an `onion' configuration with \texttt{HEALPix} maps at a resolution of $\mathtt{NSIDE}=1024$, similar to those used in \citet{Tessore2023}.

To understand the expected results from the earliest \Euclid{} data release we then cut the simulations to a pre-launch definition of the footprint for data release 1 (DR1; the first large data release from the first year of the wide survey), with the north region covering $\approx\!1330\,\mathrm{deg}^{2}$ (shown in Fig.~\ref{fig_kmeans_vs_bpm}) and the south $\approx\!1260\,\mathrm{deg}^{2}$ (shown in Fig.~\ref{fig_bpm_vs_njk}).

\section{\label{sc:methods_mask}Sky partitioning}

Delete-1 jackknife covariances are computed by partitioning the input data set into $N_{\mathrm{JK}}$ samples and running the analysis $N_{\mathrm{JK}}$ times, each with one of the $N_{\mathrm{JK}}$ samples removed. There are a number of ways data sets can be partitioned and resampled to compute jackknife covariances. In this analysis, we perform the partitioning at the map level, since this preserves spatial correlations and is a natural choice given the statistic of interest, angular power spectra, are computed from maps on the sky. This results in a spatial block jackknife. In cosmology, the sky partitioning is often carried out using a $k$-means clustering algorithm \citep[e.g.~see][]{Kwan2017}, which partitions the data set into $N_{\mathrm{JK}}$ clusters. Data points are members of the nearest cluster, creating a Voronoi cell partition structure. An example of the $k$-means partitioning is shown in the left panel of Fig.~\ref{fig_kmeans_vs_bpm}. This was performed using a $k$-means partitioning algorithm for the unit sphere.\footnote{\url{https://github.com/esheldon/kmeans_radec}} While this is a useful tool, the method has some computational drawbacks: firstly, the cluster finding algorithm is computationally intensive and secondly, and more critically, jackknife segments can have significant disparities in area. The area for each $k$-means jackknife segment, constructed from the \Euclid{} DR1-like footprint, varies with a standard deviation of $16\%$ from the mean. Jackknife samples are assumed to be equal and independent, but the extra variance in area means these regions cannot be treated as equal and depending on the statistics measured, could introduce significant extra area-dependent variance.

To tackle the drawbacks of $k$-means clustering, we develop a new method for partitioning the sky that applies the `binary space partitioning' \citep[BSP;][]{Fuchs1980} algorithm on the unit sphere. BSP works by sequentially segmenting a polygon in any dimension along hyperplanes. For our specific requirements we take as input the survey footprint as a coverage map, given in \texttt{HEALPix} format \citep{Gorski2005}, with values between 0 and 1. To compute a jackknife segmentation we first assign all pixels inside the footprint a jackknife segment ID of one and, to keep track of the number $N_{\mathrm{Seg}}$ of further subdivisions each segment needs to be divided into, we assign segment one an $N_{\mathrm{Seg}}=N_{\mathrm{JK}}$. Schematically, a single step in the sequential partitioning scheme works as follows:

\begin{enumerate}
    \item Assuming the segment to be divided needs to be divided into $N_{\mathrm{Seg}}$, we compute the weights of the two partitions to be 
    \begin{equation}
        w_{1}=\frac{N_{1}}{N_{\mathrm{Seg}}}\quad\mathrm{and}\quad w_{2}=\frac{N_{2}}{N_{\mathrm{Seg}}}\,,
    \end{equation}
    where 
    \begin{equation}
        N_{1} = \left\lfloor \frac{N_{\mathrm{Seg}}}{2} \right\rfloor\quad\mathrm{and}\quad N_{2}=N_{\mathrm{Seg}} - N_{1}\,,
    \end{equation}
    and $\lfloor x\rfloor$ is the floor function which rounds a number down to the nearest integer. This ensures the segment can be divided into an arbitrary number of segments rather than being limited to powers of two.
    \item All points on the mask are expressed in angular coordinates $(\phi, \theta)$ where $\phi\in[0,2\pi)$ is a longitudinal angle and $\theta\in[0,\pi]$ the colatitude angle.
    We compute the barycentre of the segment by computing a weighted mean of the mask positions. This is carried out by first converting the coordinates of pixels in the mask into Cartesian coordinates $\vec{r}=(x,y,z)$ (assuming the points lie on a unit sphere) and computing the weighted mean
    \begin{equation}
        \vec{r}_{\mathrm{barycentre}} = \frac{\sum_i w_{i} \vec{r}_{i}}{\sum_i w_{i}}\,,
    \end{equation}
    where the sum is over pixels in the mask and $i$ denotes a specific pixel. The Cartesian barycentre is then converted into spherical polar coordinates. 
    \item We now rotate the mask so that the barycentre lies at the north pole of the unit sphere, denoting this new coordinate system as $(\phi^{\prime},\theta^{\prime})$. We find the maximum $\theta^{\prime}$ of pixels in the mask in a wedge with width $\Delta\phi^{\prime}$. The resolution needs to be fine enough to correctly measure the shape of the segment without being dominated by noise and becoming computationally intractable -- in our analysis one hundred segments were used and this appears to be sufficient for both criteria (i.e.~$\Delta\phi = \pi/50$). The longest side is taken to be 
    \begin{equation}
        \phi^{\prime}_{\mathrm{longest-plane}} = \mathrm{argmax}\left\{\theta^{\prime}_{\max}(\phi^{\prime})+\theta^{\prime}_{\max}(\phi^{\prime}+\pi) \right\}\,,
    \end{equation}
    where the function $\mathrm{argmax}$ finds the argument, in this case $\phi^{\prime}$ that maximises the function.
    \item We perform another rotation to the plane of the longest side, which we will denote as $(\phi^{\prime\prime}, \theta^{\prime\prime})$, such that the line $\phi^{\prime}_{\mathrm{longest-plane}}$ now lies on the line $\theta^{\prime\prime}=\pi/2$ and the centre of the points on the longest side lies at $\phi^{\prime\prime}=\pi$.
    \item The segment is then divided along the longitudinal plane at $\phi^{\prime\prime}_{\mathrm{div}}$ taken at one hundred steps between the minimum and maximum $\phi^{\prime\prime}$ -- with steps $\Delta\phi^{\prime\prime} = (\phi_{\max}^{\prime\prime}-\phi_{\min}^{\prime\prime})/100$. On this first pass, the dividing hyperplane is taken to be the $\phi^{\prime\prime}_{\mathrm{div}}$ where
    \begin{equation}
        \phi^{\prime\prime}_{\mathrm{div}} = \mathrm{argmin} \left\{\Bigg\|\frac{w_{1}}{w_{2}} - \frac{\sum_{i,\phi^{\prime\prime}_{i} < \phi^{\prime\prime}_{\mathrm{div}}} w(\phi^{\prime\prime}_{i})}{\sum_{i,\phi^{\prime\prime}_{i} > \phi^{\prime\prime}_{\mathrm{div}}} w(\phi^{\prime\prime}_{\mathrm{i}})}\Bigg\|\right\}\,,
    \end{equation}
    searching for a dividing hyperplane that is closest to the intended balance $w_{1}/w_{2}$ for the two segments. The function $\mathrm{argmin}$ finds the argument, in this case $\phi^{\prime\prime}$, that minimises the function. This step is repeated with a second pass which finds $\phi^{\prime\prime}_{\mathrm{div}}$ more precisely by searching only in one hundred steps between $\phi^{\prime\prime}_{\mathrm{div}} \pm 2\,\Delta\phi^{\prime\prime}$.
    \item Points in the segment with $\phi^{\prime\prime} > \phi^{\prime\prime}_{\mathrm{div}}$ are assigned a new jackknife segment ID (one which is currently unassigned) with a $N_{\mathrm{Seg}}=N_{2}$ while those with $\phi^{\prime\prime} \leq \phi^{\prime\prime}_{\mathrm{div}}$ retain their current segment ID with a new $N_{\mathrm{Seg}}=N_{1}$.
\end{enumerate}

This sequential partitioning repeats until all segments have $N_{\mathrm{Seg}}=1$. In Fig.~\ref{fig_skysegmentor} we show an illustrative explanation of the BSP algorithm. The partitioned map produced appears in sharp contrast to the Voronoi segments produced with $k$-means clustering (see Fig.~\ref{fig_bpm_vs_njk}), with the BSP producing segments that are trapezoidal or triangular in shape. Rather importantly, the BSP scheme resolves some of the limitations of $k$-means: firstly, by construction, the segment areas are kept very close to equal in area; and secondly the algorithm scales very well to larger maps (with higher resolution) and a larger number of partitions, typically taking 10\% of the time. The segment areas from $k$-means vary significantly, with a standard deviation of $1.47\,\mathrm{deg}^{2}$ ($16\%$ around the mean) for $N_{\mathrm{JK}}=296$, while for the binary space partition this is $0.01\,\mathrm{deg}^{2}$ ($0.11\%$ around the mean) -- a decrease in the standard deviation of roughly two orders of magnitude in comparison to $k$-means. This will minimise any uncertainties from variances in the area to any measured jackknife statistics. The BSP algorithm has been packaged into the public Python package \texttt{SkySegmentor},\footnote{\url{https://skysegmentor.readthedocs.io/}} allowing for the partitioning of either \texttt{HEALPix} maps or points on the unit sphere. Furthermore, if a mask contains many disjoint regions, \texttt{SkySegmentor} can be used to find and label the disjoint regions, so that they can be segmented individually.

In this paper we use several different jackknife partition maps, using the $k$-means and BSP methods with $N_{\mathrm{JK}}=74$, $148$, $222$ and $296$  regions across the North and South \Euclid{} DR1-like footprint. These numbers are chosen to minimise the area discrepancy between the segments in the North and South, which were partitioned individually. Partition maps for the South with $N_{\mathrm{JK}}=36$ and $144$ are shown for the BSP method in Fig.~\ref{fig_bpm_vs_njk}. Note, that apart from the computational issues in constructing partition maps with $k$-means, we have found no difference in calculating jackknife covariances in the context of this paper and therefore have limited our discussion in this paper to the results from BSP.

\section{\label{sc:methods_cov}Covariance estimation}

In this section we detail the methods for computing angular power spectra for angular clustering and weak lensing from one thousand \Euclid{} DR1-like catalogues. We then describe the techniques for estimating the sample covariance and computing jackknife covariance. Lastly, we describe the linear shrinkage and jackknife bias removal techniques used in this study.

\subsection{Angular power spectra}

Angular power spectra were computed using the techniques outlined in \citet{EP-Tessore} which are available in the public python package \texttt{Heracles}.\footnote{\url{https://heracles.readthedocs.io/stable/}} We compute the auto and cross spectra $\vec{C}_{\ell}$ for angular clustering (denoted with a P), and weak-lensing $E$ and $B$ modes (denoted with a $E$ and $B$) for all tomographic bins (with the first bin denoted with a 0 and second with a 1). Note that we measure partial-sky angular power spectra here, and do not correct for the footprint. To ensure that at least some of the jackknife covariances (namely those with $N_{\mathrm{JK}}=296$) are not singular, we bin the angular power spectra into ten equal bins (whereas \citealt{EuclidSkyOverview} provisionally use 30 bins) in logarithmic space in $\ell$, between $10\leq\ell\leq1024$, resulting in a combined data vector with length $210$. Modes below $\ell<10$ are ignored as they are dominated by cosmic variance and are poorly sampled with the \Euclid{} DR1-like footprint, while modes beyond $\ell>1024$ go beyond the resolution of the GLASS simulations used to construct the mocks.

\subsection{Sample covariance}

The sample covariance is computed by first calculating the sample mean,
\begin{equation}
    \bar{\vec{C}}_{\ell}^{ff^{\prime}} = \frac{1}{N_{\mathrm{S}}}\sum_{i=1}^{N_{\mathrm{S}}}\vec{C}_{\ell,i}^{ff^{\prime}}\,,
\end{equation}
where $f$ and $f^{\prime}$ denote specific maps or fields and in combination a specific auto- or cross-spectrum either between clustering P, or weak lensing $E$- or $B$-modes for tomographic bins 0 or 1. The subscript $i$ denotes the $\vec{C}_{\ell}$ computed from mock catalogue $i$ from $N_{\mathrm{S}}$ samples. In our case this summation occurs across one thousand realisations (i.e.~$N_{\mathrm{S}}=1000$). The sample covariance is computed as
\begin{equation}
    \tens{C}^{f^{~}_{1}f^{\prime}_{1}f^{~}_{2}f^{\prime}_{2}}_{\mathrm{S}}(\ell_{1},\ell_{2})=\frac{1}{N_{\mathrm{S}}-1} \sum_{i=1}^{N_{\mathrm{S}}}
    \left(\vec{C}_{\ell_{1},i}^{f^{~}_{1}f^{\prime}_{1}}-\bar{\vec{C}}_{\ell_{1}}^{f^{~}_{1}f^{\prime}_{1}}\right)\left(\vec{C}_{\ell_{2},i}^{f^{~}_{2}f^{\prime}_{2}}-\bar{\vec{C}}_{\ell_{2}}^{f^{~}_{2}f^{\prime}_{2}}\right)^{\top}\,,
    \label{eq_samp_cov}
\end{equation}
where $\top$ denotes the transpose. In most cases we will drop the superscripts $f$ and $f^{\prime}$ and only use them when we are referring to a specific auto or cross spectrum, otherwise the reader can assume we are referring to the full data vector $\vec{C}_{\ell}$ and full sample covariance matrix $\tens{C}_{\mathrm{S}}$. The sample covariance is treated as the ground truth covariance for this study.

\subsection{Jackknife covariance}

Jackknife covariances allow us to estimate the covariance directly from the data, without needing to make assumptions about the data, such as the cosmology and galaxy bias modelling. In this study, jackknife covariances are computed from a single simulation, and repeated for ten simulations -- the latter to study the noise properties of the jackknife covariance in comparison to the sample covariance.

\begin{figure*}[htbp!]
\centering
\includegraphics[width=0.975\textwidth]{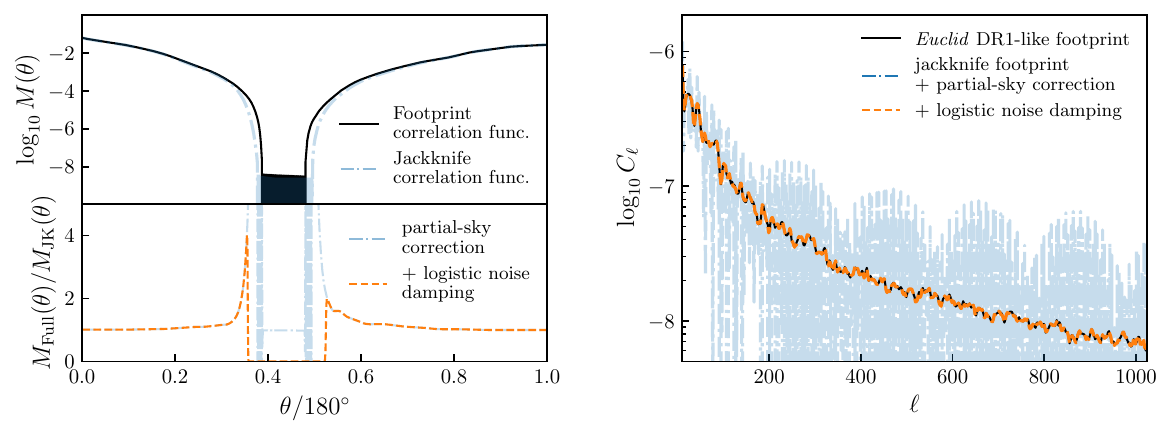}
\caption{A demonstration of the partial-sky correction for the jackknife pseudo-$C_{\ell}$. In the top-left panel, the angular correlation function for the \Euclid{} DR1-like mask (black) is compared to that of a single jackknife sample mask (blue). For angular scales between approximately $80^{\circ}$ and $100^{\circ}$, no baselines are present in the footprint or the mask correlation function; this region is dominated by numerical noise. Since the jackknife footprint is slightly smaller, its mask correlation function contains fewer baselines in this range. In the bottom-left panel, we show the partial-sky correction function as the ratio of the full footprint to the jackknife footprint correlation functions (blue). The difference in baselines leads to significant noise in this correction function. This noise is mitigated by applying a logistic damping function (shown in orange). On the right, we illustrate the effect of applying a partial-sky correction directly to the jackknife $C_{\ell}$ (blue), and compare it to a correction using the logistic damping function (orange). While a direct correction results in a very noisy angular power spectrum, the damped partial-sky correction yields a more robust and reliable estimate of the $C_{\ell}$.}
\label{fig_mask_correction}
\end{figure*}

To compute the jackknife covariance, we must first create a set of jackknife samples. These are constructed by removing parts of the data and recomputing the $\vec{C}_{\ell}$. To do this we use the partition maps described in Sect.\ref{sc:methods_mask} and shown in Figs.~\ref{fig_kmeans_vs_bpm} and \ref{fig_bpm_vs_njk}, and compute the $\vec{C}_{\ell}$ with galaxies in one of the jackknife segments removed. This is referred to as delete-1 jackknife samples, since only a single element is removed. In total this allows us to create $N_{\mathrm{JK}}$ jackknife samples $\vec{C}_{\ell}$, which we define as $\vec{C}_{\ell}^{\mathrm{JK}}$, that we can use to compute an estimate of the covariance. We first compute the jackknife mean
\begin{equation}
    \bar{\vec{C}}_{\ell}^{\mathrm{JK},ff^{\prime}} = \frac{1}{N_{\mathrm{JK}}}\sum_{i=1}^{N_{\mathrm{JK}}}\vec{C}_{\ell, i}^{\mathrm{JK},ff^{\prime}}\,,
\end{equation}
and then the jackknife covariance
\begin{equation}
\begin{split}
    \tens{C}^{f^{~}_{1}f^{\prime}_{1}f^{~}_{2}f^{\prime}_{2}}_{\mathrm{JK}}(\ell_{1},\ell_{2})=\frac{N_{\mathrm{JK}}-1}{N_{\mathrm{JK}}} \sum_{i=1}^{N_{\mathrm{JK}}}&
    \left(\vec{C}_{\ell_{1},i}^{\mathrm{JK},f^{~}_{1}f^{\prime}_{1}}-\bar{\vec{C}}_{\ell_{1}}^{\mathrm{JK},f^{~}_{1}f^{\prime}_{1}}\right)\\
    &\times\,\left(\vec{C}_{\ell_{2},i}^{\mathrm{JK},f^{~}_{2}f^{\prime}_{2}}-\bar{\vec{C}}_{\ell_{2}}^{\mathrm{JK},f^{~}_{2}f^{\prime}_{2}}\right)^{\top}\;.
\end{split}
\label{eq_jackknife_cov}
\end{equation}
This differs from Eq.~(\ref{eq_samp_cov}) only by a jackknife prefactor $(N_{\mathrm{JK}}-1)^{2} / N_{\mathrm{JK}}$, meaning we can make use of standard covariance computation libraries, which assume independent samples, and then simply multiply the output by the jackknife prefactor.

\subsection{\label{sc_method_sky_correction}Partial sky correction}

In removing a portion of the data to compute the jackknife samples we are introducing a systematic bias caused by altering the footprints for the jackknife $\vec{C}^{\mathrm{JK}}_{\ell}$. This is because the partial-sky $C_{\ell}$ computed in this analysis are affected by the survey footprint, unlike real space two-point correlation functions. Therefore the errors computed directly from the jackknife samples is not consistent with the measurements made with the full footprint. To correct the jackknife $\vec{C}^{\mathrm{JK}}_{\ell}$ for the change in footprint we apply a correction for the difference in sky by transforming the angular power spectra into real space, analogous to the techniques employed in \citet{Szapudi2001} and \citet{Chon2004}. To do this we also need to make a measurement of the angular power spectra for the footprint (used for clustering power spectra) and weight map (used for weak lensing power spectra) which we will refer to as $\vec{M}_{\ell}$ for the entire footprint and $\vec{M}_{\ell}^{\mathrm{JK}}$ for the jackknife samples. Converting from spherical harmonic space to real space requires the following transform
\begin{equation}
    C^{ff^{\prime}}(\theta) = \sum_{\ell} \frac{2\ell+1}{4\pi}\,C^{ff^{\prime}}_{\ell}\,d^{\ell}_{ss^{\prime}}(\theta)\;,
\end{equation}
where $C(\theta)$ is the real-space equivalent of the angular power spectra $C_{\ell}$, $d^{\ell}_{ss^{\prime}}$ is the Wigner D-matrix and $s$ and $s^{\prime}$ denote the spin weights of the field ($0$ for spin-$0$ fields such as the overdensity field and $2$ for spin-$2$ fields such as the cosmic shear field). For a scalar fields (i.e. spin-$0$) this reduces to the Legendre polynomials. The inverse transform is given by
\begin{equation}
    C_{\ell}^{ff^{\prime}}=2\pi\int_{0}^{\pi}C^{ff^{\prime}}(\theta)\,d^{\ell}_{ss^{\prime}}(\theta)\,\sin(\theta)\,\mathrm{d}\theta\;.
\end{equation}

To account for the change in sky coverage caused by removing a jackknife region when computing $\vec{C}_{\ell}^{\mathrm{JK}}$, we correct for the resulting mixing of harmonic modes -- introduced by the altered footprint -- by transforming to the real-space correlation function $\vec{C}_{\mathrm{JK}}(\theta)$. In real space, this mode coupling simplifies to a multiplicative correction involving the mask. Specifically, we apply the correction:
\begin{equation}
    \tilde{\vec{C}}^{ff^{\prime}}_{\mathrm{JK}}(\theta) = \frac{\vec{M}^{ff^{\prime}}(\theta)}{\vec{M}^{ff^{\prime}}_{\mathrm{JK}}(\theta)}\,\vec{C}^{ff^{\prime}}_{\mathrm{JK}}(\theta)\;,
\end{equation}
where $\vec{M}(\theta)$ and $\vec{M}_{\mathrm{JK}}(\theta)$ are the real-space analogues of the angular power spectra of the full-sky and jackknife footprint masks, respectively-that is, the real-space transforms of $\vec{M}_\ell$ and $\vec{M}^{\mathrm{JK}}_\ell$. We then convert $\tilde{\vec{C}}^{ff^{\prime}}_{\mathrm{JK}}(\theta)$ back to spherical harmonic space $\tilde{\vec{C}}^{\mathrm{JK},ff^{\prime}}_{\ell}$. However, this function is not stable when $\vec{M}_{\mathrm{JK}}(\theta)\rightarrow0$. 
To ensure $\vec{M}_{\mathrm{JK}}(\theta)\rightarrow 0$ does not cause numerical instabilities, we force $\tilde{\vec{C}}^{ff^{\prime}}_{\mathrm{JK}}(\theta) \rightarrow 0$ when $\vec{M}(\theta)\rightarrow0$. This is carried out by multiplying by a damping function
\begin{equation}
    \tilde{\vec{C}}^{ff^{\prime}}_{\mathrm{JK}}(\theta) =  \frac{\vec{M}^{ff^{\prime}}(\theta)}{\vec{M}^{ff^{\prime}}_{\mathrm{JK}}(\theta)}\,\vec{C}^{ff^{\prime}}_{\mathrm{JK}}(\theta)\,f_{\mathrm{L}}\left(\log_{10}\vec{M}^{ff^{\prime}}_{\mathrm{JK}}(\theta)\right)\;,
\end{equation}
where $f_{\mathrm{L}}$ is the logistic function
\begin{equation}
    f_{\mathrm{L}}(x,x_{\mathrm{L}},k_{\mathrm{L}}) = \frac{1}{1+\exp\big[-k_{\mathrm{L}}\,(x-x_{\mathrm{L}})\big]}\;.
\end{equation}
The variables $x_{\mathrm{L}}$ and $k_{\mathrm{L}}$ have been set to $x_{\mathrm{L}}=-5$ and $k_{\mathrm{L}}=50$; in effect this damps any signal where $\vec{M}^{ff^{\prime}}_{\mathrm{JK}}(\theta)<10^{-5}$. See Fig.~\ref{fig_mask_correction} for a demonstration of the partial-sky correction to the pseudo $C_{\ell}$.

\subsection{\label{sc_method_shrinkage}Linear shrinkage}

Estimates of the covariance from jackknife samples are often dominated by noise, a problem that can be alleviated by adding more samples and therefore more jackknife segments. While noisy estimates of the covariance are unbiased, the same is not true for its inverse and can lead to artificially tight parameter bounds \citep{Hartlap2007, Sellentin2016}. This is because the estimated covariance follows a Wishart distribution, which can be accounted for approximately through a bias correction \citep{Hartlap2007, Percival2022} or more robustly at the likelihood level \citep{Hotelling1931, Sellentin2016}. However, for \Euclid{}'s very large data vector, in addition to a noisy covariance, any reasonable number of jackknife samples will result in a covariance that is singular. This means its usage will be limited to its diagonal components and will not be usable for any likelihood or complex analysis. To address this issue we apply linear shrinkage \citep{Ledoit2004, Schafer2005, Pope2008, Simpson2016, Looijmans2024}, a technique used to dampen (or shrink) noise in an estimated covariance $\tens{C}_{\mathrm{est}}$ by combining it with a well-conditioned target covariance $\tens{C}_{\mathrm{tar}}$, for instance one that is non-singular and not noise dominated,
\begin{equation}
    \tens{C}_{\mathrm{shr}} = \lambda\,\tens{C}_{\mathrm{tar}}+(1-\lambda)\,\tens{C}_{\mathrm{est}}\;,
    \label{eq_shrinkage}
\end{equation}
where $\lambda$ is the linear shrinkage intensity and $\tens{C}_{\mathrm{shr}}$ the shrunk covariance.

To compute the linear shrinkage intensity we must first compute the matrices 
\begin{equation}
    \tens{W}_{k}(\ell_{1},\ell_{2}) = \left(\vec{C}_{\ell_{1},k}-\bar{\vec{C}}_{\ell_{1}}\right)\left(\vec{C}_{\ell_{2},k}-\bar{\vec{C}}_{\ell_{2}}\right)^{\top}\;,
\end{equation}
which we will use to estimate the variance of the covariance estimate. Note, the above computation assumes independent $\vec{C}_{\ell}$, if these are actually jackknife estimates $\vec{C}^{\mathrm{JK}}_{\ell}$ then we multiply by the jackknife prefactor,
\begin{equation}
    \tens{W}_{k}(\ell_{1},\ell_{2}) = \frac{(N_{\mathrm{JK}}-1)^{2}}{N_{\mathrm{JK}}}\,\left(\vec{C}^{\mathrm{JK}}_{\ell_{1},k}-\bar{\vec{C}}^{\mathrm{JK}}_{\ell_{1}}\right)\left(\vec{C}^{\mathrm{JK}}_{\ell_{2},k}-\bar{\vec{C}}^{\mathrm{JK}}_{\ell_{2}}\right)^{\top}\;.
\end{equation}
The mean of the matrix $\tens{W}_{k}$ is given by
\begin{equation}
    \overline{\tens{W}} = \frac{1}{N}\sum_{k=1}^{N_{\mathrm{S}}}\,\tens{W}_{k}\;,
\end{equation}
which is related to the estimated covariance (sample or jackknife)
\begin{equation}
    \tens{C}_{\mathrm{est}} = \frac{N}{N-1}\,\overline{\tens{W}}\;,
\end{equation}
where $N$ is the number of samples used to compute the estimate, either $N_{\mathrm{S}}$ for the sample covariance or $N_{\mathrm{JK}}$ for the jackknife covariance.

The covariance of the estimated matrix is generally defined as
\begin{equation}
\begin{split}
    \mathrm{Cov}\left(\tens{C}_{\mathrm{est}}^{(ij)},\tens{C}_{\mathrm{est}}^{(xy)}\right)=\frac{N}{(N-1)^{3}}\sum_{k=1}^{N}&\,\left(\tens{W}_{k}^{(ij)}-\overline{\tens{W}}^{(ij)}\right)\\
    &\times\,\left(\tens{W}_{k}^{(xy)}-\overline{\tens{W}}^{(xy)}\right)\;,
\end{split}
\end{equation}
where the superscripts $ij$ and $xy$ denote two specific elements in the estimated covariance matrix. The variance for a single element $ij$ in the covariance is given by
\begin{equation}
    \mathrm{Var}\left(\tens{C}_{\mathrm{est}}^{(ij)}\right)= \mathrm{Cov}\left(\tens{C}_{\mathrm{est}}^{(ij)},\tens{C}_{\mathrm{est}}^{(ij)}\right)\;.
\end{equation}
The optimal shrinkage intensity \citep{Schafer2005} is generally given by
\begin{equation}
\lambda^{\star}=\frac{\sum_{i,j}\left[\mathrm{Var}\left(\tens{C}_{\mathrm{est}}^{(ij)}\right) - \mathrm{Cov}\left(\tens{C}_{\mathrm{tar}}^{(ij)},\tens{C}_{\mathrm{est}}^{(ij)}\right)\right]}{\sum_{i,j}\left(\tens{C}_{\mathrm{tar}}^{(ij)}-\tens{C}_{\mathrm{est}}^{(ij)}\right)^{2}}\;,
\label{eq_lambdastar}
\end{equation}
where we sum over all $ij$ combinations (i.e.~across all elements of the matrix) and $\lambda^{\star}$ is a scalar. This specific form of shrinkage will be referred to as scalar shrinkage. However, we can compute the shrinkage intensity across blocks (unique combinations of auto- or cross-spectra, which we will refer to as block shrinkage) or without a summation at all and independently for each matrix element (referred to as matrix shrinkage). The form of Eq.~(\ref{eq_shrinkage}) means that we must ensure that the shrinkage intensity $\lambda$ lies in the range $0\leq\lambda\leq1$, which we can carry out by setting
\begin{equation}
\lambda = \begin{dcases}
    0\,,\quad&\quad\lambda^{\star} < 0\,,\\
    \lambda^{\star}\,,&\quad0\leq\lambda^{\star}\leq1\,,\\
    1\,,&\quad\lambda^{\star} >1\,,\\
\end{dcases}    
\end{equation}
where values of $\lambda^{\star}$ below 0 and above 1, can occur due to noise.

\begin{figure*}[htbp!]
\centering
\includegraphics[width=0.975\textwidth]{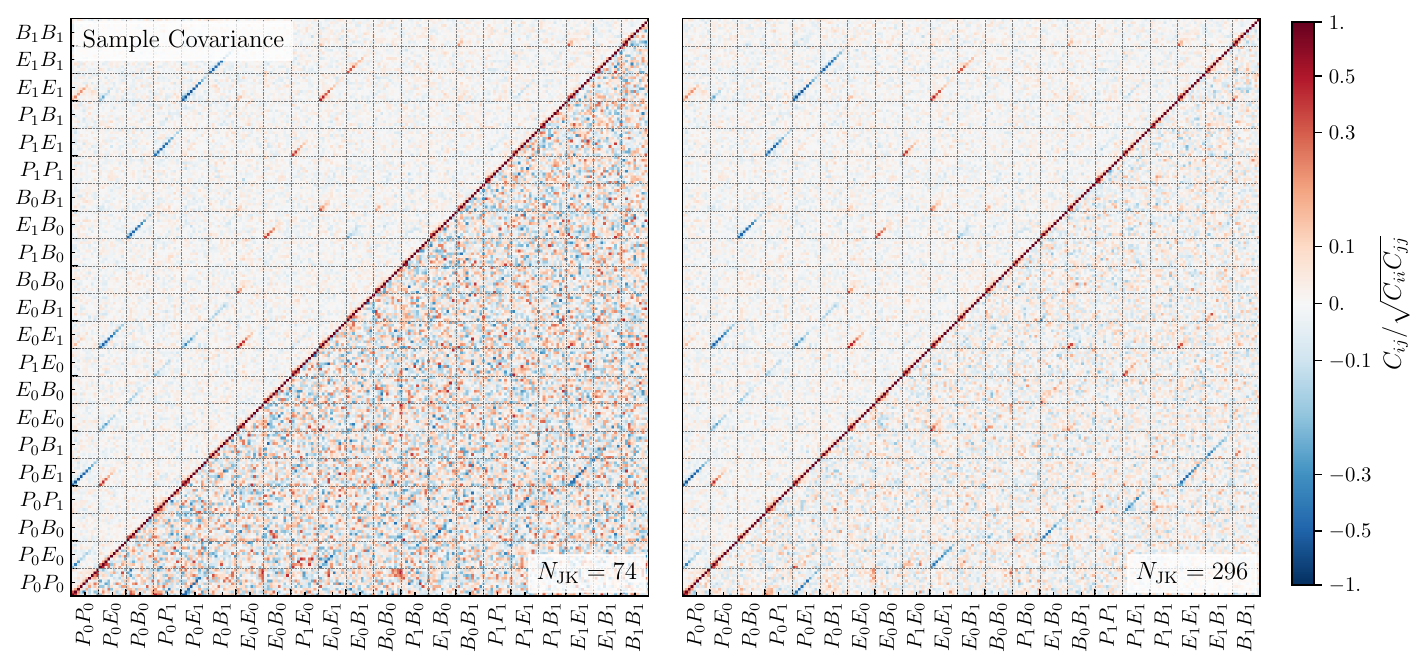}
\caption{The correlation matrix shown in relation to the number of jackknife samples $N_{\mathrm{JK}}$ (bottom right), in comparison to the sample covariance correlation matrix (top left corner). The correlation matrix is divided into blocks representing the covariance between each angular power spectra pair. On the left we show the jacknife covariances $N_{\mathrm{JK}}=74$ and on the right with $N_{\mathrm{JK}}=296$. Each angular power spectra is indicated with a label of the form $X_{i}Y_{j}$ where $X$ and $Y$ represent the angular power spectra type, either $P$ for clustering or $E$ and $B$ for $E$- and $B$-modes, while the subscripts $i$ and $j$ represent the tomographic bin. Increasing $N_{\mathrm{JK}}$ improves the non-diagonal structure of the jackknife covariance, features in the sample covariance start to show more clearly for $N_{\mathrm{JK}}=296$ where the noise levels are lower.}
\label{fig_jackknife_diagcov_vs_njk}
\end{figure*}

In this paper we shrink towards a Gaussian correlation estimate, where the Gaussian prediction for the covariance is given by
\begin{equation}
    \tens{C}_{\mathrm{G}}^{f^{~}_{1}f^{\prime}_{1}f^{~}_{2}f^{\prime}_{2}}(\ell_{1},\ell_{2}) = \left(\tilde{\vec{C}}_{\ell_{1}}^{f^{~}_{1}f^{~}_{2}}\,\tilde{\vec{C}}_{\ell_{2}}^{f^{\prime}_{1}f^{\prime}_{2}} + \tilde{\vec{C}}_{\ell_{1}}^{f^{~}_{1}f^{\prime}_{2}}\,\tilde{\vec{C}}_{\ell_{2}}^{f^{\prime}_{1}f^{~}_{2}}\right)\,\delta_{\mathrm{K}}^{\ell_{1}\ell_{2}} \;,
\end{equation}
identity matrix given by $\tens{\mathbf{I}}$, the angular power spectra given by
\begin{equation}
    \tilde{\vec{C}}_{\ell}^{ff^{\prime}} = \vec{C}_{\ell}^{ff^{\prime}} + \tens{\mathbf{I}}\cdot N^{f}\,\delta_{\mathrm{K}}^{ff^{\prime}}\;,
\end{equation}
the additive noise bias term given by $N^{f}$ and $\delta_{\mathrm{K}}^{ij}$ the Kronecker delta function. The Gaussian covariance $\tens{C}_{\mathrm{G}}$ therefore has a block-diagonal structure. We define the target covariance matrix as
\begin{equation}
    C^{ij}_{\mathrm{tar}} = \begin{dcases}
        C_{\mathrm{est}}^{(ii)}\,, \quad&i=j\,,\\
        \hat{r}_{\mathrm{G}}^{(ij)}\,\sqrt{C_{\mathrm{est}}^{(ii)}\,C_{\mathrm{est}}^{(jj)}}\,, &i\neq j\,,
    \end{dcases}
    \label{eq_ctarget}
\end{equation}
and therefore shrink towards the correlation matrix of the Gaussian covariance $\hat{r}_{\mathrm{G}}$, where $ij$ are indices used to denote elements of the matrix,
\begin{equation}
    \hat{r}_{\mathrm{G}}^{(ij)} = \frac{C_{\mathrm{G}}^{(ij)}}{\sqrt{C_{\mathrm{G}}^{(ii)}\,C_{\mathrm{G}}^{(jj)}}}\;.
\end{equation}
We chose this target covariance for its simplicity and broad applicability to any angular power spectra measurements on the sphere, including systematics. Nevertheless, our method is general, and an alternative target covariance matrix may be used.
Inserting Eq.~(\ref{eq_ctarget}) into Eq.~(\ref{eq_lambdastar}),
\begin{equation}
    \lambda^{\star} = \frac{\sum_{i\neq j} \mathrm{Var}\left(C_{\mathrm{est}}^{(ij)}\right)-\hat{r}_{\mathrm{G}}^{(ij)}f^{(ij)}}{\sum_{i\neq j} \left(C_{\mathrm{est}}^{(ij)}-\hat{r}_{\mathrm{G}}^{(ij)}\sqrt{C_{\mathrm{est}}^{(ii)}C_{\mathrm{est}}^{(jj)}}\right)^{2}}\;,
\end{equation}
where 
\begin{equation}
    f^{(ij)} = \frac{1}{2}\left(\sqrt{\frac{C_{\mathrm{est}}^{(jj)}}{C_{\mathrm{est}}^{(ii)}}}\,\mathrm{Cov}\left(C_{\mathrm{est}}^{(ii)},C_{\mathrm{est}}^{(ij)}\right)+
    \sqrt{\frac{C_{\mathrm{est}}^{(ii)}}{C_{\mathrm{est}}^{(jj)}}}\,\mathrm{Cov}\left(C_{\mathrm{est}}^{(jj)},C_{\mathrm{est}}^{(ij)}\right)\right)\;,
\end{equation}
closely following the maximum likelihood estimator for the \citet{Schafer2005} constant correlation target. As discussed previously, the shrinkage intensity is summed across the entire matrix for scalar shrinkage, across blocks for block shrinkage and without summation for matrix shrinkage -- for the latter, diagonal components of the covariance are not shrunk (i.e.~$\lambda=0$).

\subsection{\label{sc_method_bias_correction}Bias correction}

\citet{Efron1981} showed jackknife estimates of the variance tend to be biased high, which they proposed to remove with a delete-2 bias correction. In delete-2 jackknife samples, two jackknife segments are removed at a time, instead of the single segment removed in our original jackknife samples. This produces $N_{\mathrm{JK}}(N_{\mathrm{JK}}-1)/2$ unique jackknife pairs and therefore comes at quite a significant computational cost. The corrected covariance, which we will refer to as the debiased jackknife, is computed from
\begin{equation}
    \begin{split}
    \tens{C}_{\mathrm{JK}}^{\mathrm{debias}}&(\ell_{1},\ell_{2}) = \tens{C}_{\mathrm{JK}}(\ell_{1},\ell_{2})\\
    &- \frac{1}{N_{\mathrm{JK}}(N_{\mathrm{JK}}+1)}\sum_{i<i^{\prime}} \left(\vec{Q}_{ii^{\prime}}(\ell_{1})-\bar{\vec{Q}}(\ell_{1})\right)\left(\vec{Q}_{ii^{\prime}}(\ell_{2})-\bar{\vec{Q}}(\ell_{2})\right)^{\top}\;,
    \end{split}
\end{equation}
where
\begin{equation}
    \vec{Q}_{ii^{\prime}}(\ell) = N_{\mathrm{JK}}\,\vec{C}_{\ell} - \left(N_{\mathrm{JK}}-1\right)\,\left(\tilde{\vec{C}}^{\mathrm{JK}}_{\ell,i}+\tilde{\vec{C}}^{\mathrm{JK}}_{\ell,i^{\prime}}\right)+\left(N_{\mathrm{JK}}-2\right)\,\tilde{\vec{C}}^{\mathrm{JK2}}_{\ell,ii^{\prime}}
    \label{eq_d2_Q}
\end{equation}
and
\begin{equation}
    \bar{\vec{Q}}(\ell) = \frac{2}{N_{\mathrm{JK}}(N_{\mathrm{JK}}-1)}\sum_{i<i^{\prime}}\vec{Q}_{ii^{\prime}}(\ell)\;,
\end{equation}
is the mean. In Eq.~(\ref{eq_d2_Q}) the first $\vec{C}_{\ell}$ is the angular power spectra from the entire footprint, $\tilde{\vec{C}}^{\mathrm{JK}}_{\ell,i}$ the delete-1 partial sky-corrected jackknife sample with segment $i$ removed and $\tilde{\vec{C}}^{\mathrm{JK2}}_{\ell,ii^{\prime}}$ the delete-2 partial sky-corrected debiased jackknife sample with segment $i$ and $i^{\prime}$ removed. The partial sky correction for the delete-2 jackknife sample are computed in the same way outlined in Sect.~\ref{sc_method_sky_correction}.


\section{\label{sc:results}Performance}

In this section we test the performance of our internal estimates of the covariance matrix from jackknife resampling. Furthermore, we explore the effects of partial sky correction, the dependence on jackknife sample number, linear shrinkage, and jackknife debiasing.

\subsection{Dependence on partial sky correction}

In Sect.~\ref{sc_method_sky_correction} we explain how to correct the jackknife angular power spectra for differences in the footprint. To correct for the jackknife footprint we transform our data into real space to damp any signals where the correlations approach zero and would otherwise cause our correction to be numerically unstable. Partial-sky correction is crucial for correcting the mean of the angular power spectra but has little impact on the covariances which remain biased high with and without partial sky correction -- a general property of covariance estimates from jackknife resampling \cite{Efron1981}. In Appendix~\ref{app_mask_correction} we discuss the effects of partial-sky correction in more detail.

\begin{figure}[htbp!]
\centering
\includegraphics[width=0.975\columnwidth]{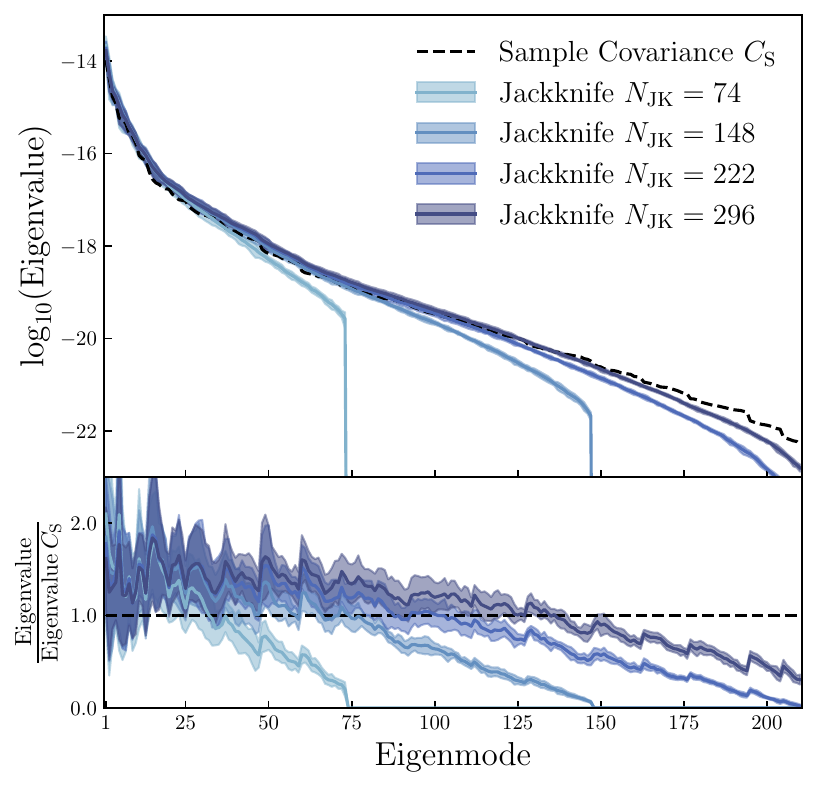}
\caption{The eigenspectrum plotted as a function of the number of jackknife samples $N_{\mathrm{JK}}$ (light to dark blue envelopes) in comparison to the sample covariance (dashed black line). In the bottom subplot we show the ratio with respect to the sample covariance. The solid lines represents the mean and the envelopes the spread (i.e.~95\% confidence intervals) from ten realisations. Increasing $N_{\mathrm{JK}}$ improves the off-diagonal structure of the jackknife covariance although we see a bias towards higher eigenvalues for the smaller eigenmodes. For $N_{\mathrm{JK}}=74$ and $148$ the covariance is singular, shown by the eigenspectrum dropping to zero for eigenmodes $> N_{\mathrm{JK}}$.}
\label{fig_jackknife_eig_vs_njk}
\end{figure}

\subsection{Dependence on partition number}

In Appendix~\ref{app_Njk} we discuss the relation between the number of jackknife partitions and the partial sky-corrected jackknife mean and standard deviation. For the mean, more partitions results in more accurate estimates of the angular power spectra, while the standard deviation (diagonals of the covariance) appears insensitive. The same is not true for the off-diagonal elements that appear to be strongly affected by the number of jackknife samples. In Fig.~\ref{fig_jackknife_diagcov_vs_njk} we compare the correlation matrix of one realisation of the jackknife covariance matrix with $N_{\mathrm{JK}}=74$ and $N_{\mathrm{JK}}=296$ to the sample covariance. The off-diagonal components of the covariance follow a block diagonal structure, with strong correlations between angular power spectra typically only occurring at equal $\ell$ modes. Increasing the number of jackknife samples improves the off-diagonal structure of the covariance, at $N_{\mathrm{JK}}=74$ the off-diagonals are dominated by noise, while at $N_{\mathrm{JK}}=296$ the noise is suppressed and we can see more of the features shown clearly in the sample covariance. This fact is made clearer in Fig.~\ref{fig_jackknife_eig_vs_njk} where we compare the eigenspectrum, the distribution of eigenvalues in the covariances (sorted from the largest eigenvalue to the smallest). Here we see a clear improvement in the covariance's off-diagonal terms which more closely match that of the sample covariance as $N_{\mathrm{JK}}$ increases. The eigenspectrum is biased high for large eigenmodes, likely due to the bias towards larger diagonal components in the covariances shown in Fig.~\ref{fig_jackknife_cov_vs_njk}.

\subsection{Shrinkage}

In the previous section we have shown that the accuracy of the jackknife covariance, particularly off-diagonal components, significantly improves when more jackknife segments are used. However, for very large data vectors, of the order of $10^{5}$ elements, any reasonable number of jackknife samples (of the order of $10^{2}$) will produce a covariance that is singular (see Fig.~\ref{fig_jackknife_eig_vs_njk}). To address this issue we turn to shrinkage methods, which combine a noisy estimate of the covariance with a well-conditioned target covariance (see Sect.~\ref{sc_method_shrinkage}).

In Appendix~\ref{app_shrinkage} we consider several shrinkage methodologies, where the shrinkage intensity is either a single value (scalar shrinkage), a single value for each angular power spectra combination (block shrinkage), or a matrix where each component of the covariance is evaluate independently (matrix shrinkage). Scalar shrinkage is found to be the most accurate in general, while also being numerically stable. Therefore, we will only apply scalar shrinkage in the following parts of this paper.

In Appendix~\ref{app_shrinkage_njk} we test the sensitivity of the shrinkage intensity $\lambda$ to $N_{\mathrm{JK}}$. We find that higher values of $N_{\mathrm{JK}}$ lead to $\lambda$ estimates that are better constrained. This improves the precisions of the shrunk covariance although the covariance, still remains biased high, since shrinking towards a Gaussian correlation matrix does not alter the biased jackknife estimates of the standard deviation.

\begin{figure*}[htbp!]
\centering
\includegraphics[width=0.975\textwidth]{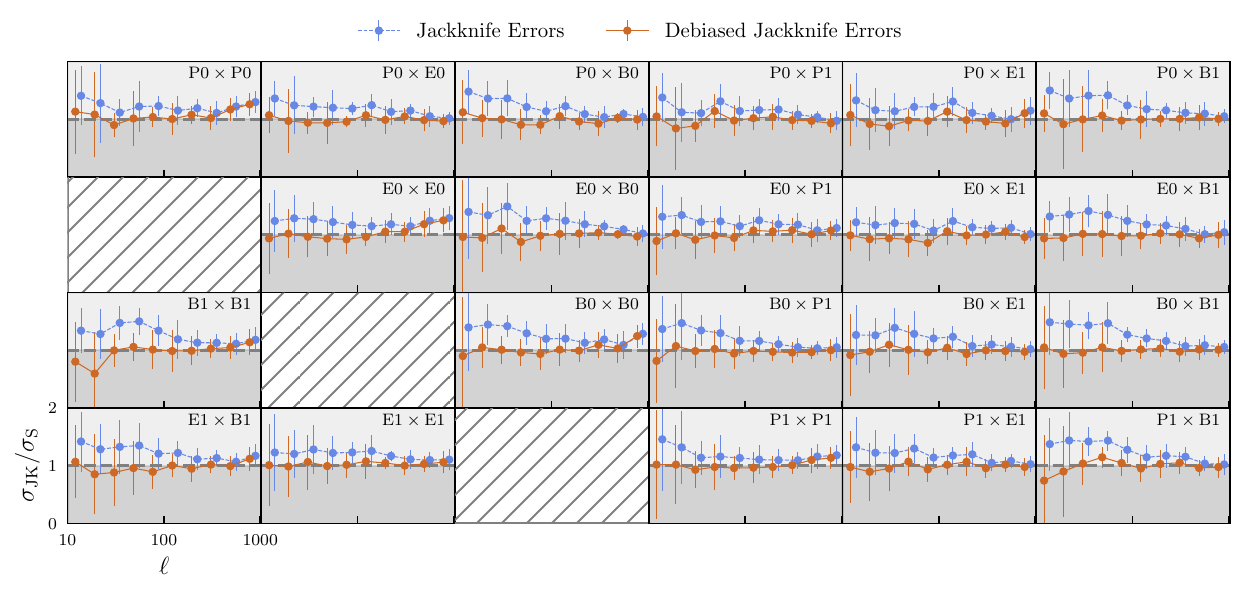}
\caption{The standard deviation for the jackknife and debiased jackknife estimates of the covariance $\sigma_{\mathrm{JK}}$, for $N_{\mathrm{JK}}=74$, are compared to the sample covariance. The lines and error bars represent the mean and spread across ten realisations. See Fig.~\ref{fig_jackknife_cov_mask} for details on the subplot layout. Debiasing removes the bias towards high diagonal components in the jackknife covariance.}
\label{fig_jackknife_cov_bias_removal}
\end{figure*}

\begin{figure}[htbp!]
\centering
\includegraphics[width=0.975\columnwidth]{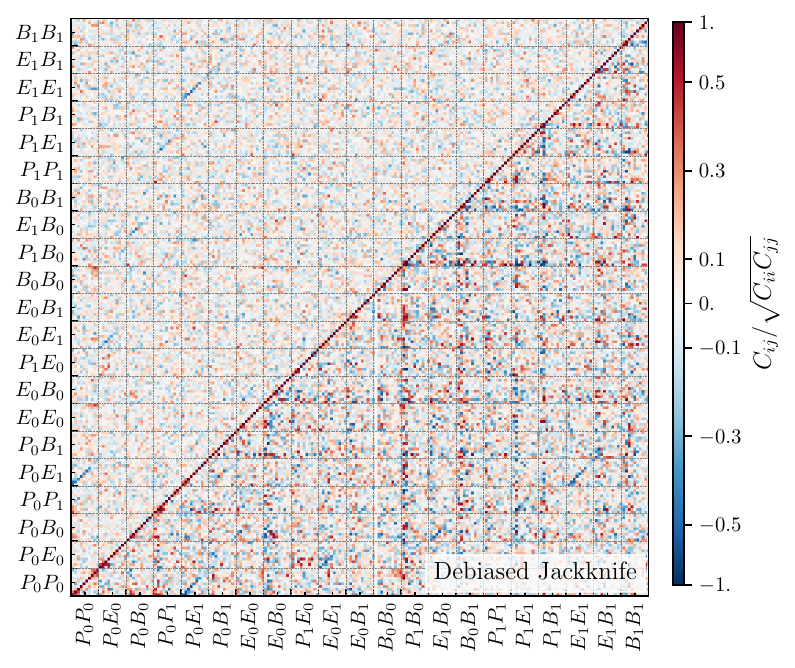}
\caption{The correlation matrix of the jackknife covariance (top left) is compared to the same matrix after jackknife debiasing (bottom right). See Fig.~\ref{fig_jackknife_diagcov_vs_njk} for details on the subplot layout. Debiasing produces a noisy covariance, especially in regions of low $\ell$ where the assumptions of the jackknife (i.e. that the samples can be treated as independent) break down. In contrast, linear shrinkage provides a covariance with lower noise-properties more closely matching the structure of the sample covariance correlation matrix.}
\label{fig_jackknife_nondiag_cov_bias_removal}
\end{figure}

\begin{figure}[htbp!]
\centering
\includegraphics[width=0.975\columnwidth]{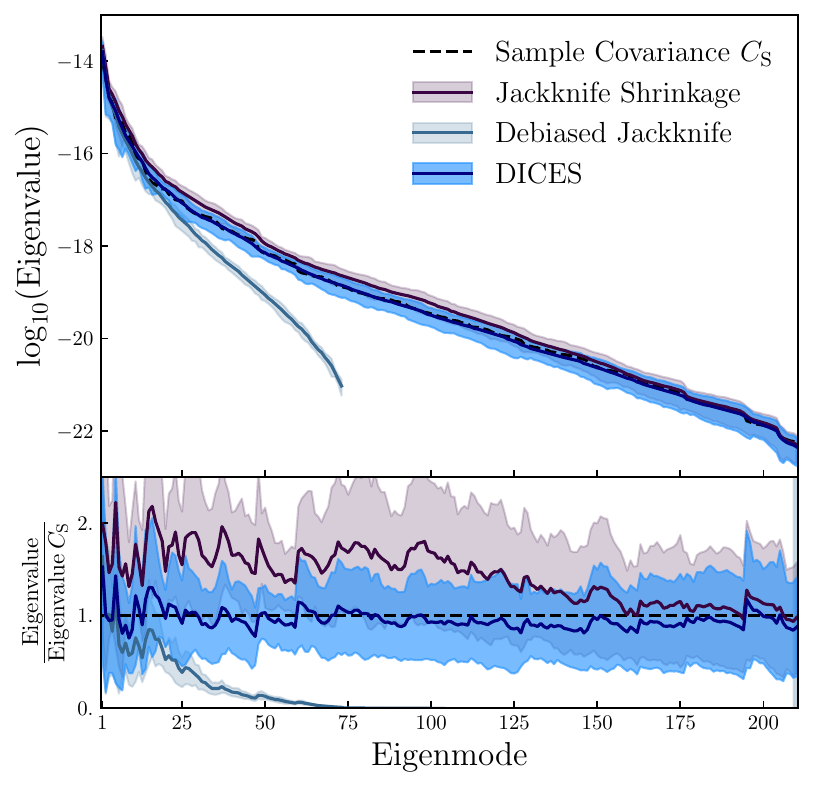}
\caption{The eigenspectrum for jackknife shrinkage, debiased jackknife, and DICES (a combination of both with the jackknife shrinkage correlation structure with debiased jackknife standard deviation) computed from $N_{\mathrm{JK}}=74$ and compared to the eigenspectrum of the sample covariance (dashed black line). In the bottom subplot we show the ratio with respect to the sample covariance. The solid lines represents the mean and the envelopes the spread (i.e.~95\% confidence interval) from ten realisations. Linear shrinkage produces a covariance estimate that is non-singular but biased high, while the debiased jackknife is singular and has a poor non-diagonal structure. The combination resolves the deficiency of both methods, providing a covariance that is non-singular and unbiased.}
\label{fig_jackknife_eig_bias_removal}
\end{figure}

\begin{figure*}
\centering
\includegraphics[width=0.975\textwidth]{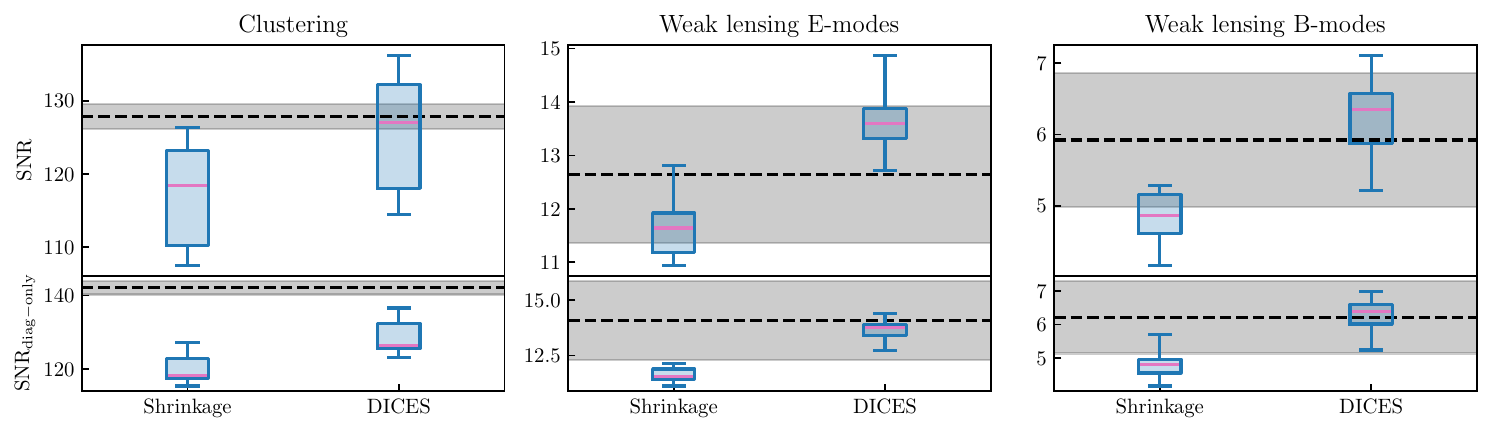}
\caption{The $\mathrm{SNR}$ (top) and $\mathrm{SNR_{diag}}$ (bottom) broken down into the clustering (left), weak-lensing $E$-mode (middle), and $B$-mode (right) components. The SNR for jackknife shrinkage and DICES are shown. The dashed black line shows the mean and the grey envelopes the 95\% confidence interval spread from ten realisations. The box-plot displays the full range with a vertical line, the box representing the interquartile range and the median indicated by a pink horizontal line. Replacing the standard deviations with the debiased jackknife diagonals improves the $\mathrm{SNR}$ and $\mathrm{SNR_{diag}}$ for clustering while for $E$ and $B$-modes they are consistent with and without this correction. For $\mathrm{SNR_{diag}}$ the bias towards low values is seen only for clustering and improves with the corrected covariance but is not completely removed.}
\label{fig_SNR_breakdown}
\end{figure*}

\begin{figure}
\centering
\includegraphics[width=0.975\columnwidth]{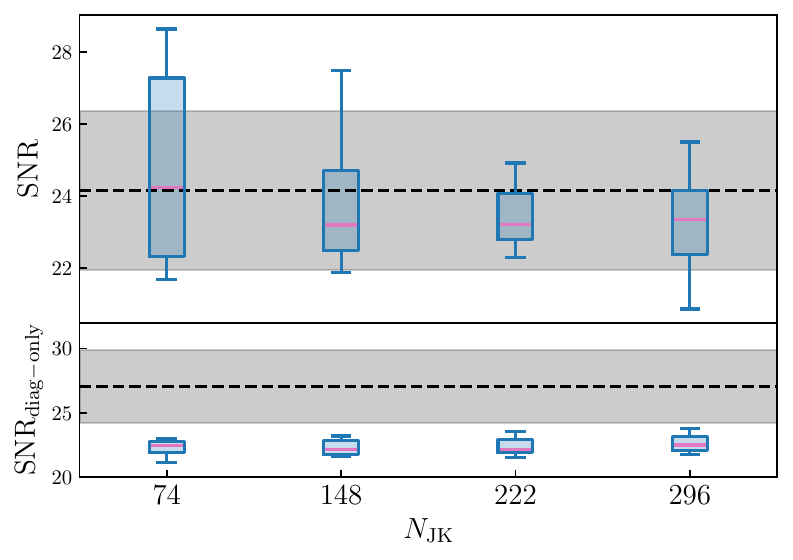}
\caption{The $\mathrm{SNR}$ (top) and $\mathrm{SNR_{diag}}$ (bottom) for the joint clustering and weak-lensing data vector shown using the shrunk jackknife covariance as a function of $N_{\mathrm{JK}}$, relative to the sample covariance. Clustering spectra are limited to $10 \leq \ell \leq 80$ to balance contributions. Increasing $N_{\mathrm{JK}}$ reduces the spread in both metrics. While $\mathrm{SNR}$ agrees with the sample covariance across all $N_{\mathrm{JK}}$, $\mathrm{SNR_{diag}}$ remains biased, though converging toward the sample estimate. See Fig.~\ref{fig_SNR_breakdown} for details.}
\label{fig_SNR_vs_njk}
\end{figure}

\begin{figure}
\centering
\includegraphics[width=0.975\columnwidth]{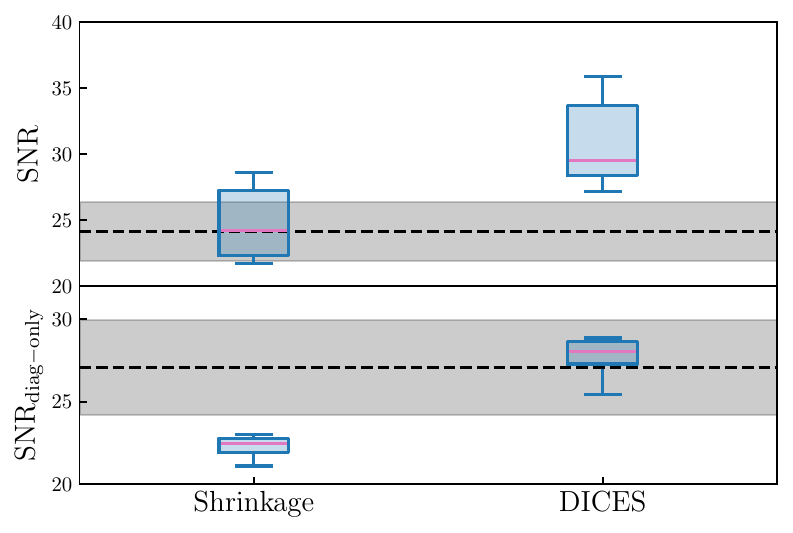}
\caption{The $\mathrm{SNR}$ (top) and $\mathrm{SNR_{diag}}$ (bottom) for the joint clustering and weak-lensing data vector plotted for the jackknife shrinkage in comparison to DICES with $N_{\mathrm{JK}}=74$. See Fig.~\ref{fig_SNR_breakdown} for details on the plot layout. To balance the contributions from clustering and weak lensing we limit all clustering auto and cross angular power spectra to $10\leq\ell\leq80$. Replacing the standard deviations with the debiased jackknife diagonals improves $\mathrm{SNR_{diag}}$ but biases $\mathrm{SNR}$ slightly high. This illustrates the bias is caused due to biases in the estimate of the correlation matrix.}
\label{fig_SNR_vs_d2}
\end{figure}

\subsection{Bias correction}

Up to this point we have seen that the diagonal components of the jackknife have tended to be biased high (see Figs.~\ref{fig_jackknife_cov_mask} and \ref{fig_jackknife_cov_vs_njk}), an issue which cannot be resolved with our choice of shrinkage  towards the correlation matrix. To address this limitation, in Sect.~\ref{sc_method_bias_correction} we describe how to remove the jackknife bias with a method that uses delete-2 jackknife samples. In Fig.~\ref{fig_jackknife_cov_bias_removal} we implement the delete-2 bias removal, and show the diagonal components for $N_{\mathrm{JK}}$ in relation to the sample covariance, demonstrating that the correction is able to remove the bias seen in the original jackknife covariance. However, at high-$\ell$ the bias for the auto spectra is slightly high, this appears to be correlated with the biases seen in the mean of the jackknife samples shown in Figs.~\ref{fig_jackknife_mean_mask} and \ref{fig_jackknife_mean_vs_njk}. 

In Fig.~\ref{fig_jackknife_nondiag_cov_bias_removal} we compare the original jackknife covariance with $N_{\mathrm{JK}}$ with scalar shrinkage and the debiased jackknife, showing that, although the debiasing is very good at correcting the diagonal components, the off-diagonals are very noisy. Noise appears to be concentrated at low $\ell$. This is likely due to a breakdown of the jackknife assumptions; for instance the samples can be assumed to be independent, since low-$\ell$ measurements are highly correlated. This means we cannot use $\tens{C}_{\mathrm{JK}}^{\mathrm{debias}}$ by itself since the correlation structure is poor; see Fig.~\ref{fig_jackknife_eig_bias_removal} showing the eigenspectrum that demonstrates this more clearly. To get the benefits of both methods, the unbiased diagonals of $\tens{C}_{\mathrm{JK}}^{\mathrm{debias}}$ and the correlation structure of $\tens{C}_{\mathrm{shr}}$, we combine the two in the following way
\begin{equation}
    \tens{C}_{\mathrm{DICES}}^{(ij)} = \frac{\tens{C}_{\mathrm{shr}}^{(ij)}}{\sqrt{\tens{C}_{\mathrm{shr}}^{(ii)}\tens{C}_{\mathrm{shr}}^{(jj)}}}\sqrt{\tens{C}_{\mathrm{JK}}^{\mathrm{debias},(ii)}\tens{C}_{\mathrm{JK}}^{\mathrm{debias},(jj)}}\;.
\end{equation}
This keeps the correlation structure of the shrunk covariance but replaces the diagonals with the debiased jackknife standard deviation. We call this technique DICES (Debiased Internal Covariance Estimation with Shrinkage). In Fig.~\ref{fig_jackknife_eig_bias_removal} we compare the eigenspectrum of the sample covariance, in comparison to the shrunk covariance, debiased jackknife and DICES. The combination provides a covariance matrix with the correct off-diagonal structure as the sample covariance and removes the bias towards high eigenvalues seen in the ordinary jackknife while also removing the poor off-diagonal structure seen in the debiased jackknife. For this reason we advocate using DICES for internal covariances in future \Euclid{} angular power spectrum measurement.

\begin{figure}[htbp!]
\centering
\includegraphics[width=0.975\columnwidth]{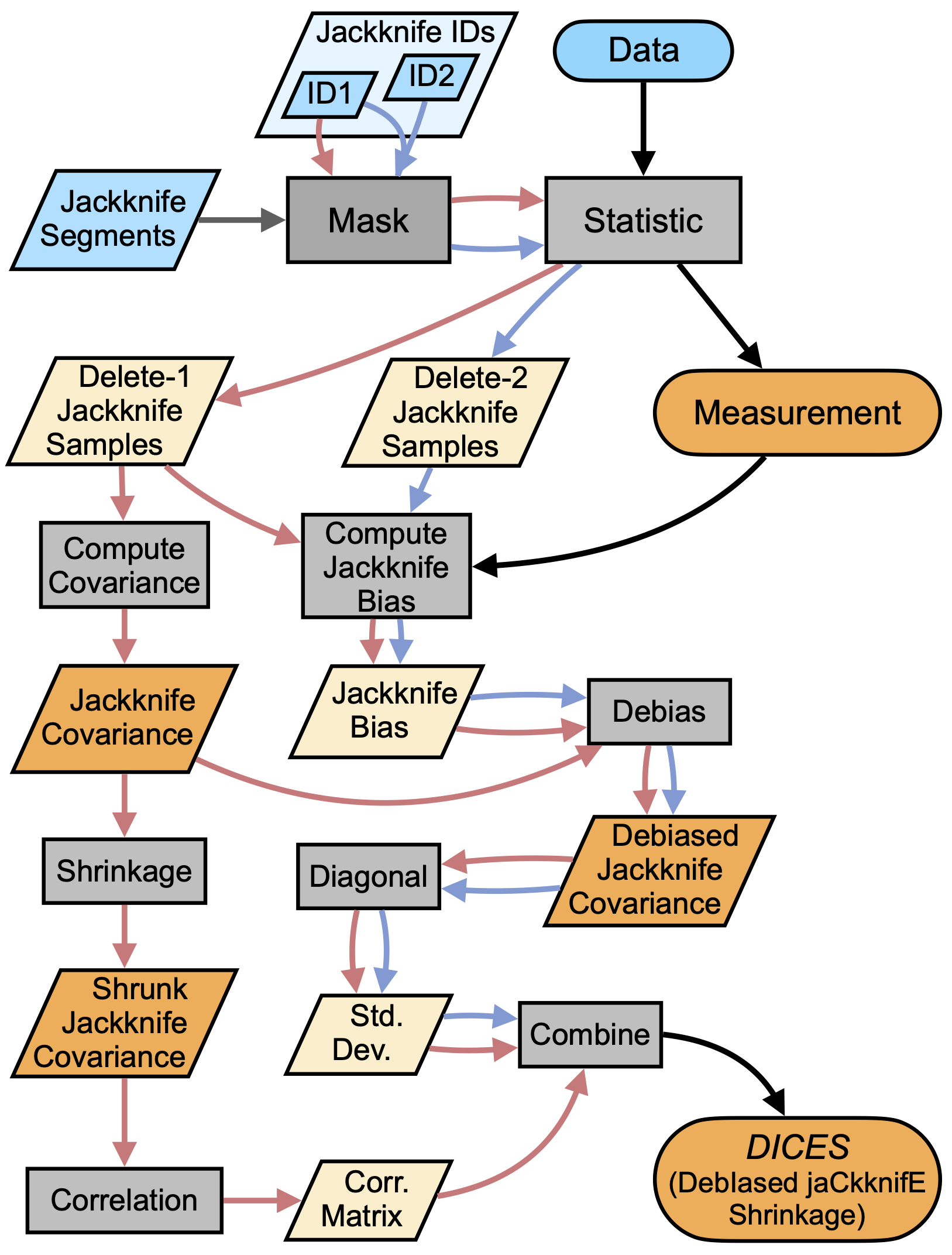}
\caption{A schematic flow chart showing how to compute internal covariance estimates using DICES. Inputs are coloured in blue, while (internal) output products are coloured in (yellow) orange. Processes are coloured in grey. The red arrows indicate processes initiated with delete-1 jackknife samples, while blue arrows indicate processes involving delete-2 jackknife samples.}
\label{fig_schematic_dices}
\end{figure}

\begin{figure*}[htbp!]
\centering
\includegraphics[width=0.975\textwidth]{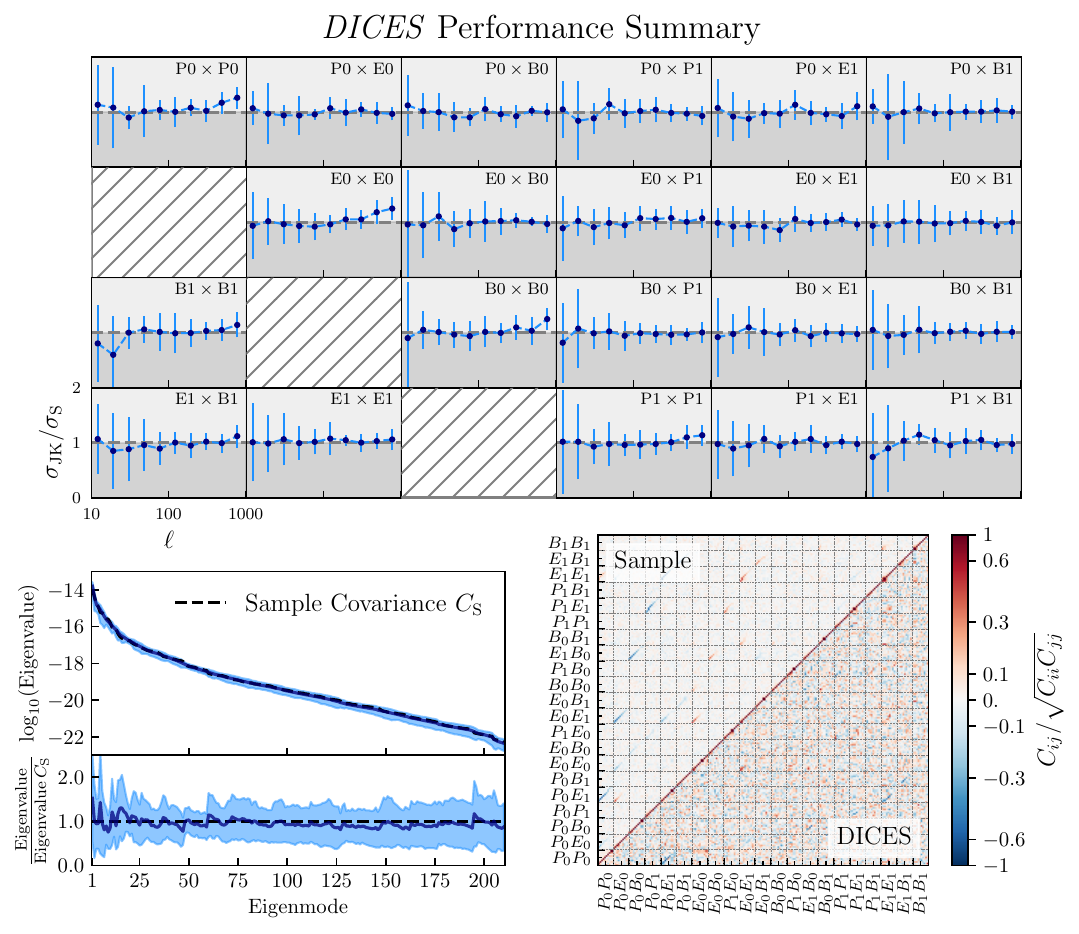}
\caption{A summary of the performance of the internal covariance estimate DICES in comparison to the sample covariance. In the top panel we show the unbiased estimates of the standard deviation from DICES in comparison to the standard deviation of the sample covariance. On the bottom plots we compare DICES with the sample covariance showing that DICES reproduces an unbiased estimate of the eigenspectrum (left) and retains the correlation structure of the covariance (right).}
\label{fig_summary_dices}
\end{figure*}

\section{\label{sc:results_cov_accuracy}Accuracy of internal covariance estimate}

We test the accuracy of our covariance estimate using a cumulative signal-to-noise ratio (SNR),
\begin{equation}
    \mathrm{SNR} = \sqrt{\vec{C}_{\ell}^{\top}\, \tens{C}^{-1} \vec{C}_{\ell}}\;,
\end{equation}
where $\tens{C}$ is any estimate of the covariance matrix, or alternatively assuming all non-diagonal components are zero
\begin{equation}
    \mathrm{SNR_{diag}} = \sqrt{\vec{C}_{\ell}^{2}\cdot\vec{\sigma}^{-2}}\;,
\end{equation}
where $\vec{\sigma}$ is a vector containing the square root of the diagonals of $\tens{C}$. Computations of SNR are similar in form to a $\chi^{2}$, used to compute a likelihood. Moreover, the SNR is equivalent to the Fisher information of a single global amplitude parameter, assuming Gaussian-distributed data \citep{tegmark97}. Therefore, the SNR weighs the covariance elements similarly to how they will be used in regression and inference.

In Fig.~\ref{fig_SNR_breakdown} we try to establish what aspect of the data vector dominates the measured $\mathrm{SNR}$. We compute the $\mathrm{SNR}$ and $\mathrm{SNR_{diag}}$ for ten realisations with the corresponding shrunk jackknife covariance and DICES covariance. We carry this out by only including angular power spectra for clustering (i.e. $P$ modes) and weak lensing $E$- and $B$-modes separately. Here we see that DICES for $E$- and $B$-modes is in both cases consistent with the sample covariance. On the other hand, the bias seen in clustering is removed when we use DICES. For the diagonal-only $\mathrm{SNR_{diag}}$ the values are biased low in both cases but the bias is improved with DICES. This illustrates that the covariance is strongly dependent on the accuracy of the clustering component. To balance the contributions between clustering and weak lensing $E$- and $B$-modes discussed below we limit the clustering angular power spectra to $10\leq \ell\leq 80$, which provides a SNR of approximately $300$.

In Fig.~\ref{fig_SNR_vs_njk} we compare $\mathrm{SNR}$ and $\mathrm{SNR_{diag}}$ for the joint clustering and weak lensing data vector computed with the sample covariance and the shrunk jackknife covariance as a function of $N_{\mathrm{JK}}$. Increasing $N_{\mathrm{JK}}$ improves the precision of the covariance, leading to $\mathrm{SNR}$ with a smaller spread. For the full $\mathrm{SNR}$ this is always consistent with the sample covariance; however, when limited to only the diagonal components the $\mathrm{SNR_{diag}}$ is underestimated, due to the jackknife bias. In Fig.~\ref{fig_SNR_vs_d2} we compare the $\mathrm{SNR}$ for the shrunk jackknife to DICES (i.e.~$\tens{C}_{\mathrm{DICES}}$). Here we see that with the corrected diagonals the full $\mathrm{SNR}$ for DICES is biased slightly; however, the accuracy of $\mathrm{SNR_{diag}}$ improves when we use DICES instead of the shrunk covariance. This slight overestimation in the $\mathrm{SNR}$ is therefore likely caused by biases in the shrinkage estimation of the correlation matrix.

To quantify the overall improvement in the covariance estimate, we measure the relative error (i.e.~average fractional deviation) between elements of the estimated covariance to the sample covariance,
\begin{equation}
    \varepsilon = \frac{\sum_{i,j}\bigg|\tens{C}_{\mathrm{est}}^{(ij)}-\tens{C}_{\mathrm{S}}^{(ij)}\bigg|}{\sum_{i,j}\bigg|\tens{C}_{\mathrm{S}}^{(ij)}\bigg|}\;.
    \label{eq_relative_error}
\end{equation}
Altogether, we find that DICES improves the deviations of the jackknife covariance by $33\%$ and the jackknife correlation structure by $48\%$.

While in this paper we have only explored DICES with $N_{\mathrm{JK}}=74$ to keep computational cost down, we expect that the biases in $\mathrm{SNR}$ are due in part to the noisy estimates of the correlation matrix from such small jackknife samples, which will reduce when $N_{\mathrm{JK}}$ is increased.


\section{\label{sc:conclusion}Discussion}

In this paper, we have developed a technique for computing accurate internal covariances for clustering and weak lensing angular power spectra. We call the method DICES, standing for Debiased Internal Covariance Estimation with Shrinkage, a combination of linear shrinkage of the correlation matrix and a debiased jackknife estimate. In Fig.~\ref{fig_schematic_dices} we outline the steps for computing the DICES internal covariance estimate, and in Fig.~\ref{fig_summary_dices} we summarise its performance.

In developing DICES we have outlined a new methodology for partitioning regions on the sky to high accuracy, a method based on applying the binary space partition algorithm \citep{Fuchs1980} on the unit sphere and made publicly available in the \texttt{SkySegmentor}\footnote{\url{https://skysegmentor.readthedocs.io/}} package. We have explored the dependence of jackknife covariance estimates on partial sky coverage and the number of jackknife samples, finding the former to have little to no impact on the covariance, while the off-diagonal structure of the covariance is significantly improved with increasing jackknife samples. We find scalar shrinkage, towards a Gaussian predicted correlation matrix, to produce reliable non-singular covariance estimates even for cases of large data vectors. 

Finally, we estimate the jackknife bias via \citet{Efron1981} that we use to compute a debiased jackknife covariance. We show that the diagonal standard deviations are consistent with the sample covariance, but the off-diagonals are dominated by noise.  In combining the debiased jackknife covariance with the shrunk jackknife we are able to produce an internal covariance that is both debiased and with off-diagonals that are noise-reduced and consistent with the sample covariance.

In all cases we find that increasing the number of jackknife samples $N_{\mathrm{JK}}$ is always preferred. Even with shrinkage applied, the precision of the covariance improves with more jackknife samples. For this reason the Euclid Wide Survey should use as many jackknife samples as is computationally feasible. The main limitation is the computation of the jackknife bias, which requires the computation of $N_{\mathrm{JK}}(N_{\mathrm{JK}}-1)/2$ jackknife samples. Since validation will require these covariances to be computed over many iterations, $N_{\mathrm{JK}}$ of order $100$ will enable the computation to be made relatively fast while data release products could be produced with a much larger number of jackknife samples.

The DICES methodology outlined in this paper will enable robust measurements of the \Euclid{} angular power spectra covariances directly from data, allowing for accurate angular power spectra covariances that are model-independent and for the measurement of systematic errors critical for validation. Both shrinkage and debiasing change the noise properties of the covariance matrix estimate, meaning neither the \citet{Hartlap2007} correction nor the \citet{Sellentin2016} likelihood correction can be applied. Future work could look towards improving our methodology, including understanding the small bias at high $\ell$ seen in the jackknife mean and jackknife covariances, the inclusion of $E$- and $B$-mode mixing during partial sky correction, and improvements in the covariance estimate at low $\ell$ where we expect the jackknife assumptions to break down (see \citealt{Shirasaki2017,Lacasa2017}). Furthermore, future analysis could explore the breakdown of some of the assumptions made in this paper-namely, that the summary statistics (angular power spectra) are dominated by Gaussian errors and noise, with non-Gaussian terms being small; and that the covariance matrices are parameter-independent. Fortunately, we find no evidence that these issues will drastically affect the covariance estimate, and therefore accurate internal covariances for \Euclid{} DR1 will be possible using the DICES methodology. For future data releases, the increase in area will allow further optimisation of the shrinkage target, shrinkage intensity and the exploration of non-linear shrinkage \citep{Joachimi2017}. The DICES methodology is made publicly available through the \texttt{heracles.py} \footnote{https://github.com/heracles-ec/heracles} package of the Euclid collaboration. A tutorial on how to implement the DICES methodology using the methods of \texttt{heracles.py} can be found \href{https://github.com/heracles-ec/heracles/blob/main/examples/jackknife-covariance.ipynb}{here}.

\begin{acknowledgements}
KN, NT, JRZ, and BJ acknowledge support by the UK Space Agency through grants ST/W002574/1 and ST/X00208X/1. AL acknowledges support by the Swedish National Space Agency (Rymdstyrelsen) through the Career Grant Project Dnr.~2024-00171.
\AckEC  
\end{acknowledgements}

\bibliography{mybib}

@ARTICLE{EuclidSkyOverview,
author = {{Euclid Collaboration: Mellier}, Y. and {Abdurro'uf} and {Acevedo~Barroso}, J.A. and others},
	title = {Euclid - I. Overview of the Euclid mission},
	DOI= "10.1051/0004-6361/202450810",
	url= "https://doi.org/10.1051/0004-6361/202450810",
	journal = {A\&A},
	year = 2025,
	volume = 697,
	pages = "A1",
}

@ARTICLE{EP-Tessore,
       author = {{Euclid Collaboration: Tessore}, N. and {Joachimi}, B. and {Loureiro}, A. and others},
        title = "{Euclid preparation: LIX. Angular power spectra from discrete observations}",
      journal = {\aap},
         year = 2025,
        month = feb,
       volume = {694},
          eid = {A141},
        pages = {A141},
          doi = {10.1051/0004-6361/202452018},
archivePrefix = {arXiv},
       eprint = {2408.16903},
 primaryClass = {astro-ph.CO},
       adsurl = {https://ui.adsabs.harvard.edu/abs/2025A&A...694A.141E}
}

@ARTICLE{Scaramella-EP1,
       author = {{Euclid Collaboration: Scaramella}, R. and {Amiaux}, J. and {Mellier}, Y. and others},
        title = "{Euclid preparation. I. The Euclid Wide Survey}",
      journal = {\aap},
         year = 2022,
        month = jun,
       volume = {662},
          eid = {A112},
        pages = {A112},
          doi = {10.1051/0004-6361/202141938},
archivePrefix = {arXiv},
       eprint = {2108.01201},
 primaryClass = {astro-ph.CO},
       adsurl = {https://ui.adsabs.harvard.edu/abs/2022A&A...662A.112E}
}

@ARTICLE{Abbott2022,
author = {{Abbott}, T.~M.~C. and {Aguena}, M. and {Alarcon}, A. and {Allam}, S. and {Alves}, O. and {Amon}, A. and {Andrade-Oliveira}, F. and {Annis}, J. and {Avila}, S. and {Bacon}, D. and {Baxter}, E. and {Bechtol}, K. and {Becker}, M.~R. and {Bernstein}, G.~M. and {Bhargava}, S. and {Birrer}, S. and {Blazek}, J. and {Brandao-Souza}, A. and {Bridle}, S.~L. and {Brooks}, D. and {Buckley-Geer}, E. and {Burke}, D.~L. and {Camacho}, H. and {Campos}, A. and {Carnero Rosell}, A. and {Carrasco Kind}, M. and {Carretero}, J. and {Castander}, F.~J. and {Cawthon}, R. and {Chang}, C. and {Chen}, A. and {Chen}, R. and {Choi}, A. and {Conselice}, C. and {Cordero}, J. and {Costanzi}, M. and {Crocce}, M. and {da Costa}, L.~N. and {da Silva Pereira}, M.~E. and {Davis}, C. and {Davis}, T.~M. and {De Vicente}, J. and {DeRose}, J. and {Desai}, S. and {Di Valentino}, E. and {Diehl}, H.~T. and {Dietrich}, J.~P. and {Dodelson}, S. and {Doel}, P. and {Doux}, C. and {Drlica-Wagner}, A. and {Eckert}, K. and {Eifler}, T.~F. and {Elsner}, F. and {Elvin-Poole}, J. and {Everett}, S. and {Evrard}, A.~E. and {Fang}, X. and {Farahi}, A. and {Fernandez}, E. and {Ferrero}, I. and {Fert{\'e}}, A. and {Fosalba}, P. and {Friedrich}, O. and {Frieman}, J. and {Garc{\'\i}a-Bellido}, J. and {Gatti}, M. and {Gaztanaga}, E. and {Gerdes}, D.~W. and {Giannantonio}, T. and {Giannini}, G. and {Gruen}, D. and {Gruendl}, R.~A. and {Gschwend}, J. and {Gutierrez}, G. and {Harrison}, I. and {Hartley}, W.~G. and {Herner}, K. and {Hinton}, S.~R. and {Hollowood}, D.~L. and {Honscheid}, K. and {Hoyle}, B. and {Huff}, E.~M. and {Huterer}, D. and {Jain}, B. and {James}, D.~J. and {Jarvis}, M. and {Jeffrey}, N. and {Jeltema}, T. and {Kovacs}, A. and {Krause}, E. and {Kron}, R. and {Kuehn}, K. and {Kuropatkin}, N. and {Lahav}, O. and {Leget}, P. -F. and {Lemos}, P. and {Liddle}, A.~R. and {Lidman}, C. and {Lima}, M. and {Lin}, H. and {MacCrann}, N. and {Maia}, M.~A.~G. and {Marshall}, J.~L. and {Martini}, P. and {McCullough}, J. and {Melchior}, P. and {Mena-Fern{\'a}ndez}, J. and {Menanteau}, F. and {Miquel}, R. and {Mohr}, J.~J. and {Morgan}, R. and {Muir}, J. and {Myles}, J. and {Nadathur}, S. and {Navarro-Alsina}, A. and {Nichol}, R.~C. and {Ogando}, R.~L.~C. and {Omori}, Y. and {Palmese}, A. and {Pandey}, S. and {Park}, Y. and {Paz-Chinch{\'o}n}, F. and {Petravick}, D. and {Pieres}, A. and {Plazas Malag{\'o}n}, A.~A. and {Porredon}, A. and {Prat}, J. and {Raveri}, M. and {Rodriguez-Monroy}, M. and {Rollins}, R.~P. and {Romer}, A.~K. and {Roodman}, A. and {Rosenfeld}, R. and {Ross}, A.~J. and {Rykoff}, E.~S. and {Samuroff}, S. and {S{\'a}nchez}, C. and {Sanchez}, E. and {Sanchez}, J. and {Sanchez Cid}, D. and {Scarpine}, V. and {Schubnell}, M. and {Scolnic}, D. and {Secco}, L.~F. and {Serrano}, S. and {Sevilla-Noarbe}, I. and {Sheldon}, E. and {Shin}, T. and {Smith}, M. and {Soares-Santos}, M. and {Suchyta}, E. and {Swanson}, M.~E.~C. and {Tabbutt}, M. and {Tarle}, G. and {Thomas}, D. and {To}, C. and {Troja}, A. and {Troxel}, M.~A. and {Tucker}, D.~L. and {Tutusaus}, I. and {Varga}, T.~N. and {Walker}, A.~R. and {Weaverdyck}, N. and {Wechsler}, R. and {Weller}, J. and {Yanny}, B. and {Yin}, B. and {Zhang}, Y. and {Zuntz}, J. and {DES Collaboration}},
title = "{Dark Energy Survey Year 3 results: Cosmological constraints from galaxy clustering and weak lensing}",
journal = {\prd},
year = 2022,
month = jan,
volume = {105},
number = {2},
eid = {023520},
pages = {023520},
doi = {10.1103/PhysRevD.105.023520},
archivePrefix = {arXiv},
eprint = {2105.13549},
primaryClass = {astro-ph.CO},
adsurl = {https://ui.adsabs.harvard.edu/abs/2022PhRvD.105b3520A}
}

@ARTICLE{Alonso2019,
author = {{Alonso}, David and {Sanchez}, Javier and {Slosar}, An{\v{z}}e and {LSST Dark Energy Science Collaboration}},
title = "{A unified pseudo-C$_{{\ensuremath{\ell}}}$ framework}",
journal = {\mnras},
year = 2019,
month = apr,
volume = {484},
number = {3},
pages = {4127-4151},
doi = {10.1093/mnras/stz093},
archivePrefix = {arXiv},
eprint = {1809.09603},
primaryClass = {astro-ph.CO},
adsurl = {https://ui.adsabs.harvard.edu/abs/2019MNRAS.484.4127A}
}

@ARTICLE{garciagarcia2019,
       author = {{Garc{\'\i}a-Garc{\'\i}a}, Carlos and {Alonso}, David and {Bellini}, Emilio},
        title = "{Disconnected pseudo-C$_{l}$ covariances for projected large-scale structure data}",
      journal = {\jcap},
         year = 2019,
        month = nov,
       volume = "11",
        pages = "043",
          doi = {10.1088/1475-7516/2019/11/043},
archivePrefix = {arXiv},
       eprint = {1906.11765},
 primaryClass = {astro-ph.CO},
       adsurl = {https://ui.adsabs.harvard.edu/abs/2019JCAP...11..043G}
}

@ARTICLE{Nicola2021,
       author = {{Nicola}, Andrina and {Garc{\'\i}a-Garc{\'\i}a}, Carlos and {Alonso}, David and {Dunkley}, Jo and {Ferreira}, Pedro G. and {Slosar}, An{\v{z}}e and {Spergel}, David N.},
        title = "{Cosmic shear power spectra in practice}",
      journal = {\jcap},
         year = 2021,
        month = mar,
       volume = "03",
        pages = "067",
          doi = {10.1088/1475-7516/2021/03/067},
archivePrefix = {arXiv},
       eprint = {2010.09717},
 primaryClass = {astro-ph.CO},
       adsurl = {https://ui.adsabs.harvard.edu/abs/2021JCAP...03..067N}
}

@ARTICLE{Chon2004,
author = {{Chon}, Gayoung and {Challinor}, Anthony and {Prunet}, Simon and {Hivon}, Eric and {Szapudi}, Istv{\'a}n},
title = "{Fast estimation of polarization power spectra using correlation functions}",
journal = {\mnras},
year = 2004,
month = may,
volume = {350},
number = {3},
pages = {914-926},
doi = {10.1111/j.1365-2966.2004.07737.x},
archivePrefix = {arXiv},
eprint = {astro-ph/0303414},
primaryClass = {astro-ph},
adsurl = {https://ui.adsabs.harvard.edu/abs/2004MNRAS.350..914C}
}

@article{Efron1981,
  title={The jackknife estimate of variance},
  author={Efron, Bradley and Stein, Charles},
  journal={The Annals of Statistics},
  pages={586--596},
  year={1981},
  publisher={JSTOR}
}

@ARTICLE{Escoffier2016,
author = {{Escoffier}, S. and {Cousinou}, M. -C. and {Tilquin}, A. and {Pisani}, A. and {Aguichine}, A. and {de la Torre}, S. and {Ealet}, A. and {Gillard}, W. and {Jullo}, E.},
title = "{Jackknife resampling technique on mocks: an alternative method for covariance matrix estimation}",
journal = {arXiv e-prints},
year = 2016,
month = jun,
eid = {arXiv:1606.00233},
pages = {arXiv:1606.00233},
doi = {10.48550/arXiv.1606.00233},
archivePrefix = {arXiv},
eprint = {1606.00233},
primaryClass = {astro-ph.CO},
adsurl = {https://ui.adsabs.harvard.edu/abs/2016arXiv160600233E}
}

@ARTICLE{Favole2021,
author = {{Favole}, Ginevra and {Granett}, Benjamin R. and {Silva Lafaurie}, Javier and {Sapone}, Domenico},
title = "{Does jackknife scale really matter for accurate large-scale structure covariances?}",
journal = {\mnras},
year = 2021,
month = aug,
volume = {505},
number = {4},
pages = {5833-5845},
doi = {10.1093/mnras/stab1720},
archivePrefix = {arXiv},
eprint = {2004.13436},
primaryClass = {astro-ph.CO},
adsurl = {https://ui.adsabs.harvard.edu/abs/2021MNRAS.505.5833F}
}

@article{Fuchs1980,
author = {Fuchs, Henry and Kedem, Zvi M. and Naylor, Bruce F.},
title = {On visible surface generation by a priori tree structures},
year = {1980},
issue_date = {July 1980},
publisher = {Association for Computing Machinery},
address = {New York, NY, USA},
volume = {14},
number = {3},
issn = {0097-8930},
url = {https://doi.org/10.1145/965105.807481},
doi = {10.1145/965105.807481},
journal = {SIGGRAPH Comput. Graph.},
month = {jul},
pages = {124–133},
numpages = {10}
}

@ARTICLE{Friedrich2016,
author = {{Friedrich}, O. and {Seitz}, S. and {Eifler}, T.~F. and {Gruen}, D.},
title = "{Performance of internal covariance estimators for cosmic shear correlation functions}",
journal = {\mnras},
year = 2016,
month = mar,
volume = {456},
number = {3},
pages = {2662-2680},
doi = {10.1093/mnras/stv2833},
archivePrefix = {arXiv},
eprint = {1508.00895},
primaryClass = {astro-ph.CO},
adsurl = {https://ui.adsabs.harvard.edu/abs/2016MNRAS.456.2662F}
}

@ARTICLE{Gorski2005,
author = {{G{\'o}rski}, K.~M. and {Hivon}, E. and {Banday}, A.~J. and {Wandelt}, B.~D. and {Hansen}, F.~K. and {Reinecke}, M. and {Bartelmann}, M.},
title = "{HEALPix: A Framework for High-Resolution Discretization and Fast Analysis of Data Distributed on the Sphere}",
journal = {\apj},
year = 2005,
month = apr,
volume = {622},
number = {2},
pages = {759-771},
doi = {10.1086/427976},
archivePrefix = {arXiv},
eprint = {astro-ph/0409513},
primaryClass = {astro-ph},
adsurl = {https://ui.adsabs.harvard.edu/abs/2005ApJ...622..759G}
}

@ARTICLE{Hall2022,
       author = {{Hall}, Alex and {Taylor}, Andy},
        title = "{Non-Gaussian likelihood of weak lensing power spectra}",
      journal = {\prd},
         year = 2022,
        month = jun,
       volume = {105},
       number = {12},
          eid = {123527},
        pages = {123527},
          doi = {10.1103/PhysRevD.105.123527},
archivePrefix = {arXiv},
       eprint = {2202.04095},
 primaryClass = {astro-ph.CO},
       adsurl = {https://ui.adsabs.harvard.edu/abs/2022PhRvD.105l3527H}
}

@ARTICLE{Hartlap2007,
author = {{Hartlap}, J. and {Simon}, P. and {Schneider}, P.},
title = "{Why your model parameter confidences might be too optimistic. Unbiased estimation of the inverse covariance matrix}",
journal = {\aap},
year = 2007,
month = mar,
volume = {464},
number = {1},
pages = {399-404},
doi = {10.1051/0004-6361:20066170},
archivePrefix = {arXiv},
eprint = {astro-ph/0608064},
primaryClass = {astro-ph},
adsurl = {https://ui.adsabs.harvard.edu/abs/2007A&A...464..399H}
}

@ARTICLE{Heymans2021,
author = {{Heymans}, Catherine and {Tr{\"o}ster}, Tilman and {Asgari}, Marika and {Blake}, Chris and {Hildebrandt}, Hendrik and {Joachimi}, Benjamin and {Kuijken}, Konrad and {Lin}, Chieh-An and {S{\'a}nchez}, Ariel G. and {van den Busch}, Jan Luca and {Wright}, Angus H. and {Amon}, Alexandra and {Bilicki}, Maciej and {de Jong}, Jelte and {Crocce}, Martin and {Dvornik}, Andrej and {Erben}, Thomas and {Fortuna}, Maria Cristina and {Getman}, Fedor and {Giblin}, Benjamin and {Glazebrook}, Karl and {Hoekstra}, Henk and {Joudaki}, Shahab and {Kannawadi}, Arun and {K{\"o}hlinger}, Fabian and {Lidman}, Chris and {Miller}, Lance and {Napolitano}, Nicola R. and {Parkinson}, David and {Schneider}, Peter and {Shan}, HuanYuan and {Valentijn}, Edwin A. and {Verdoes Kleijn}, Gijs and {Wolf}, Christian},
title = "{KiDS-1000 Cosmology: Multi-probe weak gravitational lensing and spectroscopic galaxy clustering constraints}",
journal = {\aap},
year = 2021,
month = feb,
volume = {646},
eid = {A140},
pages = {A140},
doi = {10.1051/0004-6361/202039063},
archivePrefix = {arXiv},
eprint = {2007.15632},
primaryClass = {astro-ph.CO},
adsurl = {https://ui.adsabs.harvard.edu/abs/2021A&A...646A.140H}
}

@ARTICLE{Hikage2019,
author = {{Hikage}, Chiaki and {Oguri}, Masamune and {Hamana}, Takashi and {More}, Surhud and {Mandelbaum}, Rachel and {Takada}, Masahiro and {K{\"o}hlinger}, Fabian and {Miyatake}, Hironao and {Nishizawa}, Atsushi J. and {Aihara}, Hiroaki and {Armstrong}, Robert and {Bosch}, James and {Coupon}, Jean and {Ducout}, Anne and {Ho}, Paul and {Hsieh}, Bau-Ching and {Komiyama}, Yutaka and {Lanusse}, Fran{\c{c}}ois and {Leauthaud}, Alexie and {Lupton}, Robert H. and {Medezinski}, Elinor and {Mineo}, Sogo and {Miyama}, Shoken and {Miyazaki}, Satoshi and {Murata}, Ryoma and {Murayama}, Hitoshi and {Shirasaki}, Masato and {Sif{\'o}n}, Crist{\'o}bal and {Simet}, Melanie and {Speagle}, Joshua and {Spergel}, David N. and {Strauss}, Michael A. and {Sugiyama}, Naoshi and {Tanaka}, Masayuki and {Utsumi}, Yousuke and {Wang}, Shiang-Yu and {Yamada}, Yoshihiko},
title = "{Cosmology from cosmic shear power spectra with Subaru Hyper Suprime-Cam first-year data}",
journal = {\pasj},
year = 2019,
month = apr,
volume = {71},
number = {2},
eid = {43},
pages = {43},
doi = {10.1093/pasj/psz010},
archivePrefix = {arXiv},
eprint = {1809.09148},
primaryClass = {astro-ph.CO},
adsurl = {https://ui.adsabs.harvard.edu/abs/2019PASJ...71...43H}
}

@article{Hotelling1931,
  title={The Generalization of Student's Ratio},
  author={Hotelling, Harold},
  journal={The Annals of Mathematical Statistics},
  volume={2},
  number={3},
  pages={360--378},
  year={1931},
  publisher={Institute of Mathematical Statistics}
}

@ARTICLE{Joachimi2017,
author = {{Joachimi}, Benjamin},
title = "{Non-linear shrinkage estimation of large-scale structure covariance}",
journal = {\mnras},
year = 2017,
month = mar,
volume = {466},
number = {1},
pages = {L83-L87},
doi = {10.1093/mnrasl/slw240},
archivePrefix = {arXiv},
eprint = {1612.00752},
primaryClass = {astro-ph.IM},
adsurl = {https://ui.adsabs.harvard.edu/abs/2017MNRAS.466L..83J}
}

@ARTICLE{Kwan2017,
author = {{Kwan}, J. and {S{\'a}nchez}, C. and {Clampitt}, J. and {Blazek}, J. and {Crocce}, M. and {Jain}, B. and {Zuntz}, J. and {Amara}, A. and {Becker}, M.~R. and {Bernstein}, G.~M. and {Bonnett}, C. and {DeRose}, J. and {Dodelson}, S. and {Eifler}, T.~F. and {Gaztanaga}, E. and {Giannantonio}, T. and {Gruen}, D. and {Hartley}, W.~G. and {Kacprzak}, T. and {Kirk}, D. and {Krause}, E. and {MacCrann}, N. and {Miquel}, R. and {Park}, Y. and {Ross}, A.~J. and {Rozo}, E. and {Rykoff}, E.~S. and {Sheldon}, E. and {Troxel}, M.~A. and {Wechsler}, R.~H. and {Abbott}, T.~M.~C. and {Abdalla}, F.~B. and {Allam}, S. and {Benoit-L{\'e}vy}, A. and {Brooks}, D. and {Burke}, D.~L. and {Rosell}, A. Carnero and {Carrasco Kind}, M. and {Cunha}, C.~E. and {D'Andrea}, C.~B. and {da Costa}, L.~N. and {Desai}, S. and {Diehl}, H.~T. and {Dietrich}, J.~P. and {Doel}, P. and {Evrard}, A.~E. and {Fernandez}, E. and {Finley}, D.~A. and {Flaugher}, B. and {Fosalba}, P. and {Frieman}, J. and {Gerdes}, D.~W. and {Gruendl}, R.~A. and {Gutierrez}, G. and {Honscheid}, K. and {James}, D.~J. and {Jarvis}, M. and {Kuehn}, K. and {Lahav}, O. and {Lima}, M. and {Maia}, M.~A.~G. and {Marshall}, J.~L. and {Martini}, P. and {Melchior}, P. and {Mohr}, J.~J. and {Nichol}, R.~C. and {Nord}, B. and {Plazas}, A.~A. and {Reil}, K. and {Romer}, A.~K. and {Roodman}, A. and {Sanchez}, E. and {Scarpine}, V. and {Sevilla-Noarbe}, I. and {Smith}, R.~C. and {Soares-Santos}, M. and {Sobreira}, F. and {Suchyta}, E. and {Swanson}, M.~E.~C. and {Tarle}, G. and {Thomas}, D. and {Vikram}, V. and {Walker}, A.~R. and {DES Collaboration}},
title = "{Cosmology from large-scale galaxy clustering and galaxy-galaxy lensing with Dark Energy Survey Science Verification data}",
journal = {\mnras},
year = 2017,
month = feb,
volume = {464},
number = {4},
pages = {4045-4062},
doi = {10.1093/mnras/stw2464},
archivePrefix = {arXiv},
eprint = {1604.07871},
primaryClass = {astro-ph.CO},
adsurl = {https://ui.adsabs.harvard.edu/abs/2017MNRAS.464.4045K}
}

@ARTICLE{Lacasa2017,
author = {{Lacasa}, Fabien and {Kunz}, Martin},
title = "{Inadequacy of internal covariance estimation for super-sample covariance}",
journal = {\aap},
year = 2017,
month = aug,
volume = {604},
eid = {A104},
pages = {A104},
doi = {10.1051/0004-6361/201730784},
archivePrefix = {arXiv},
eprint = {1703.03337},
primaryClass = {astro-ph.CO},
adsurl = {https://ui.adsabs.harvard.edu/abs/2017A&A...604A.104L}
}

@article{Ledoit2004,
title={A well-conditioned estimator for large-dimensional covariance matrices},
author={Ledoit, Olivier and Wolf, Michael},
journal={Journal of multivariate analysis},
volume={88},
number={2},
pages={365--411},
year={2004},
publisher={Elsevier}
}

@ARTICLE{Looijmans2024,
author = {{Looijmans}, Marnix J. and {Wang}, Mike Shengbo and {Beutler}, Florian},
title = "{A comparison of shrinkage estimators of the cosmological precision matrix}",
journal = {arXiv e-prints},
year = 2024,
month = feb,
eid = {arXiv:2402.13783},
pages = {arXiv:2402.13783},
doi = {10.48550/arXiv.2402.13783},
archivePrefix = {arXiv},
eprint = {2402.13783},
primaryClass = {astro-ph.CO},
adsurl = {https://ui.adsabs.harvard.edu/abs/2024arXiv240213783L}
}

@ARTICLE{Loureiro2022,
author = {{Loureiro}, A. and {Whittaker}, L. and {Spurio Mancini}, A. and {Joachimi}, B. and {Cuceu}, A. and {Asgari}, M. and {St{\"o}lzner}, B. and {Tr{\"o}ster}, T. and {Wright}, A.~H. and {Bilicki}, M. and {Dvornik}, A. and {Giblin}, B. and {Heymans}, C. and {Hildebrandt}, H. and {Shan}, H. and {Amara}, A. and {Auricchio}, N. and {Bodendorf}, C. and {Bonino}, D. and {Branchini}, E. and {Brescia}, M. and {Capobianco}, V. and {Carbone}, C. and {Carretero}, J. and {Castellano}, M. and {Cavuoti}, S. and {Cimatti}, A. and {Cledassou}, R. and {Congedo}, G. and {Conversi}, L. and {Copin}, Y. and {Corcione}, L. and {Cropper}, M. and {Da Silva}, A. and {Douspis}, M. and {Dubath}, F. and {Duncan}, C.~A.~J. and {Dupac}, X. and {Dusini}, S. and {Farrens}, S. and {Ferriol}, S. and {Fosalba}, P. and {Frailis}, M. and {Franceschi}, E. and {Fumana}, M. and {Garilli}, B. and {Gillis}, B. and {Giocoli}, C. and {Grazian}, A. and {Grupp}, F. and {Haugan}, S.~V.~H. and {Holmes}, W. and {Hormuth}, F. and {Jahnke}, K. and {K{\"u}mmel}, M. and {Kermiche}, S. and {Kiessling}, A. and {Kilbinger}, M. and {Kitching}, T. and {Kuijken}, K. and {Kunz}, M. and {Kurki-Suonio}, H. and {Ligori}, S. and {Lilje}, P.~B. and {Lloro}, I. and {Mansutti}, O. and {Marggraf}, O. and {Markovic}, K. and {Marulli}, F. and {Massey}, R. and {Meneghetti}, M. and {Meylan}, G. and {Moresco}, M. and {Morin}, B. and {Moscardini}, L. and {Munari}, E. and {Niemi}, S.~M. and {Padilla}, C. and {Paltani}, S. and {Pasian}, F. and {Pedersen}, K. and {Pettorino}, V. and {Pires}, S. and {Poncet}, M. and {Popa}, L. and {Raison}, F. and {Rhodes}, J. and {Rix}, H. and {Roncarelli}, M. and {Saglia}, R. and {Schneider}, P. and {Secroun}, A. and {Serrano}, S. and {Sirignano}, C. and {Sirri}, G. and {Stanco}, L. and {Starck}, J.~L. and {Tallada-Cresp{\'\i}}, P. and {Taylor}, A.~N. and {Tereno}, I. and {Toledo-Moreo}, R. and {Torradeflot}, F. and {Valentijn}, E.~A. and {Wang}, Y. and {Welikala}, N. and {Weller}, J. and {Zamorani}, G. and {Zoubian}, J. and {Andreon}, S. and {Baldi}, M. and {Camera}, S. and {Farinelli}, R. and {Polenta}, G. and {Tessore}, N.},
title = "{KiDS and Euclid: Cosmological implications of a pseudo angular power spectrum analysis of KiDS-1000 cosmic shear tomography}",
journal = {\aap},
year = 2022,
month = sep,
volume = {665},
eid = {A56},
pages = {A56},
doi = {10.1051/0004-6361/202142481},
archivePrefix = {arXiv},
eprint = {2110.06947},
primaryClass = {astro-ph.CO},
adsurl = {https://ui.adsabs.harvard.edu/abs/2022A&A...665A..56L}
}

@ARTICLE{Mohammad2022,
author = {{Mohammad}, Faizan G. and {Percival}, Will J.},
title = "{Creating jackknife and bootstrap estimates of the covariance matrix for the two-point correlation function}",
journal = {\mnras},
year = 2022,
month = jul,
volume = {514},
number = {1},
pages = {1289-1301},
doi = {10.1093/mnras/stac1458},
archivePrefix = {arXiv},
eprint = {2109.07071},
primaryClass = {astro-ph.CO},
adsurl = {https://ui.adsabs.harvard.edu/abs/2022MNRAS.514.1289M}
}

@ARTICLE{Percival2022,
       author = {{Percival}, Will J. and {Friedrich}, Oliver and {Sellentin}, Elena and {Heavens}, Alan},
        title = "{Matching Bayesian and frequentist coverage probabilities when using an approximate data covariance matrix}",
      journal = {\mnras},
         year = 2022,
        month = mar,
       volume = {510},
       number = {3},
        pages = {3207-3221},
          doi = {10.1093/mnras/stab3540},
archivePrefix = {arXiv},
       eprint = {2108.10402},
 primaryClass = {astro-ph.IM},
       adsurl = {https://ui.adsabs.harvard.edu/abs/2022MNRAS.510.3207P}
}

@ARTICLE{Planck2020,
author = {{Planck Collaboration} and {Aghanim}, N. and {Akrami}, Y. and {Ashdown}, M. and {Aumont}, J. and {Baccigalupi}, C. and {Ballardini}, M. and {Banday}, A.~J. and {Barreiro}, R.~B. and {Bartolo}, N. and {Basak}, S. and {Battye}, R. and {Benabed}, K. and {Bernard}, J. -P. and {Bersanelli}, M. and {Bielewicz}, P. and {Bock}, J.~J. and {Bond}, J.~R. and {Borrill}, J. and {Bouchet}, F.~R. and {Boulanger}, F. and {Bucher}, M. and {Burigana}, C. and {Butler}, R.~C. and {Calabrese}, E. and {Cardoso}, J. -F. and {Carron}, J. and {Challinor}, A. and {Chiang}, H.~C. and {Chluba}, J. and {Colombo}, L.~P.~L. and {Combet}, C. and {Contreras}, D. and {Crill}, B.~P. and {Cuttaia}, F. and {de Bernardis}, P. and {de Zotti}, G. and {Delabrouille}, J. and {Delouis}, J. -M. and {Di Valentino}, E. and {Diego}, J.~M. and {Dor{\'e}}, O. and {Douspis}, M. and {Ducout}, A. and {Dupac}, X. and {Dusini}, S. and {Efstathiou}, G. and {Elsner}, F. and {En{\ss}lin}, T.~A. and {Eriksen}, H.~K. and {Fantaye}, Y. and {Farhang}, M. and {Fergusson}, J. and {Fernandez-Cobos}, R. and {Finelli}, F. and {Forastieri}, F. and {Frailis}, M. and {Fraisse}, A.~A. and {Franceschi}, E. and {Frolov}, A. and {Galeotta}, S. and {Galli}, S. and {Ganga}, K. and {G{\'e}nova-Santos}, R.~T. and {Gerbino}, M. and {Ghosh}, T. and {Gonz{\'a}lez-Nuevo}, J. and {G{\'o}rski}, K.~M. and {Gratton}, S. and {Gruppuso}, A. and {Gudmundsson}, J.~E. and {Hamann}, J. and {Handley}, W. and {Hansen}, F.~K. and {Herranz}, D. and {Hildebrandt}, S.~R. and {Hivon}, E. and {Huang}, Z. and {Jaffe}, A.~H. and {Jones}, W.~C. and {Karakci}, A. and {Keih{\"a}nen}, E. and {Keskitalo}, R. and {Kiiveri}, K. and {Kim}, J. and {Kisner}, T.~S. and {Knox}, L. and {Krachmalnicoff}, N. and {Kunz}, M. and {Kurki-Suonio}, H. and {Lagache}, G. and {Lamarre}, J. -M. and {Lasenby}, A. and {Lattanzi}, M. and {Lawrence}, C.~R. and {Le Jeune}, M. and {Lemos}, P. and {Lesgourgues}, J. and {Levrier}, F. and {Lewis}, A. and {Liguori}, M. and {Lilje}, P.~B. and {Lilley}, M. and {Lindholm}, V. and {L{\'o}pez-Caniego}, M. and {Lubin}, P.~M. and {Ma}, Y. -Z. and {Mac{\'\i}as-P{\'e}rez}, J.~F. and {Maggio}, G. and {Maino}, D. and {Mandolesi}, N. and {Mangilli}, A. and {Marcos-Caballero}, A. and {Maris}, M. and {Martin}, P.~G. and {Martinelli}, M. and {Mart{\'\i}nez-Gonz{\'a}lez}, E. and {Matarrese}, S. and {Mauri}, N. and {McEwen}, J.~D. and {Meinhold}, P.~R. and {Melchiorri}, A. and {Mennella}, A. and {Migliaccio}, M. and {Millea}, M. and {Mitra}, S. and {Miville-Desch{\^e}nes}, M. -A. and {Molinari}, D. and {Montier}, L. and {Morgante}, G. and {Moss}, A. and {Natoli}, P. and {N{\o}rgaard-Nielsen}, H.~U. and {Pagano}, L. and {Paoletti}, D. and {Partridge}, B. and {Patanchon}, G. and {Peiris}, H.~V. and {Perrotta}, F. and {Pettorino}, V. and {Piacentini}, F. and {Polastri}, L. and {Polenta}, G. and {Puget}, J. -L. and {Rachen}, J.~P. and {Reinecke}, M. and {Remazeilles}, M. and {Renzi}, A. and {Rocha}, G. and {Rosset}, C. and {Roudier}, G. and {Rubi{\~n}o-Mart{\'\i}n}, J.~A. and {Ruiz-Granados}, B. and {Salvati}, L. and {Sandri}, M. and {Savelainen}, M. and {Scott}, D. and {Shellard}, E.~P.~S. and {Sirignano}, C. and {Sirri}, G. and {Spencer}, L.~D. and {Sunyaev}, R. and {Suur-Uski}, A. -S. and {Tauber}, J.~A. and {Tavagnacco}, D. and {Tenti}, M. and {Toffolatti}, L. and {Tomasi}, M. and {Trombetti}, T. and {Valenziano}, L. and {Valiviita}, J. and {Van Tent}, B. and {Vibert}, L. and {Vielva}, P. and {Villa}, F. and {Vittorio}, N. and {Wandelt}, B.~D. and {Wehus}, I.~K. and {White}, M. and {White}, S.~D.~M. and {Zacchei}, A. and {Zonca}, A.},
title = "{Planck 2018 results. VI. Cosmological parameters}",
journal = {\aap},
year = 2020,
month = sep,
volume = {641},
eid = {A6},
pages = {A6},
doi = {10.1051/0004-6361/201833910},
archivePrefix = {arXiv},
eprint = {1807.06209},
primaryClass = {astro-ph.CO},
adsurl = {https://ui.adsabs.harvard.edu/abs/2020A&A...641A...6P}
}

@ARTICLE{Pope2008,
author = {{Pope}, Adrian C. and {Szapudi}, Istv{\'a}n},
title = "{Shrinkage estimation of the power spectrum covariance matrix}",
journal = {\mnras},
year = 2008,
month = sep,
volume = {389},
number = {2},
pages = {766-774},
doi = {10.1111/j.1365-2966.2008.13561.x},
archivePrefix = {arXiv},
eprint = {0711.2509},
primaryClass = {astro-ph},
adsurl = {https://ui.adsabs.harvard.edu/abs/2008MNRAS.389..766P}
}

@ARTICLE{Norberg2009,
author = {{Norberg}, P. and {Baugh}, C.~M. and {Gazta{\~n}aga}, E. and {Croton}, D.~J.},
title = "{Statistical analysis of galaxy surveys - I. Robust error estimation for two-point clustering statistics}",
journal = {\mnras},
year = 2009,
month = jun,
volume = {396},
number = {1},
pages = {19-38},
doi = {10.1111/j.1365-2966.2009.14389.x},
archivePrefix = {arXiv},
eprint = {0810.1885},
primaryClass = {astro-ph},
adsurl = {https://ui.adsabs.harvard.edu/abs/2009MNRAS.396...19N}
}

@article{Schafer2005,
title={A shrinkage approach to large-scale covariance matrix estimation and implications for functional genomics},
author={Sch{\"a}fer, Juliane and Strimmer, Korbinian},
journal={Statistical applications in genetics and molecular biology},
volume={4},
number={1},
year={2005},
publisher={De Gruyter}
}

@ARTICLE{Simpson2016,
author = {{Simpson}, F. and {Blake}, C. and {Peacock}, J.~A. and {Baldry}, I.~K. and {Bland-Hawthorn}, J. and {Heavens}, A.~F. and {Heymans}, C. and {Loveday}, J. and {Norberg}, P.},
title = "{Galaxy and mass assembly: Redshift space distortions from the clipped galaxy field}",
journal = {\prd},
year = 2016,
month = jan,
volume = {93},
number = {2},
eid = {023525},
pages = {023525},
doi = {10.1103/PhysRevD.93.023525},
archivePrefix = {arXiv},
eprint = {1505.03865},
primaryClass = {astro-ph.CO},
adsurl = {https://ui.adsabs.harvard.edu/abs/2016PhRvD..93b3525S}
}

@ARTICLE{Szapudi2001,
author = {{Szapudi}, Istv{\'a}n and {Prunet}, Simon and {Pogosyan}, Dmitry and {Szalay}, Alexander S. and {Bond}, J. Richard},
title = "{Fast Cosmic Microwave Background Analyses via Correlation Functions}",
journal = {\apjl},
year = 2001,
month = feb,
volume = {548},
number = {2},
pages = {L115-L118},
doi = {10.1086/319105},
adsurl = {https://ui.adsabs.harvard.edu/abs/2001ApJ...548L.115S}
}

@ARTICLE{Ross2017,
author = {{Ross}, Ashley J. and {Beutler}, Florian and {Chuang}, Chia-Hsun and {Pellejero-Ibanez}, Marcos and {Seo}, Hee-Jong and {Vargas-Maga{\~n}a}, Mariana and {Cuesta}, Antonio J. and {Percival}, Will J. and {Burden}, Angela and {S{\'a}nchez}, Ariel G. and {Grieb}, Jan Niklas and {Reid}, Beth and {Brownstein}, Joel R. and {Dawson}, Kyle S. and {Eisenstein}, Daniel J. and {Ho}, Shirley and {Kitaura}, Francisco-Shu and {Nichol}, Robert C. and {Olmstead}, Matthew D. and {Prada}, Francisco and {Rodr{\'\i}guez-Torres}, Sergio A. and {Saito}, Shun and {Salazar-Albornoz}, Salvador and {Schneider}, Donald P. and {Thomas}, Daniel and {Tinker}, Jeremy and {Tojeiro}, Rita and {Wang}, Yuting and {White}, Martin and {Zhao}, Gong-bo},
title = "{The clustering of galaxies in the completed SDSS-III Baryon Oscillation Spectroscopic Survey: observational systematics and baryon acoustic oscillations in the correlation function}",
journal = {\mnras},
year = 2017,
month = jan,
volume = {464},
number = {1},
pages = {1168-1191},
doi = {10.1093/mnras/stw2372},
archivePrefix = {arXiv},
eprint = {1607.03145},
primaryClass = {astro-ph.CO},
adsurl = {https://ui.adsabs.harvard.edu/abs/2017MNRAS.464.1168R}
}

@ARTICLE{Sellentin2016,
author = {{Sellentin}, Elena and {Heavens}, Alan F.},
title = "{Parameter inference with estimated covariance matrices}",
journal = {\mnras},
year = 2016,
month = feb,
volume = {456},
number = {1},
pages = {L132-L136},
doi = {10.1093/mnrasl/slv190},
archivePrefix = {arXiv},
eprint = {1511.05969},
primaryClass = {astro-ph.CO},
adsurl = {https://ui.adsabs.harvard.edu/abs/2016MNRAS.456L.132S}
}

@ARTICLE{Shirasaki2017,
author = {{Shirasaki}, Masato and {Takada}, Masahiro and {Miyatake}, Hironao and {Takahashi}, Ryuichi and {Hamana}, Takashi and {Nishimichi}, Takahiro and {Murata}, Ryoma},
title = "{Robust covariance estimation of galaxy-galaxy weak lensing: validation and limitation of jackknife covariance}",
journal = {\mnras},
year = 2017,
month = sep,
volume = {470},
number = {3},
pages = {3476-3496},
doi = {10.1093/mnras/stx1477},
archivePrefix = {arXiv},
eprint = {1607.08679},
primaryClass = {astro-ph.CO},
adsurl = {https://ui.adsabs.harvard.edu/abs/2017MNRAS.470.3476S}
}

@ARTICLE{Tessore2023,
author = {{Tessore}, Nicolas and {Loureiro}, Arthur and {Joachimi}, Benjamin and {von Wietersheim-Kramsta}, Maximilian and {Jeffrey}, Niall},
title = "{GLASS: Generator for Large Scale Structure}",
journal = {The Open Journal of Astrophysics},
year = 2023,
month = mar,
volume = {6},
eid = {11},
pages = {11},
doi = {10.21105/astro.2302.01942},
archivePrefix = {arXiv},
eprint = {2302.01942},
primaryClass = {astro-ph.CO},
adsurl = {https://ui.adsabs.harvard.edu/abs/2023OJAp....6E..11T}
}

@ARTICLE{tegmark97,
       author = {{Tegmark}, Max and {Taylor}, Andy N. and {Heavens}, Alan F.},
        title = "{Karhunen-Lo{\`e}ve Eigenvalue Problems in Cosmology: How Should We Tackle Large Data Sets?}",
      journal = {\apj},
         year = 1997,
        month = may,
       volume = {480},
       number = {1},
        pages = {22-35},
          doi = {10.1086/303939},
archivePrefix = {arXiv},
       eprint = {astro-ph/9603021},
 primaryClass = {astro-ph},
       adsurl = {https://ui.adsabs.harvard.edu/abs/1997ApJ...480...22T}
}

%

\begin{appendix}
\onecolumn 
  
\section{Mask correction}
\label{app_mask_correction}

\begin{figure*}[htbp!]
\centering
\includegraphics[width=\textwidth]{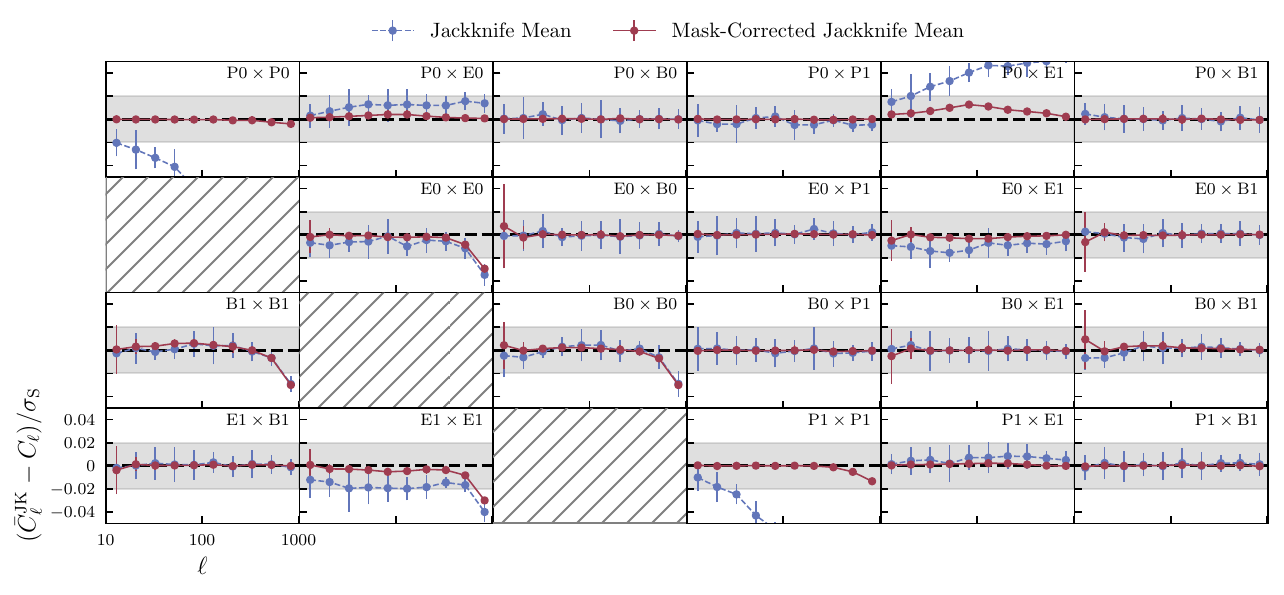}
\caption{The jackknife mean ($\bar{\vec{C}}_{\ell}^{\mathrm{JK}}$) before (in blue) and after (in red) correcting for the jackknife footprint is shown in comparison to the original angular power spectra computed for the entire footprint $\vec{C}_{\ell}$. The lines and error bars represent the mean and spread across ten realisations. The difference is scaled as a function of the sample covariance standard deviation ($\sigma_{\mathrm{S}}$ i.e.~the square root of the diagonals $\tens{C}_{\mathrm{S}}$) with grey bands representing 2\% standard deviations from the $\vec{C}_{\ell}$ for the entire footprint. The figure is divided into subplots for each angular auto and cross spectra pair, where $P$ denotes overdensity, and $E$- and $B$-modes from weak lensing and the numbers $0$ and $1$ denote the tomographic bin. The sky correction is important for correcting the auto spectra for angular clustering but otherwise the correction has a minimal effect with the exception of a few cross-spectra (in particular $\mathrm{P}0\times \mathrm{E}1$).}
\label{fig_jackknife_mean_mask}
\end{figure*}

\begin{figure*}[htbp!]
\centering
\includegraphics[width=\textwidth]{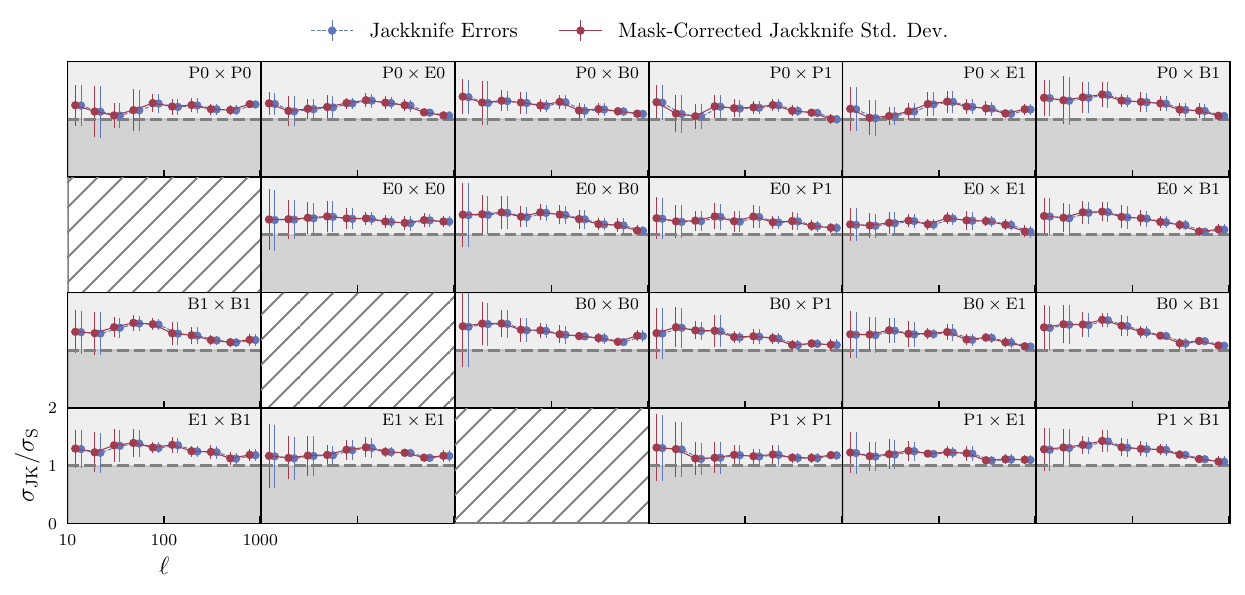}
\caption{The standard deviation for jackknife estimates of the covariance $\sigma_{\mathrm{JK}}$ (dashed horizontal grey line) before (in blue) and after (in red) correcting for the jackknife footprint is shown in comparison to the sample covariance standard deviation $\sigma_{\mathrm{S}}$. The lines and error bars represent the mean and spread across ten realisations. The standard deviation for the jackknife covariance is unaffected by the sky correction.}
\label{fig_jackknife_cov_mask}
\end{figure*}

In Fig.~\ref{fig_jackknife_mean_mask} we show that the partial-sky correction (measured for $N_{\mathrm{JK}}=296$) removes a bias in the jackknife sample mean. For the auto-spectra, the bias can be quite significant, but the correction generally mitigates this potential systematic error, leaving the jackknife mean to within $2\%$ of the standard deviation of the original $\vec{C}_{\ell}$. However, there is a small systematic drop in the amplitudes at large $\ell$ for the auto-spectra which is not currently understood; however this offset is only seen in the mean (and is small, a $<2\%$ offset) and does not impact the covariance estimate, discussed later. In Fig.~\ref{fig_jackknife_cov_mask} we show the sky correction does not affect the covariance, which remains biased high with and without partial sky correction; this is a general property of covariance estimates from jackknife resampling \citep{Efron1981}.

\section{Sensitivity to jackknife partition number}
\label{app_Njk}

\begin{figure*}[htbp!]
\centering
\includegraphics[width=\textwidth]{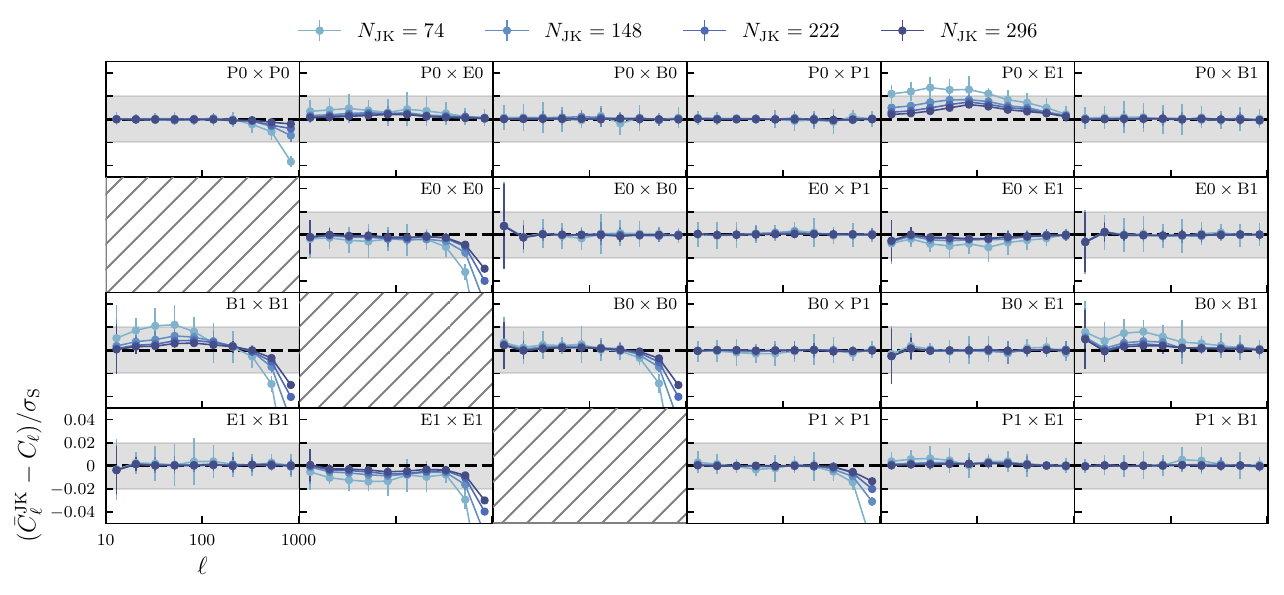}
\caption{The sky-corrected jackknife mean $\bar{\vec{C}}_{\ell}^{\mathrm{JK}}$ is plotted as a function of the number of jackknife samples $N_{\mathrm{JK}}$. The lines and error bars represent the mean and spread across ten realisations. The difference is scaled as a function of the sample covariance standard deviation ($\sigma_{\mathrm{S}}$ i.e.~the square root of the diagonals of $\tens{C}_{\mathrm{S}}$). Increasing $N_{\mathrm{JK}}$ improves the bias of the auto-spectra particularly at large $\ell$, but except for a few cases (in particular $\mathrm{P}0\times\mathrm{E}1$), cross-spectra are generally unbiased for all $N_{\mathrm{JK}}$.
}
\label{fig_jackknife_mean_vs_njk}
\end{figure*}

\begin{figure*}[htbp!]
\centering
\includegraphics[width=\textwidth]{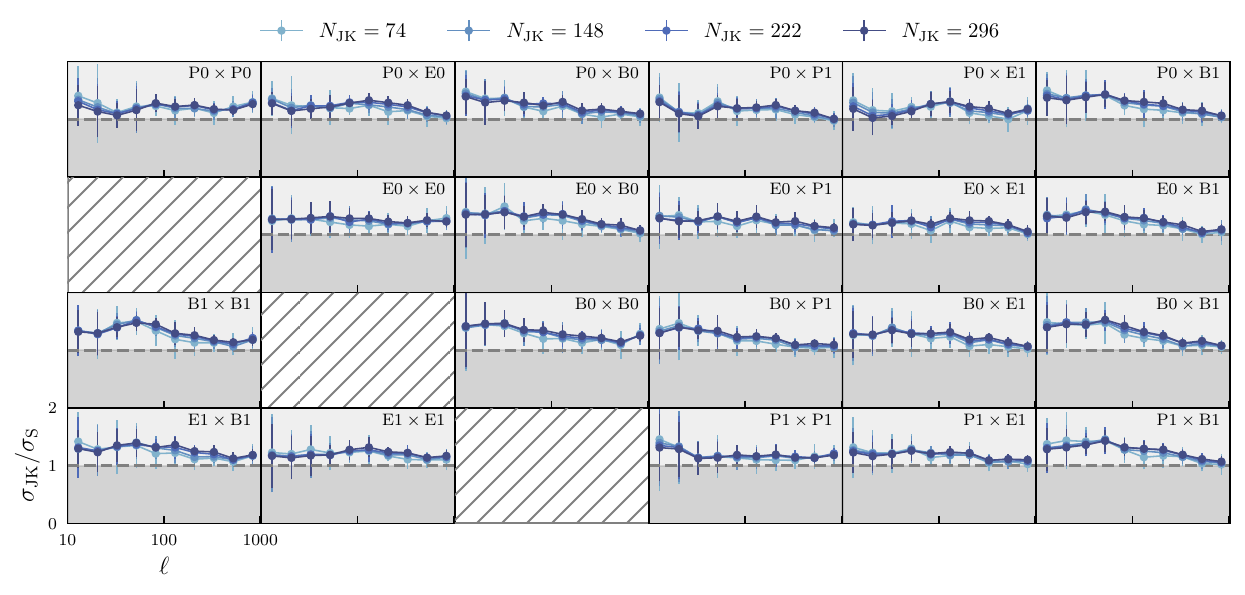}
\caption{The standard deviation for jackknife estimates of the covariance $\sigma_{\mathrm{JK}}$ is plotted as a function of the number of jackknife samples $N_{\mathrm{JK}}$ in comparison to the sample covariance standard deviation $\sigma_{\mathrm{S}}$. The lines and error bars represent the mean and spread across ten realisations. The number of jackknife samples does not appear to be changing the estimates of the diagonal components of the jackknife covariance.}
\label{fig_jackknife_cov_vs_njk}
\end{figure*}

In Fig.~\ref{fig_jackknife_mean_vs_njk} we test the relation between the partial sky-corrected jackknife mean and the number of jackknife partitions $N_{\mathrm{JK}}$, showing that on the whole the jackknife angular power spectra are unbiased, with the exception of the auto-spectra which show a damping at high $\ell$. This damping reduces as the number of jackknife samples increases. The exact cause of this issue is unclear but is small, no more than $5\%$ of the standard deviation. On the other hand, the cross-spectra are generally unbiased with the exception of the cross-spectra $\mathrm{P0\times E1}$ (i.e.~the correlation between the overdensity for the first tomographic bin with the $E$-modes of the second bin) which is biased high; again this is a small bias of no more than a few percent in comparison to the overall standard deviation. In Fig.~\ref{fig_jackknife_cov_vs_njk} we compare the diagonal components of the covariance as a function of the number of jackknife samples, showing that the jackknife standard deviation is robust to changes in the number of jackknife samples. In all cases we see that the jackknife standard deviation remains biased high in comparison to the sample covariance. Although the diagonal elements of the covariance are not dependent on $N_{\mathrm{JK}}$ for the values of $N_{\mathrm{JK}}$ studied, they will presumably be dependent for smaller $N_{\mathrm{JK}}$.

\section{Shrinkage method comparison}
\label{app_shrinkage}

\begin{figure*}[htbp]
\centering
\includegraphics[width=\textwidth]{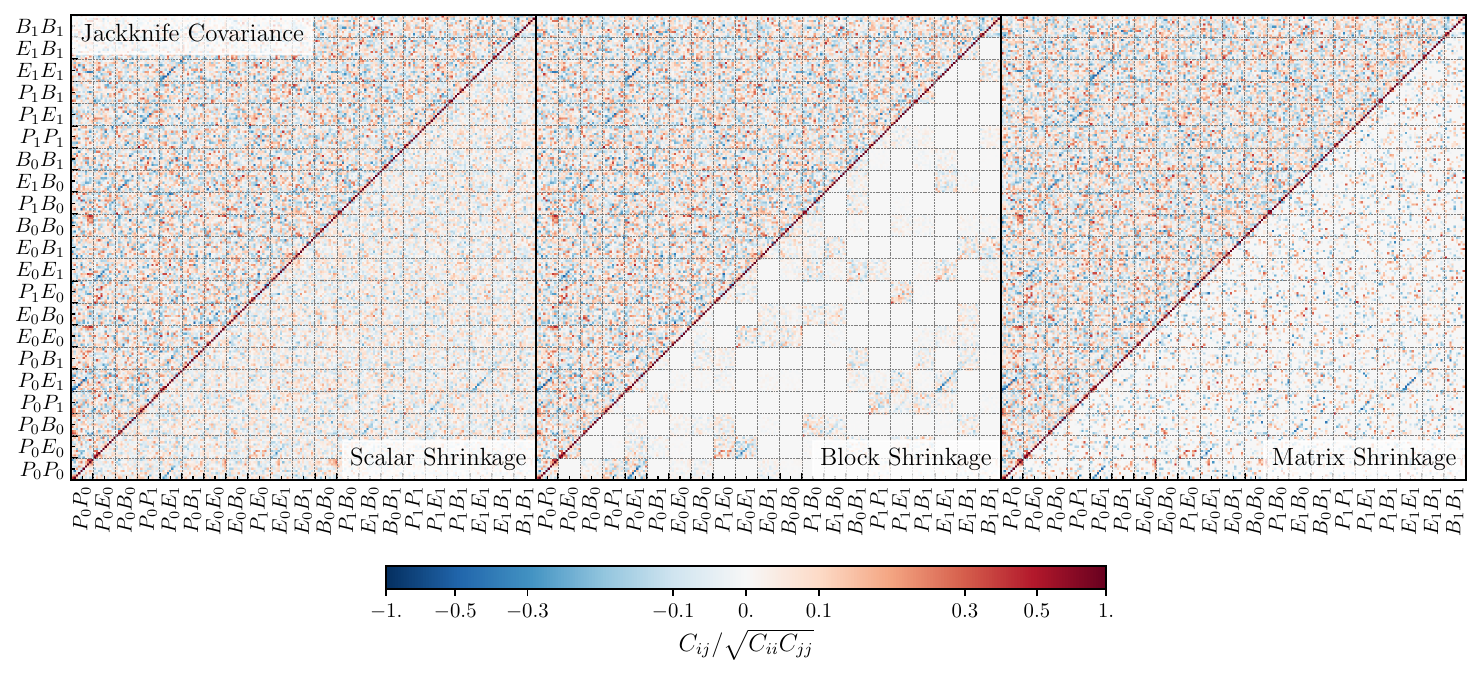}
\caption{The correlation matrix of the jackknife covariance with $N_{\mathrm{JK}}=74$ (top left) is compared to the three different shrinkage methods: scalar shrinkage (left), block shrinkage (middle) and matrix shrinkage (right). See Fig.~\ref{fig_jackknife_diagcov_vs_njk} for details on the subplot layout. In all case shrinkage dampens the noise-level of the covariance, however for matrix shrinkage this coupled with spurious noise.}
\label{fig_cov_shrinkage_method}
\end{figure*}

In Fig.~\ref{fig_cov_shrinkage_method} we compare the jackknife covariance from $N_{\mathrm{JK}}=74$ with scalar shrinkage, block shrinkage, and matrix shrinkage. The methods differ in the way the shrinkage intensity is computed; for scalar shrinkage the shrinkage intensity $\lambda$ is a single number estimated from the entire covariance matrix and is therefore the best constrained, for block shrinkage $\lambda$ is evaluated for each block, and for matrix shrinkage $\lambda$ is evaluated separately for each element of the matrix. The figure shows that in all cases the noise levels of the covariance are significantly reduced with linear shrinkage. However, for matrix shrinkage the noise levels appear to be enhanced and preserved for some elements, while block shrinkage dampens (i.e. moves towards zero) entire blocks of the covariance.

\begin{figure}[htbp!]
\centering
\includegraphics[width=0.5\columnwidth]{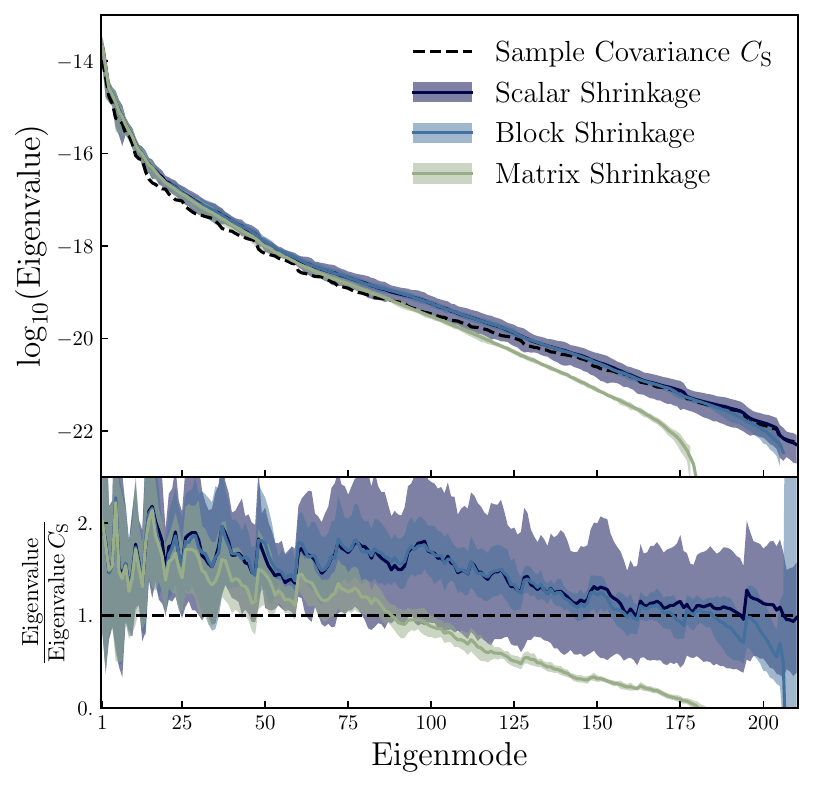}
\caption{The eigenspectrum for three shrinkage methods is compared to the sample covariance (dashed black line). Linear shrinkage is performed on the jackknife covariance with $N_{\mathrm{JK}}=74$, using scalar, block, and matrix shrinkage.  In the bottom subplot we show the ratio with respect to the sample covariance. The solid lines represent the mean and the envelopes the spread (i.e.~95\% confidence interval) from ten realisations. Scalar shrinkage, with a single value for the shrinkage intensity, produces a covariance matrix that is non-singular and follows the sample covariance eigenspectrum for all eigenmodes, albeit with larger scatter. Block shrinkage is sometimes singular (eigenmodes $>200$ are sometimes zero) while matrix shrinkage is always singular (eigenmodes $>180$ are always zero). All methods are biased high for small eigenmodes, due to a bias towards high diagonals in the jackknife covariance which are not altered since we shrink towards a Gaussian correlation prediction.}
\label{fig_eig_shrinkage_method}
\end{figure}

In Fig.~\ref{fig_eig_shrinkage_method} the eigenspectrum for each shrinkage method is compared to the eigenspectrum of the sample covariance. We show that in all cases shrinkage improves the off-diagonal structure of the covariance, with scalar shrinkage ensuring a non-singular well-conditioned matrix in all cases, block shrinkage is generally well conditioned but is singular in some cases (eigenmodes $>\!200$ are close to zero), while matrix shrinkage (although better conditioned than the original jackknife covariance shown in Fig.~\ref{fig_jackknife_eig_vs_njk}) is singular for eigenmodes $>\!180$ and starts to dip below the sample covariance at eigenmodes $>\!100$. Methods for which the shrinkage intensity is computed from fewer summations, and which are therefore more localised (such as block and matrix shrinkage), have smaller uncertainties, while scalar shrinkage has fairly large uncertainties. This is likely due to the weaker damping of noise in scalar shrinkage in comparison to block and matrix shrinkage. However, since scalar shrinkage produces non-singular covariances in all cases, it is the most reliable and well conditioned. For this reason, in the rest of the paper we will only consider scalar shrinkage but future studies could consider improving the reliability of the other methods.

\section{Shrinkage sensitivity to jackknife partition number}
\label{app_shrinkage_njk}

\begin{figure}[htbp!]
\centering
\includegraphics[width=0.5\columnwidth]{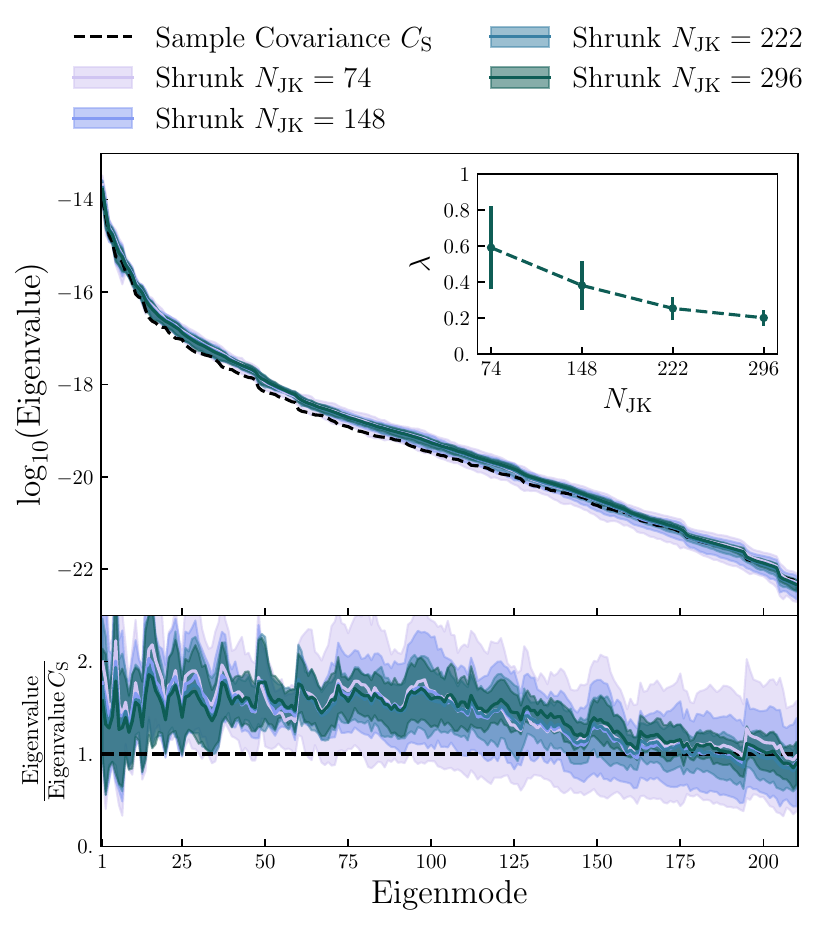}
\caption{The eigenspectrum for linear shrinkage as a function of $N_{\mathrm{JK}}$ is compared to the eigenspectrum of the sample covariance (black dashed line). In the bottom subplot we
show the ratio with respect to the sample covariance. The solid lines represents the mean and the envelopes the spread (i.e.~95\% confidence interval) from ten realisations. In the inset subplot we show the mean and standard deviation for the scalar shrinkage intensity $\lambda$ as a function of $N_{\mathrm{JK}}$, showing that increasing $N_{\mathrm{JK}}$ leads to a smaller (non-zero) and better constrained $\lambda$. This more constrained $\lambda$ leads to a more precise covariance, for instance the eigenspectrum shows less spread between realisations.}
\label{fig_eig_shrinkage_vs_njk}
\end{figure}

In Fig.~\ref{fig_eig_shrinkage_vs_njk} we compare the eigenspectrum of the shrunk covariance as a function of $N_{\mathrm{JK}}$, showing that increasing the number of jackknife samples has a diminishing impact on the jackknife covariance estimate. However, the shrinkage intensity decreases and becomes better constrained (smaller variance) when $N_{\mathrm{JK}}$ increases. The increase in $N_{\mathrm{JK}}$ yields no noticeable difference in the eigenspectrum mean but improves the variance seen in the ten realisations and therefore improves the precision. We still see a bias in the eigenvalues which appears to originate from the bias in the standard deviation shown previously. Note that, since we shrink towards the correlation matrix of the Gaussian covariance expectation, the diagonals of the covariance will not change, meaning the bias towards higher standard deviations shown in Figs.~\ref{fig_jackknife_cov_mask} and \ref{fig_jackknife_cov_vs_njk} remains.

\end{appendix}

\end{document}